\documentclass[10pt]{article}

\usepackage[english]{babel}
\usepackage[utf8]{inputenc}
\usepackage[round]{natbib} 

\usepackage[nohead]{geometry} 
\usepackage{setspace} 

\usepackage{hyperref} 
\usepackage{xpatch}
\usepackage{url} 

\usepackage{amsmath} 
\usepackage{amssymb} 
\usepackage{amsfonts} 
\usepackage{latexsym} 
\usepackage{amsthm} 
\usepackage{mathrsfs} 
\usepackage{mathtools} 

\usepackage{booktabs} 
\usepackage{multirow} 
\usepackage{multicol} 
\usepackage{enumitem} 

\usepackage{graphics} 
\usepackage{caption} 
\usepackage{comment} 
\usepackage{accents}

\usepackage{authblk} 
\usepackage[table,dvipsnames]{xcolor} 

\hypersetup{
	colorlinks = {true}, 
	linkcolor = {red}, 
	urlcolor = {black}, 
	citecolor = {blue}
}

\makeatletter
\xpatchcmd\NAT@citex
{%
	\@citea\NAT@hyper@{%
		\NAT@nmfmt{\NAT@nm}%
		\hyper@natlinkbreak{\NAT@aysep\NAT@spacechar}{\@citeb\@extra@b@citeb}%
		\NAT@date
	}%
}
{%
	\@citea
	\NAT@nmfmt{\NAT@nm}%
	\NAT@aysep\NAT@spacechar
	\NAT@hyper@{\NAT@date}%
}
{}{}
\xpatchcmd\NAT@citex
{%
	\@citea\NAT@hyper@{%
		\NAT@nmfmt{\NAT@nm}%
		\hyper@natlinkbreak{\NAT@spacechar\NAT@@open\if*#1*\else#1\NAT@spacechar\fi}%
		{\@citeb\@extra@b@citeb}%
		\NAT@date
	}%
}
{
	\@citea
	\NAT@nmfmt{\NAT@nm}%
	\NAT@spacechar\NAT@@open\if*#1*\else#1\NAT@spacechar\fi
	\NAT@hyper@{\NAT@date}%
}
{}{}
\makeatother	

\usepackage[commentsnumbered, ruled, vlined]{algorithm2e}

\SetCommentSty{myfont}

\SetKwInput{KwData}{Data}
\SetKwInput{KwInput}{Input}
\SetKwInput{KwOutput}{Output}
\SetKwInput{KwInit}{Initialize}

\counterwithout{equation}{section}

\usepackage{lipsum} 
\usepackage{xargs} 
\usepackage[colorinlistoftodos, prependcaption, textsize = small]{todonotes}

\def\keywords{{\vspace{0.25cm}\noindent\textbf{Keywords:~}}}

\def\path{./utils}

\newtheorem{proposition}{Proposition}
\newtheorem{result}{Result}
\newtheorem{remark}{Remark}
\newtheorem{example}{Example}

\def\eps{\epsilon}
\def\veps{\varepsilon}

\def\vphi{\varphi}

\def\bbeta{\boldsymbol{\beta}}

\def\bdelta{\boldsymbol{\delta}}

\def\btheta{\boldsymbol{\theta}}
\def\bvtheta{\boldsymbol{\vartheta}}

\def\blambda{\boldsymbol{\lambda}}
\def\bmu{\boldsymbol{\mu}}
\def\bnu{\boldsymbol{\nu}}
\def\bxi{\boldsymbol{\xi}}

\def\brho{\boldsymbol{\rho}}

\def\bsigma{\boldsymbol{\sigma}}

\def\bomega{\boldsymbol{\omega}}

\def\bSigma{\boldsymbol{\Sigma}}

\def\bPsi{\boldsymbol{\Psi}}
\def\bOmega{\boldsymbol{\Omega}}

\def\b0{\mathbf{0}}

\def\bc{\mathbf{c}}

\def\bg{\mathbf{g}}

\def\bp{\mathbf{p}}

\def\bx{\mathbf{x}}
\def\by{\mathbf{y}}
\def\bz{\mathbf{z}}

\def\bC{\mathbf{C}}

\def\bH{\mathbf{H}}

\def\bM{\mathbf{M}}

\def\bR{\mathbf{R}}
\def\bS{\mathbf{S}}

\def\bW{\mathbf{W}}
\def\bX{\mathbf{X}}

\def\bZ{\mathbf{Z}}

\def\bsu{\boldsymbol{u}}

\def\bsx{\boldsymbol{x}}

\def\bbI{\mathbb{I}}

\def\bbN{\mathbb{N}}

\def\cB{\mathcal{B}}
\def\cC{\mathcal{C}}

\def\cH{\mathcal{H}}

\def\cO{\mathcal{O}}

\def\cQ{\mathcal{Q}}

\def\cY{\mathcal{Y}}

\def\scb{\mathsf{b}}

\def\scd{\textsc{d}}

\def\scf{\textsc{f}}

\def\tr{\textrm{\textup{r}}}

\def\d{\textrm{\textup{d}}} 
\def\H{\mathsf{H}} 

\def\N{\mathbb{N}} 
\def\R{\mathbb{R}} 
\def\C{\mathbb{C}} 

\def\digamma{\textrm{\textup{digamma}}}	

\def\argmin{\mathop{\textrm{\textup{argmin}}}}

\def\trace{\textrm{\textup{trace}}}
\def\tr{\textrm{\textup{tr}}}

\def\sign{\textrm{\textup{sign}}}

\def\stack{\mathop{\textrm{\textup{stack}}}}
\def\diag{\textrm{\textup{diag}}}

\def\blockdiag{\textrm{\textup{blockdiag}}}

\def\parrow{\;\mathop{\rightarrow}^{\textrm{\textup{p}}}\;} 

\def\E{\mathbb{E}} 
\def\Var{\mathbb{V}\textrm{\textup{ar}}} 

\def\KL{\textrm{\textup{KL}}} 

\def\N{\textrm{\textup{N}}} 
\def\IG{\textrm{\textup{IG}}} 
\def\HC{\textrm{\textup{HC}}} 

\def\Be{\textrm{\textup{Be}}} 

\definecolor{Green}{rgb}{0.0, 0.56, 0.0}
\definecolor{Azure}{RGB}{72, 68, 188}

\def\<{\langle} 
\def\>{\rangle} 

\makeatletter
\newsavebox{\@brx}
\newcommand{\llangle}[1][]{%
  \savebox{\@brx}{\(\m@th{#1\langle}\)}%
  \mathopen{\copy\@brx\mkern2mu\kern-0.9\wd\@brx\usebox{\@brx}}%
}
\newcommand{\rrangle}[1][]{%
  \savebox{\@brx}{\(\m@th{#1\rangle}\)}%
  \mathclose{\copy\@brx\mkern2mu\kern-0.9\wd\@brx\usebox{\@brx}}%
}
\makeatother

\def\dsty{\displaystyle}
\def\tsty{\textstyle}

\def\rarrow{\rightarrow}

\def\const{\textup{const}}
\def\rest{\textup{rest}}

\def\half{{\textstyle\frac{1}{2}}\,}

\def\elbo{\underline{\ell}}
\def\qsim{\sim_q}

\newcommand{\xhalf}[1]{{\textstyle\frac{#1}{2}\,}}
\newcommand{\xfrac}[2]{{\textstyle\frac{#1}{#2}\,}}

\def\SVC{{\scriptscriptstyle\mathrm{SVC}}}
\def\SVR{{\scriptscriptstyle\mathrm{SVR}}}
\def\QR{{\scriptscriptstyle\mathrm{QR}}}
\def\ER{{\scriptscriptstyle\mathrm{ER}}}

\def\HR{{\scriptscriptstyle\mathrm{HR}}}
\def\HC{{\scriptscriptstyle\mathrm{HC}}}

\def\A{{\scriptscriptstyle\mathrm{A}}}
\def\M{{\scriptscriptstyle\mathrm{M}}}
\def\C{{\scriptscriptstyle\mathrm{C}}}

\def\H{{\scriptscriptstyle{H}}}

\def\scb{{\scriptscriptstyle\mathrm{B}}}

\def\UF{{\textsc{uf}}}
\def\FF{{\textsc{ff}}}
\def\PF{{\textsc{pf}}}
\def\RF{{\textsc{rf}}}

\newcommand*{\dt}[1]{%
	\accentset{\mbox{\large\bfseries.}}{#1}}

\def\emailcristian{\href{mailto:cristian.castiglione@unibocconi.it}{cristian.castiglione@unibocconi.it}}

\title{\LARGE\bf
	Non-conjugate variational Bayes for pseudo-likelihood mixed effect models
}

\author[1]{Cristian Castiglione\thanks{Corresponding author: \emailcristian}}
\author[2]{Mauro Bernardi}

\affil[1]{
	\textit{Bocconi Institute for Data Science and Analytics, Bocconi University}\newline
	\textit{Via R\"{o}ntgen 1, 20136 Milano, Italy}.
}

\affil[2]{
	\textit{Department of Statistical Sciences, University of Padova}\newline
	\textit{Via Cesare Battisti 241, 35121 Padova, Italy}.
}

\date{}

\begin{document}

\maketitle
\date{}


\begin{abstract}
\noindent
We propose a unified, yet simple to code, non-conjugate variational Bayes algorithm for posterior approximation of generic Bayesian generalized mixed effect models. Specifically, we consider regression models identified by a linear predictor, eventually transformed using a bijective link, where the prediction misfit is measured using, possibly non-differentiable, loss functions. Examples include generalized linear models, quasi-likelihood models, and robust regression. To address the limitations of non-conjugate settings, we employ an efficient message passing optimization strategy under a Gaussian variational approximation of the posterior. The resulting algorithms automatically account for non-conjugate priors and non-smooth losses, without requiring model-specific data-augmented representations. Besides the general formulation, we provide closed-form updates for popular model specifications, including quantile regression and support vector machines. Overall, theoretical and empirical results highlight the effectiveness of the proposed method, demonstrating its computational efficiency and approximation accuracy as an alternative to existing Bayesian techniques.

\end{abstract}

\section{Introduction}%
\label{sec:introduction}

The increasing prevalence of large-scale and high-velocity data, often originating from multiple sources, presents both significant opportunities and challenges in modern Bayesian statistics.
In contemporary real-world data applications, the computational demands of Markov chain Monte Carlo simulation methods \citep{Gelman2013}, the standard approach to Bayesian inference, often exceed practical time constraints and available memory resources. 
During the past two decades, significant efforts have therefore been devoted to developing alternative estimation methods that avoid posterior simulation. 
In this regard, optimization-based algorithms have played a crucial role, offering a highly efficient approach while providing a reasonable approximation of the posterior distribution.
Within this family, key examples include the Laplace approximation \citep{Tierney1986} and its integrated nested generalization \citep{Rue2009}, as well as variational inference methods such as mean-field variational Bayes \citep{Ormerod2010}, expectation propagation \citep{Minka2013}, stochastic variational inference \citep{Hoffman2013} and black-box variational inference \citep{Ranganath2014, Kucukelbir2017}. 
These approaches are used primarily to estimate hierarchical Bayesian models with regular likelihood functions. 
Gradient-based methods, including the Laplace approximation and black-box variational inference, are hindered by the strict requirement for differentiable log-likelihoods, limiting their applicability in key problems such as quantile regression \citep{Koenker1978} and support vector machines \citep{Vapnik1998}.  In contrast, mean-field variational Bayes, expectation propagation, and stochastic variational inference are well suited for hierarchical models whose posterior distribution factorizes in the product of conditionally conjugate factors.

Approximating a posterior density becomes particularly challenging when the joint distribution lacks smoothness or conjugacy. 
In such cases, well-designed data augmentation strategies can help by introducing latent variables, allowing intricate distributions to be expressed as marginals of expanded joint models \citep{Wand2011a}. 
This data-augmentation technique often restores regularity and conjugacy in the expanded space, significantly simplifying computations.  
However, as highlighted in, e.g., \cite{Lewandowski2010}, \cite{Neville2014}, \cite{Duan2018}, and \cite{Johndrow2019}, finding a suitable stochastic representation is not always feasible and can introduce substantial computational challenges. These issues often manifest themselves as slower convergence and reduced approximation accuracy in the original parameter space, potentially offset by the advantages of data augmentation. 
Additionally, the lack of general recipes for designing augmentation strategies limits their applicability to broader inferential problems. 
Classical examples include Poisson and Gamma regression models, which lack convenient data-augmented representations. 
These challenges are even more pronounced when the parameters of interest do not identify a known probability distribution but instead emerge from a minimum-risk criterion, such as in support vector machines \citep{Vapnik1998}, quantile and expectile regression \citep{Koenker2005}. 
In such cases, the non-regular nature of the loss functions can hinder standard Bayesian approximation techniques, leading to suboptimal performance or outright failure.

This work introduces a streamlined semi-parametric variational method with a unified approach to efficiently approximate the posterior distribution in Bayesian mixed regression models \citep{McCulloch2008}. 
We focus on regression and classification models where the response variable is predicted through a linear predictor, optionally transformed via a bijective link function. To further generalize the framework, we allow the prediction misfit to be evaluated using either a loss function or a pseudo-likelihood \citep{Bissiri2016}. 
This formulation encompasses a wide range of models, including generalized linear models, generalized estimating equations \citep{McCullagh1989}, and robust mixed-effects models \citep{Richardson1997}. 
We refer to this extended class as pseudo-likelihood generalized linear mixed models (pGLMMs), which generalize classical GLMMs by removing the restriction to exponential family distributions. 
Unlike most existing methods in this field, our approach is designed to handle non-regular loss functions that are either non-differentiable or non-conjugate, where standard Bayesian approximation techniques either fail or suffer from severe drawbacks, such as slow convergence and poor approximation quality.

Using insights from well-established results from non-conjugate variational message passing \citep{Knowles2011, Wand2014, Rohde2016}, we develop efficient deterministic and stochastic optimization algorithms for variational inference in pGLMMs. 
This formulation has several advantages: it applies to a broad class of statistically relevant models, it accommodates non-differentiable loss or log-likelihood functions, and it enables efficient closed-form updates in many cases. 
More importantly, we theoretically prove that the proposed method outperforms, in Kullback-Leibler divergence, the state-of-the-art mean field variational Bayes approximations based on conjugate data-augmented representations while maintaining the same level of computational efficiency. 
An additional easily recognizable advantage of our approach is its ability to integrate the efficiency and modularity of mean-field variational Bayes \citep{Ormerod2010, Blei2017} with the flexibility of Gaussian variational approximations \citep{Wand2014}, which allows us to effectively handle parameters that do not admit conjugate distributions. 
A complete implementation of our methods is available as \texttt{Julia} package on the first author's GitHub page\footnote{Github repository: \href{https://github.com/CristianCastiglione/BayesGLMM.jl}{\texttt{CristianCastiglione/BayesGLMM.jl}}}.

Similar strategies have been explored in the past, such as semiparametric variational Bayes \citep{Rohde2016}, for various applications, including frequentist generalized linear mixed models \citep{Ormerod2012}, Bayesian generalized linear mixed models \citep{Tan2013, Tan2014}, semiparametric models for overdispersed count data \citep{Luts2015}, semiparametric heteroscedastic regression \citep{Menictas2015}, and high-dimensional semiparametric regression \citep{Wand2017}. 
Notably, many of these studies, which are based upon exponential family likelihoods or extensions thereof, can be framed as special cases of the proposed unified variational framework. 
However, our proposal extends beyond these examples, offering greater flexibility for a wider range of regression models.

The article is organized as follows. 
Section~\ref{sec:approximate_belief_updating} introduces the pGLMM family, describes the proposed variational method, characterizes its properties, and presents several examples for non-regular models, including quantile regression and support vector machines.
Section~\ref{sec:extensions} presents some extensions to improve the scalability of the proposed approach in high-dimensional problems.
Section~\ref{sec:simulation_studies} presents an extensive simulation study designed to evaluate the quality and efficiency of the proposed approximation with respect to the gold standard methods in the literature.
Section~\ref{sec:real_data_application} discusses a real data example on quantile additive regression applied to the electric power consumption data in the US.
Section~\ref{sec:discussion} is devoted to a closing discussion.

\section{Variational mixed effect models}%
\label{sec:approximate_belief_updating}

Taking a generalized Bayesian point of view \citep{Bissiri2016}, we here consider Bayesian mixed models of the form
\begin{equation}
	\label{eq:model}
	\begin{aligned}
		& y_i | \eta_i \sim \pi(y_i | \eta_i) \propto \exp\{ - \psi(y_i, \eta_i) \}, \quad \eta_i = \bx_i^\top \bbeta + \sum_{h = 1}^{\H} \bz_{ih}^\top \bsu_h, \quad i = 1, \dots, n, \\
		& \bbeta \sim \N_p \big( \b0, \sigma_\beta^2 \,\bR_\beta^{-1} \big),
		\quad \bsu_h | \sigma_h^2 \sim \N_{d_h} \big( \b0, \sigma_h^2 \bR_h^{-1} \big),
		\quad \sigma_h^2 \sim \IG(A_h, B_h), \quad h = 1, \dots, H,
	\end{aligned}
\end{equation}
where $\psi (\cdot, \cdot)$ is a continuous, typically convex, loss function, measuring the misfit between the $i$th observation and the corresponding prediction.
The linear predictor $\eta_i$ takes an additive structure accounting for the contribution of fixed and random effects, with $\bbeta \in \R^{p}$ being the vector of fixed-effect parameters corresponding to covariate vector $\bsx_i \in \R^p$, and $\bsu_h \in \R^{d_h}$ being the $h$th vector of random-effect parameters corresponding to covariate vector $\bz_{ih} \in \R^{d_h}$.
The regression parameters $\bbeta$ and $\bsu_h$ are endowed with independent Gaussian prior distributions, while the $h$th random-effect variance parameter $\sigma_h^2 > 0$ follows an Inverse-Gamma prior distribution. 
Here, $\bR_\beta$ and $\bR_h$ are non-stochastic positive semi-definite matrices, and $\sigma_\beta^2, A_h, B_h > 0$ are fixed user-specified prior parameters.
We also define $\bsu^\top = (\bsu_1^\top, \dots, \bsu_\H^\top)$ and $\bz_i^\top = (\bz_{i1}^\top, \dots, \bz_{i\H}^\top)$ as the vectors concatenating all the random-effect parameters and covariates, respectively.
Similarly, we denote by $\bc_i^\top = (\bx_i^\top, \bz_{i1}^\top, \dots, \bz_{i\H}^\top)$ the completed covariate vectors, and we introduce the design matrix $\bC$ stacking all $\bc_i^\top$ by row.

We do not assume strong differentiability conditions for $\psi(\cdot, \cdot)$, allowing us to encompass both regular and non-regular models within a unified framework, including generalized linear models, quantile regression, and support vector machines. 
In this context, \emph{generalized linear mixed models} (GLMM) based on the dispersion exponential family can be obtained by selecting an appropriate form for $\psi(\cdot, \cdot)$. 
Therefore, to highlight the distinction between traditional GLMMs and our more general specification, we refer to the model in \eqref{eq:model} as the \emph{pseudo-likelihood generalized linear mixed model} (pGLMM).

Denoting by $\btheta = (\bbeta^\top, \bsu^\top, \sigma_1^2, \dots, \sigma_\H^2)^\top$ the parameter vector indexing model \eqref{eq:model}, the generalized posterior of a pGLMM, denoted by $\pi(\btheta | \by) \propto \pi(\btheta) \,\pi(\by | \btheta)$, does not allow for analytic normalization and closed form posterior summaries, except in trivial cases.
Then, Bayesian inference must be performed via numerical methods, such as posterior sampling via MCMC or deterministic approximations, such as Laplace or variational approximations.

\subsection{Non-conjugate variational inference}
\label{subsec:non_conjugate_variational_inference}

Variational inference \citep{Ormerod2010, Blei2017} is a probabilistic approximation approach which allows for drawing inferential conclusions by replacing the true posterior distribution $\pi(\btheta | \by)$ with a more tractable density function $q^*(\btheta)$, which minimizes some appropriate divergence measure over a convenient family of distributions $\cQ$.
A common choice in the variational Bayes literature is to minimize the \emph{Kullback-Leibler divergence} $\KL \{ q(\btheta) \parallel \pi(\btheta | \by) \} = \E_q \big[ \log\{ q(\btheta) / \pi(\btheta | \by) \} \big]$, or equivalently, maximize the \emph{evidence lower bound}, $\elbo \{ \by; q(\btheta) \} = \E_q \big[ \log\{ \pi(\by, \btheta) / q(\btheta) \} \big]$, where $\E_q(\cdot)$ denotes the expectation calculated with respect to the variational density $q(\btheta)$. 

Numerical tractability of variational approximations can be achieved by imposing convenient constraints on the variational family of distributions $\cQ$. 
Common choices are product restrictions, say $q(\btheta) = \prod_{h = 1}^{\H} q(\btheta_h)$, and parametric restrictions, say $q(\btheta_h) = q(\btheta_h; \blambda_h)$ is a parametric density indexed by $\blambda_h$, often belonging to the exponential family.
Under such a general family of restrictions, \emph{non-conjugate variational message passing} \citep{Knowles2011} can be used to climb the evidence lower bound via the iterative application of the fixed-point updating $\blambda_h^{(t+1)} \gets \big[ \E_q^{(t)} \{ \nabla_{\blambda_h}^2 \log q^{(t)}(\btheta) \} \big]^{-1} \,\big[ \nabla_{\blambda_h} \E_q^{(t)} \{ \log \pi(\by, \btheta) \} \big]$, where $t$ is the iteration counter. 
For conditionally conjugate models, variational message passing is equivalent to the exact mean field solution $q^{(t+1)}(\btheta_h) \propto \exp\big[ \,\E_{-\theta_h}^{(t)} \{ \log \pi(\btheta_h | \rest ) \} \big]$, where $\pi(\btheta_h | \rest)$ is the full-conditional distribution of $\btheta_h$ and $ \E_{-\theta_h}^{(t)}(\cdot)$ denotes the expectation taken with respect to $q_{-h}^{(t)}(\btheta) = \prod_{k \neq h} q^{(t)}(\btheta_k)$; see, e.g., \cite{Hoffman2013}, \cite{Tan2013}, \cite{Wand2014}.
Variational message passing thus provides a general scheme for conjugate and non-conjugate variational learning in complex Bayesian models.

In case of multivariate Gaussian variational approximation, the canonical parameter $\blambda_h$ may be written as a bijective transformation of the mean $\bmu_h$ and variance $\bSigma_h$, that is: $\blambda_h = (\blambda_{h1}, \blambda_{h2})$, with $\blambda_{h1} = \bSigma_h^{-1} \bmu_h$ and $\blambda_{h2} = - \bSigma_h^{-1} / 2$.
Then, defining $\bg_h^{(t)} = \nabla_{\bmu_h} \E_q^{(t)} \{ \log \pi(\by, \btheta) \}$ and $\bH_h^{(t)} = \nabla_{\bmu_h}^2 \E_q^{(t)} \{ \log \pi(\by, \btheta) \}$ and following \cite{Wand2014}, the optimal variational message passing update for $\blambda_h$ may be written as
\begin{equation}
	\label{eq:gaussian_canonical_natural_gradient}
	\begin{aligned}
		\text{(natural parameter)} \qquad & 
		\blambda_{h1}^{(t+1)} \gets \bg_h^{(t)} - \bH_h^{(t)} \bmu_h^{(t)}, \quad
		\blambda_{h2}^{(t+1)} \gets \bH_h^{(t)} / 2, \\
		\text{(mean-variance)} \qquad & 
		\bmu_h^{(t+1)} \gets \bmu_h^{(t)} - \big[ \bH_h^{(t)} \big]^{-1} \bg_h^{(t)}, \quad
		\bSigma_h^{(t+1)} \gets - \big[ \bH_h^{(t)} \big]^{-1}.
	\end{aligned}
\end{equation}
Stochastic message passing schemes can also be considered to accommodate for subsampling schemes and online optimization \citep{Khan2017a, Khan2018b}.
In these cases, the optimization must be performed in the natural parametrization using updating formula $\blambda_h^{(t+1)} \gets (1 - \rho_t) \blambda_h^{(t)} + \rho_t \hat\blambda_h^{(t+1)}$, with $\hat\blambda_h^{(t+1)}$ being a stochastic unbiased estimate of \eqref{eq:gaussian_canonical_natural_gradient}, and $\rho_t$ being a learning rate parameter satisfying the Robbins-Monro convergence conditions \citep{Robbins1951}, say $\sum_{t = 0}^{\infty} \rho_t = \infty$ and $\sum_{t = 0}^{\infty} \rho_t^2 < \infty$.
Mean and variance can be easily obtained by reparametrization.

\subsection{Variational approximation and algorithm}%
\label{subsec:variational_approximation}

Here, we propose to approximate the pGLMM posterior via variational Bayes. 
Doing so, we impose the following variational restriction
\begin{equation}
	\label{eq:variational_restrictions}
	\cQ = \bigg\{ q(\btheta) : \; q(\btheta) = q(\bbeta, \bsu) \,\prod_{h = 1}^{\H} q(\sigma_h^2), \;\; (\bbeta, \bsu) \qsim \N(\bmu, \bSigma), \;\; \sigma_h^2 \qsim \IG(\alpha_h, \beta_h) \bigg\},
\end{equation}
where $\dsty (\bbeta, \bsu) \qsim \N(\bmu, \bSigma)$ means that $q(\bbeta, \bsu)$ is the density function of a Gaussian distribution with mean $\bmu$ and covariance matrix $\bSigma$. Similarly, $q(\sigma_h^2)$ is the density of an Inverse-Gamma distribution with shape $\alpha_h$ and rate $\beta_h$.
The optimal values of the variational parameters, say $\bmu$, $\bSigma$, $\alpha_h$ and $\beta_h$, can then be obtained by maximizing the evidence lower bound using non-conjugate variational message passing via natural gradient ascent.

To derive the explicit expression for the evidence lower bound derivatives, we define
\begin{equation}	
	\label{eq:psi_function_definition}
	\Psi_r(y, \xi, \nu^2)
	= \frac{\partial^r}{\partial \xi^r} \int_{-\infty}^{+\infty} \psi(y, \xi + \nu z) \,\phi(z) \,\d{z}, \quad r \in \bbN,
\end{equation}
where $\phi(\cdot)$ is the density of a univariate standard Gaussian distribution.
In particular, notice that $\Psi_0(y_i, \xi_i, \nu_i^2) = \E_q\{ \psi(y_i, \eta_i) \}$, where $\xi_i = \bc_i^\top \bmu$ and $\nu_i^2 = \bc_i^\top \bSigma \,\bc_i$ are the approximated mean and variance of $\dsty \eta_i \qsim \N(\xi_i, \nu_i^2)$.
Thus, throughout we will refer to $\Psi_r(y, \xi, \nu^2)$ as the $r$th derivative of the \emph{variational loss} function.

\begin{proposition}
	\label{prop:evidence_lower_bound}
	Under variational restrictions \eqref{eq:variational_restrictions}, the evidence lower bound for the posterior distribution of model \eqref{eq:model} is given by
	\begin{equation}
		\label{eq:evidence_lower_bound_2}
		\begin{aligned}
			\elbo \{ \by; q(\btheta) \} =\;
			& - \sum_{i = 1}^{n} \Psi_0(y_i, \bc_i^\top, \bc_i^\top \bSigma \,\bc_i) 
			- \sum_{h = 0}^{\H} \frac{\alpha_h}{2 \beta_h} \Big[
			\bmu_h^\top \bR_h \,\bmu_h
			+ \tr(\bR_h \bSigma_{hh}) + \log|\bR_h| \Big] \\
			& + \frac{1}{2} \log|\bSigma|
			+ \sum_{h = 1}^{\H} \bigg[
			\log \frac{\Gamma(\alpha_h)}{\Gamma(A_h)}
			+ \log \frac{B_h^{A_h}}{\beta_h^{\alpha_h}}
			- (B_h - \beta_h) \,\frac{\alpha_h}{\beta_h} \bigg] + \const,
		\end{aligned}
	\end{equation}
	where $\bmu_h = \E_q(\bsu_h)$ and $\bSigma_{hh} = \Var_q(\bsu_h)$, $\Gamma(\cdot)$ is the Gamma function, $|\cdot|$ is the determinant, $\tr(\cdot)$ is the trace, and $\const$ denotes a constant not depending on the variational parameters.
\end{proposition}

\noindent All the derivations and proofs are deferred to the Appendix accompanying the paper.
The optimal variational parameters for $q^*(\sigma_h^2)$ can be derived using standard conjugate mean field computations.
The optimal distribution $q^*(\bbeta, \bsu)$ is more delicate, since conjugacy is not guaranteed under general specifications of the pseudo-likelihood.
Thus, in Proposition \ref{prop:optimal_natural_gradient_updates}, we rely on non-conjugate variational message passing and, in particular, on update \eqref{eq:gaussian_canonical_natural_gradient} to obtain the optimal natural gradient update of the variational parameters.

\begin{proposition}
	\label{prop:optimal_natural_gradient_updates}
	Let $\bPsi_r^{(t)} = \Psi_r(\by, \bxi^{(t)}, \bnu^{2(t)})$, for $r = 0, 1, 2$, be the element-wise evaluation of $\Psi_r(\cdot, \cdot, \cdot)$ after the $t$th iteration of the algorithm. Define $\bp^{(t)} = - \bPsi_1^{(t)} / \bPsi_2^{(t)} + \bxi^{(t)}$, $\bW^{(t)} = \diag \big[ \bPsi_2^{(t)} \big]$ and $\bar\bR^{(t)} = \diag \big[ \,\sigma_\beta^{-2} \bR_\beta, \,(\alpha_1^{(t)} / \beta_1^{(t)}) \bR_1, \dots, \,(\alpha_\H^{(t)} / \beta_\H^{(t)}) \bR_\H \,\big]$.
	Then, the optimal message passing update for $q^{(t+1)} (\bbeta, \bsu)$ is given by
	\begin{align*}
		\text{(natural parameter)} \qquad & \blambda_1^{(t+1)} \gets \bC^\top \bW^{(t)} \bp^{(t)}, \quad
		\blambda_2^{(t+1)} \gets - \big[ \bar\bR^{(t)} + \bC^\top \bW^{(t)} \bC \big] / 2, \\
		\text{(mean-variance)} \qquad & \bmu^{(t+1)} \gets \bSigma^{(t+1)} \bC^\top \bW^{(t)} \bp^{(t)}, \quad
		\bSigma^{(t+1)} \gets \big[ \bar\bR^{(t)} + \bC^\top \bW^{(t)} \bC \big]^{-1}.
	\end{align*}
	The optimal message passing updates for $q^{(t+1)}(\sigma_h^2)$ is given by $\alpha_h^{(t+1)} \gets A_h + d_h / 2$ and $\beta_h^{(t+1)} \gets B_h + \big\{ \bmu_h^{(t)\top \bR_h \,\bmu_h^{(t)}} + \tr \big[ \bR_h \bSigma_{hh}^{(t)} \big] \big\} / 2$. 
\end{proposition}

\noindent The recursive refinement of the variational parameters in Proposition \ref{prop:optimal_natural_gradient_updates} gives rise to the optimization routine summarized in Algorithm \ref{alg:ncvb_uf_alg} in the Appendix.
Denoting $K_r = p^r + \sum_{h = 1}^{\H} d_h^r$, for $r \in \bbN$, Algorithm 1 requires $\cO(n K_1^2 + K_1^3)$ floating point operations and $\cO(n K_1 + K_1^2)$ memory allocations per iteration.

\subsection{Variational loss computation}
\label{subsec:psi_function_computation}

The message passing updates in Proposition \ref{prop:optimal_natural_gradient_updates} only depend on the loss function $\psi(\cdot, \cdot)$ via the univariate integrals in $\Psi_r(\cdot, \cdot, \cdot)$, for $r = 0, 1, 2$.
Such a simplification is guaranteed by the univariate functional dependence linking the loss function $\psi(\cdot, \cdot)$ and the linear predictor $\eta_i$, which is unidimensional and has univariate Gaussian variational posterior.
Hence, an efficient and robust evaluation of the $\Psi_r(\cdot, \cdot, \cdot)$ is fundamental for a stable numerical implementation of the proposed variational message passing routine.

\begin{proposition}
	\label{prop:psi_function_properties}
	Let $\psi(y, \cdot)$ be a $q$-integrable function, i.e. $\E_q |\psi(y,\eta)| < \infty$, for any $y \in \cY$. 
	Then, (a) the map $(\xi,\nu) \mapsto \Psi_0(y, \xi, \nu^2)$ is infinitely differentiable; (b) $\lim_{\nu \downarrow 0} \,\sup_{\xi \in \R} |\Psi_0(y, \xi, \nu^2) - \psi(y, \xi)| = 0$; if $\eta \mapsto \psi(y, \eta)$ is convex, then (c) $(\xi,\nu) \mapsto \Psi_0(y, \xi, \nu^2)$ is convex and (d) $\psi(y, \xi) \leq \Psi_0(y, \xi, \nu^2)$.
	Moreover, defining $\psi_r(y, \cdot)$ as the $r$th weak derivative of $\psi(y, \cdot)$, and $He_r(\cdot)$ as the $r$th probabilistic Hermite polynomial, (e) we have
	\begin{equation}
		\Psi_r(y, \xi, \nu^2) = \nu^{-(r-k)} \int_{-\infty}^{+\infty} \psi_{k}(y, \xi + \nu z) \,He_{r-k}(z) \,\phi(z) \,\d{z}, 
		\quad k \in \{0, \dots, r\}.
	\end{equation}
\end{proposition}

\noindent Proposition \ref{prop:psi_function_properties} guaranties existence, convexity and smoothness of $\Psi_r(\cdot, \cdot, \cdot)$ under mild regularity conditions, and provides $r$ equivalent representations for $\Psi_r(\cdot, \cdot, \cdot)$ under increasingly weaker differentiability conditions on $\psi(\cdot, \cdot)$.
In particular, for $r = 0, 1, 2$, we have
\begin{align*}
	& \Psi_0(y, \xi, \nu^2) = \E_q\{ \psi_0(y, \xi + \nu z) \}, \\ 
	& \Psi_1(y, \xi, \nu^2) = \E_q\{ \psi_1(y, \xi + \nu z) \} = \nu^{-1} \E_q\{ z \,\psi_0(y, \xi + \nu z) \}, \\
	& \Psi_2(y, \xi, \nu^2) = \E_q\{ \psi_2(y, \xi + \nu z) \} = \nu^{-1} \E_q\{ z \,\psi_1(y, \xi + \nu z) \} = \nu^{-2} \E_q\{ (z^2-1) \,\psi_0(y, \xi + \nu z) \}.
\end{align*}
Leveraging this convenient result, we derive analytical expressions for several regression and classification models, even those that do not necessarily satisfy second-order differentiability.

\begin{example}{\bf(Quantile regression)}
	\label{ex:quantile_regression_loss}
	Let $\psi(y,\eta) = \half |y-\eta| + (\tau - \half) (y-\eta)$, $y \in \R$, be the quantile regression loss \citep{Koenker2005}, for $\tau \in (0,1)$. Then, denoting $z = (y - \xi) / \nu$, we have $\Psi_0(y, \xi, \nu^2) = \nu \big[z (\Phi(z) - 1 + \tau) + \phi(z) \big]$, $\Psi_1(y, \xi, \nu^2) = 1 - \tau - \Phi(z)$ and $\Psi_2(y, \xi, \nu^2) = \nu^{-1} \phi(z)$.
\end{example}

\begin{example}{\bf(Support vector regression)}
	\label{ex:support_vector_regression_loss}
	Let $\psi(y, \eta) = \max(0, |y - \eta| - \eps)$, $y \in \R$, be the support vector regression loss \citep{Vapnik1998}, for an insensitivity parameter $\veps > 0$. Then, denoting $z_\eps^- = (y - \xi - \eps) / \nu$ and $z_\eps^+ = (y - \xi + \eps) / \nu$, we have $\Psi_0(y, \xi, \nu^2) = 2 \nu \big[ z_\eps^- \Phi(z_\eps^-) + z_\eps^+ (\Phi(z_\eps^+) - 1) + \phi(z_\eps^-) + \phi(z_\eps^+) \big]$, $\Psi_1(y, \xi, \nu^2) = 2 \big[ 1 - \Phi(z_\eps^+) - \Phi(z_\eps^-) \big]$ and $\Psi_2(y, \xi, \nu^2) = 2 \nu^{-1} \big[ \phi(z_\eps^+) + \phi(z_\eps^-) \big]$.
\end{example}

\begin{example}{\bf(Support vector classification)}
	\label{ex:support_vector_classification_loss}
	Let $\psi(y, \eta) = \max(0, 1 - y \eta)$, $y \in \{-1,+1\}$, be the support vector classification loss \citep{Vapnik1998}. Then, denoting $z = (1 - y \xi) / \nu$, we have $\Psi_0(y, \xi, \nu^2) = 2 \nu \big[ z \Phi(z) + \phi(z) \big]$, $\Psi_1(y, \xi, \nu^2) = - 2 y \Phi(z)$ and $\Psi_2(y, \xi, \nu^2) = 2 \nu^{-1} \phi(z)$.
\end{example}

\noindent Whenever analytic solutions for  $\Psi_r(y, \xi, \nu^2)$ are not available, as in the cases of logistic and probit regression, where $\psi(y, \eta) = - y \eta + \log(1 + e^\eta)$ and $\psi(y, \eta) = - \log \Phi((2 y - 1) \eta)$, stable and efficient numerical approximations can be obtained using general-form univariate integration methods. 
These include adaptive Gauss-Hermite quadrature  \citep{Ormerod2012, Tan2013}, Clenshaw-Curtis quadrature \citep{Knowles2011}, or Monahan-Stefanski approximation \cite{Nolan2017}.

\begin{figure}[t]
	\centering
	\includegraphics[width = \textwidth]{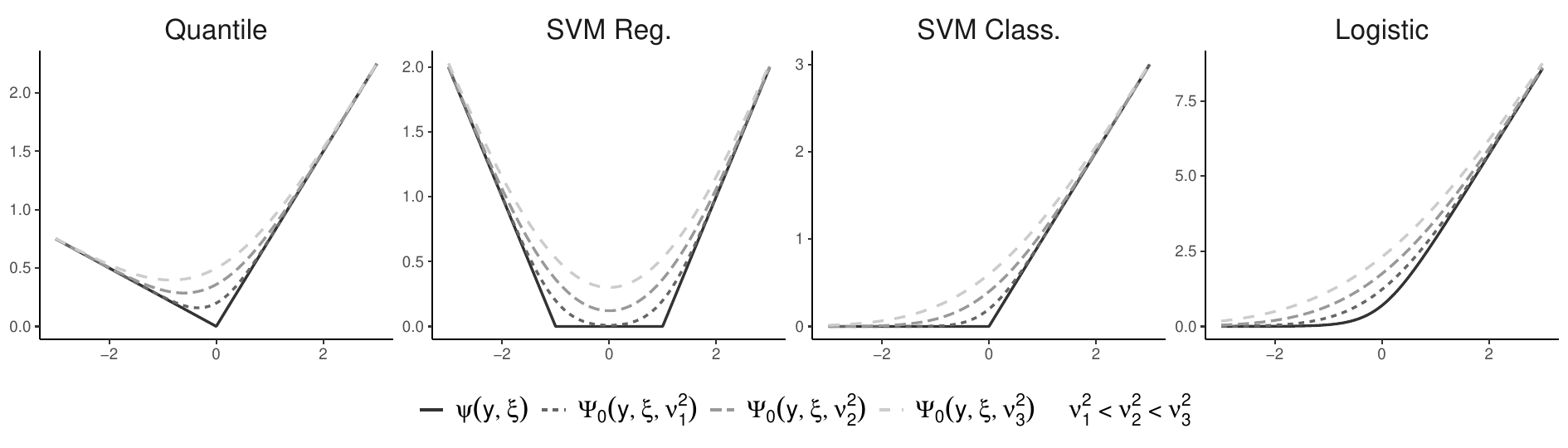}
	\caption{%
		\label{fig:variational_loss_functions}%
		Variational loss functions under four model specifications. From left to right: quantile regression ($\tau = 0.75$), support vector regression ($\eps = 0.5$), support vector classification and logistic regression.
	}
\end{figure}
In the Appendix, we also derive closed form expressions of the variational loss for other common statistical models, including Poisson, Gamma, Expectile and Huber regression.
Figure \ref{fig:variational_loss_functions} shows the variational loss functions discussed in Examples \ref{ex:quantile_regression_loss}--\ref{ex:support_vector_classification_loss} for different levels of $\nu^2$. 
Comparing the original and variational loss functions, we can clearly observe the smoothing, convergence and mojorization properties of the variational loss stated in Proposition \ref{prop:psi_function_properties}.
Specifically, the smoothing effect introduced by the Gaussian convolution in \eqref{eq:psi_function_definition} is central to our work, as it enables the seamless handling of both regular and non-regular models within the same computational framework. This approach ensures the existence and differentiability of the variational loss under fairly general assumptions on the original loss.

\subsection{Comparison with data-augmented variational Bayes}%
\label{subsec:connection_with_other_methods}

An alternative, commonly used, variational approximation method for non-conjugate regression models is data augmented variational Bayes. 
This consists of approximating an augmented posterior distribution in place of the true posterior, using conjugate variational approaches under convenient mean field restrictions; see, e.g., \cite{Wand2011a} and~\cite{McLean2019}.
Such an approach has been explored for the estimation of support vector machine classifiers~\citep{Luts2014a}, probit regression~\citep{Armagan2011, Fasano2022}, logistic regression~\citep{Durante2019} and logistic mixed models~\citep{Goplerud2022, Goplerud2024}.
Despite its widespread application, variational data augmentation is a model-specific technique that may not be applicable to generic likelihood specifications, such as in the case of generalized linear models with arbitrary link functions.
Moreover, it is prone to slow convergence~\citep{Lewandowski2010, Duan2018, Johndrow2019} and, often, poor approximation quality~\citep{Neville2014, Fasano2022}.
Non-conjugate variational methods, instead, directly approximate the target posterior, without introducing additional auxiliary variables.
Therefore, it is of interest to theoretically investigate the relationship, in the Kullback-Leibler metric, between augmented and non-augmented variational approximations belonging to the same approximating family.

To this end, we denote by $\pi(\btheta | \by) \propto \pi(\btheta) \prod_{i = 1}^{n} \pi(y_i | \btheta)$ and $\pi(\bomega, \btheta | \by) \propto \pi(\btheta) \prod_{i = 1}^{n} \pi(y_i, \omega_i | \btheta)$ the \emph{target} and \emph{augmented} posterior distributions, respectively, for some vector of local variables $\bomega = (\omega_1, \dots, \omega_n)^\top \in \Omega^n$, such that $\pi(y_i | \btheta) = \int_\Omega \pi(y_i, \omega_i | \btheta) \,\d\omega_i$.
We denote by $q_\M^*(\btheta)$ and $q_\A^*(\bomega, \btheta)$ the corresponding variational approximations, which minimize $\KL\{ q(\btheta) \parallel \pi(\by | \btheta) \}$ and $\KL\{ q(\bomega, \btheta) \parallel \pi(\bomega, \btheta | \by) \}$ under the generic variational restrictions $q_\M(\btheta) \in \cQ_\M$ and $q_\A(\bomega, \btheta) \in \cQ_\A$, such that $\cQ_\A = \big\{ q(\bomega, \btheta) = q(\btheta) q(\bomega  | \btheta) : q(\btheta) \in \cQ_\A, \;q(\bomega | \btheta) \in \cQ_\C \big\}$, for some appropriate functional spaces $\cQ_\M$ and $\cQ_\C$. Here, the subscripts M, C and A stand for \emph{marginal}, \emph{conditional}, and \emph{augmented}.

\begin{proposition}%
	\label{prop:kullback_leibler_inequalities}
	For any $q_\M^*(\btheta) \in \cQ_\M$ and $q_\A^*(\bomega, \btheta) \in \cQ_\A$, we have $\KL \{ q_\M^* (\btheta) \parallel \pi(\btheta|\by) \} \leq \KL \{ q_\A^* (\btheta) \parallel \pi(\btheta|\by) \} \leq \KL \{ q_\A^* (\bomega, \btheta) \parallel \pi(\bomega, \btheta | \by) \}.$ Moreover, if $\pi(\bomega | \btheta, \by) \in \cQ_\C$, $q_\A^*(\bomega | \btheta) = \pi(\bomega | \btheta, \by)$ and $q_\A^*(\btheta) = q_\M^*(\btheta)$ almost everywhere, and $\KL \{ q_\M^* (\btheta) \parallel \pi(\btheta|\by) \} = \KL \{ q_\A^* (\btheta) \parallel \pi(\btheta|\by) \} = \KL \{ q_\A^* (\bomega, \btheta) \parallel \pi(\bomega, \btheta | \by) \}$.
\end{proposition}

\noindent Proposition~\ref{prop:kullback_leibler_inequalities} states the suboptimality in Kullback-Leibler divergence of any augmented variational approximation compared to an appropriate marginal approximation, under the marginal compatibility assumption $q_\M^*(\btheta), q_\A^*(\btheta) \in \cQ_\M$. 
Such a result holds true for any choice of $\cQ_\M$ and $\cQ_\C$, and for any model specification, not being restricted to particular modeling assumptions.
In particular, if we consider the pGLMM family in~\eqref{eq:model}, where $\btheta = (\bbeta, \bsu, \sigma_1^2, \dots, \sigma_\H^2)$, convenient augmented representations (if any) yield closed form full-conditionals for $\bbeta$, $\bsu$, $\sigma_h^2$ and $\omega_i$, allowing the implementation of conjugate variational approximations under mean field restrictions $\cQ_\M = \big\{ q(\btheta) : q(\btheta) = q(\bbeta, \bsu) \prod_{h = 1}^{\H} q(\sigma_h^2) \big\}$ and $\cQ_\C = \big\{ q(\bomega | \btheta) : q(\bomega | \btheta) = \prod_{i = 1}^{n} q(\omega_i) \big\}$.
Then, if the optimal approximations $q_\A^*(\bbeta, \bsu)$ and $q_\A^*(\sigma_h^2)$ take Gaussian and Inverse-Gamma forms, respectively, as it is often the case \citep{Wand2011a, McLean2019}, employing the marginal approach described in Section \ref{sec:approximate_belief_updating}, we are guaranteed to obtain an improved approximation of the joint posterior density relative to augmented mean field.
Inequality in Proposition \ref{prop:kullback_leibler_inequalities} is also supported by the numerical experiments presented in Sections \ref{sec:simulation_studies} and \ref{sec:real_data_application}, which confirm a higher fidelity of marginal non-conjugate variational approximations over augmented conjugate ones.

\section{Extensions for high-dimensional problems}
\label{sec:extensions}

One iteration of the variational update in Proposition \ref{prop:optimal_natural_gradient_updates} has a computational complexity of $\cO(n K_1^2 + K_1^3)$, which is cubic in the total number of parameters and linear in the sample size.
In high-dimensional settings, such an approach may result in prohibitively expensive calculations at each iteration, thus limiting its applicability in some real data problems.
Then, to accommodate for highly parametrized models and massive data problems, we here discuss two alternative approaches, which permits to speed up the computations and reduce the computational complexity of the proposed approximation algorithm.
The first one relies on alternative factorizations of the variational density, the second accommodate for stochastic and online optimization.

\subsection{Alternative variational factorizations}%
\label{subsec:alternative_variational_factorizations}

A common approach to deal with highly parametrized models in the variational literature is to introduce additional restrictions on the variational family.
Two relevant examples are
\begin{equation}
	\label{eq:mean_field_assumption}
	\begin{aligned}
		\text{Partially factorized:} & \quad \cQ_\PF = \big\{ q(\btheta) \in \cQ_\UF : \; q(\bbeta, \bsu) = q(\bbeta | \bsu) \,{\tsty\prod_{h = 1}^{\H}} q(\bsu_h) \big\}, \\
		\text{Fully factorized:} & \quad \cQ_\FF = \big\{ q(\btheta) \in \cQ_\UF : \; q(\bbeta, \bsu) = q(\bbeta) \,{\tsty\prod_{h = 1}^{\H}} q(\bsu_h) \big\}.
	\end{aligned}
\end{equation}
Here, we denote by $\cQ_\UF$ the \emph{unfactorized} approximation discussed in \eqref{eq:variational_restrictions}, $\cQ_\PF$ denotes the \emph{partially factorized} approximation recently introduced by \cite{Goplerud2024}, while $\cQ_\FF$ represents a standard block-wise \emph{fully factorized} approximation commonly used in the literature of variational additive and mixed models; see, e.g., \cite{Tan2013} and \cite{Menictas2023}.
Factorizations \eqref{eq:mean_field_assumption} induce increasing sparsity on the variance matrix $\bSigma$ of $\bbeta$ and $\bsu$, thus diminishing the effective number of variational parameters to be estimated.

To address alternative variational factorizations of the approximating density within the same non-conjugate framework discussed in Section \ref{sec:approximate_belief_updating}, we propose introducing a projection step. This step maps an unfactorized variational density, lying on $\cQ_{\UF}$, 
into a restricted form, denoted generically as $\cQ_\RF \subseteq \cQ_\UF$, which may correspond to $\cQ_{\PF}$, $\cQ_{\FF}$, or other alternative structures.
Denoting by $\bp_\RF^{(t)}$ and $\bW_\RF^{(t)}$ the current values of the variational pseudo-data and weights under $\cQ_\RF$, the proposed projected variational update corresponds to
\begin{align*}
	\text{Update step in $\cQ_\UF$}: & \quad 
	\blambda_{1, \UF}^{(t+1)} \gets \bC^\top \bW_\RF^{(t)} \bp_\RF^{(t)}, \quad
	\blambda_{2, \UF}^{(t+1)} \gets \bar\bR^{(t)} + \bC^\top \bW_\RF^{(t)} \bC, \\
	\text{Projection onto $\cQ_\RF$}: & \quad q_\RF^{(t+1)}(\bbeta, \bsu) \gets \argmin_{q \in \cQ_\RF} \;\KL \big\{ q(\bbeta, \bsu) \,\big\|\, q_\UF ^{(t+1)}(\bbeta, \bsu) \big\},
\end{align*}
where $q_\UF^{(t+1)}(\bbeta, \bsu) \in \cQ_\UF$ is the Gaussian density with natural parameters $\blambda_{1,\UF}^{(t+1)}$ and $\blambda_{2,\UF}^{(t+1)}$.

Exploiting the Gaussianity of both $q_\UF(\bbeta, \bsu)$ and $q_\RF(\bbeta, \bsu)$ and the sparsity structure induced by \eqref{eq:mean_field_assumption}, we can obtain the restricted approximations via memory-efficient closed form updates in a coordinate ascent fashion without explicitly forming $\blambda_{1,\UF}^{(t+1)}$, $\blambda_{1,\UF}^{(t+1)}$, $\bmu_\UF^{(t+1)}$ and $\bSigma_\UF^{(t+1)}$.
For brevity, we do not report here the derivations and explicit solutions of the coordinate ascent schemes under $\cQ_\PF$ and $\cQ_\FF$, which are described in Algorithms~\ref{alg:ncvb_ff_alg}--\ref{alg:ncvb_pf_alg} in the Appendix.
Both of these routines permit us to boil down the computational complexity of variational message passing from $\cO(n K_1^2 + K_1^3)$ to $\cO(n K_2 + K_3)$ at the cost of reducing the precision of the variational approximation.

\subsection{Stochastic and online variational approximations}%
\label{subsec:stochastic_and_online_variational_approximation}

Introducing restricted variational approximations, as discussed in the previous section, is particularly helpful for handling high-dimensional models.
However, in massive or online data problems, batch optimization may result in expensive calculations even under restricted variational families, since a complete scan of the dataset must be performed at each iteration.
In these cases, it might be convenient to consider stochastic implementations which only use a subset of the whole dataset at each update, say a minibatch $B_t \subset \{ 1, \dots, n \}$ of dimension $n_\scb$, resulting in cheaper computations.
To this end, we can employ a stochastic variational routine \citep{Hoffman2013, Hoffman2015} which cycles over Robbins-Monroe updates \citep{Robbins1951} of the natural parameters, eventually followed by a projection step to obtain restricted variational approximations of the posterior. 
For the unfactorized approximation, the proposed stochastic updates take the form
\begin{align*}
	& \hat\blambda_{1, \scb}^{(t+1)} \gets (n/n_\scb) \bC_{\scb}^\top \bW_\scb^{(t)} \bp_\scb^{(t)}, \quad 
	\hat\blambda_{2, \scb}^{(t+1)} \gets - \big[ (n/n_\scb) \,\bC_{\scb}^\top \bW_\scb^{(t)} \bC_{\scb} + \bar\bR^{(t)} \big] / 2, \\
	& \blambda_{1}^{(t+1)} \gets (1 - \rho_t) \blambda_{1}^{(t)} + \rho_t \hat\blambda_{1, \scb}^{(t+1)}, \quad
	\blambda_{2}^{(t+1)} \gets (1 - \rho_t) \blambda_{2}^{(t)} + \rho_t \hat\blambda_{2, \scb}^{(t+1)},
\end{align*}
with $\hat\blambda_{1, \scb}^{(t+1)}$ and $\hat\blambda_{2, \scb}^{(t+1)}$ being stochastic unbiased estimates of the natural parameters, 
where $\bp_\scb^{(t)}$, $\bW_\scb^{(t)}$ and $\bC_\scb$ denote sub-sampled quantities only involving observations included in the current minibatch $B_t$.
The implied mean and variance updates can be easily obtained by reparametrization.
Generalizations of such a scheme can account for parameter-specific adaptive learning rates, modifying the update as $\blambda^{(t+1)} \gets \blambda^{(t)} + \brho_t \odot (\hat\blambda^{(t+1)} - \blambda^{(t)})$, where $\odot$ denotes the element-wise product and $\brho_t$ is a learning rate vector which can be dynamically adapted using, e.g., AdaGrad \citep{Duchi2011}, AdaDelta \citep{Zeiler2012}, or the recently proposed Snngm and Nagm methods by \cite{Tan2025}.

For a complete derivation and pseudo-code for stochastic non-conjugate variational message passing algorithms under factorizations \eqref{eq:mean_field_assumption}, we refer the reader to Algorithms \ref{alg:sncvb_uf_alg}--\ref{alg:sncvb_pf_alg} in the Appendix.
The stochastic implementation of the unfactorized variational approximation takes $\cO(n_\scb K_1^2 + K_1^3)$ floating-point operations per iterations. While the stochastic implementations of partially and fully factorized approximations require $\cO(n_\scb K_2 + K_3)$ operations.
In both cases, a single update of the posterior is independent on the sample size $n$, making such implementations highly scalable in massive data problems.

\section{Simulation studies}%
\label{sec:simulation_studies}

We here assess the empirical performances of the proposed method in synthetic data problems, where the truth is known.
The models we consider for this analysis are: quantile regression, support vector regression and support vector classification.
The \emph{true} posterior distribution is estimated via Markov chain Monte Carlo (MCMC), while variational approximations are obtained via augmented mean field variational Bayes (MFVB) and the proposed non-conjugate variational Bayes (NCVB) method.
For quantile regression, we considered the Gibbs sampler proposed by \cite{Kozumi2011} and the MFVB algorithm by \cite{Wand2011a}.
For support vector machines, we considered the Gibbs sampler proposed by \cite{Polson2011} and the MFVB algorithm by \cite{Luts2014a}.
All the considered methods and simulations are implemented in \texttt{Julia} (Version 1.11.3) and run on the first author's \texttt{Dell~XPS~15} laptop with 4.7 gigahertz processor and 32 gigabytes of random access memory. Reproducible code can be found in the Appendix.

We consider three data generating mechanisms: a heteroscedastic Gaussian model, $y_{ij} | x_{ij} \sim \N(\mu_{ij}, \sigma_{ij}^2)$; a homoscedastic $t$-distributed model, $y_{ij} | x_{ij} \sim \text{t}(\mu_{ij}, \sigma, \upsilon)$; and a Bernoulli model, $y_{ij} | x_{ij} \sim \Be(\pi_{ij})$.
A random intercept specification is considered for $\mu_{ij}$, $\sigma_{ij}$ and $\pi_{ij}$, say $\mu_{ij} = \beta_0 + \beta_1 x_{ij} + u_j$, $\log (\sigma_{ij}) = \gamma_0 + \gamma_1 x_{ij} + w_j$ and $\log \{\pi_{ij}/(1-\pi_{ij})\} = \mu_{ij}$. 
The indices $i = 1, \dots, n$ and $j = 1, \dots, d$ identify, respectively, the $i$th subject belonging to the $j$th group in a stratified study.
The fixed effect parameters $\beta_0, \beta_1$ and $\gamma_0, \gamma_1$ are generated according to a $\N(0,1/2)$ distribution, the random intercepts $u_1, \dots, u_d$ and $w_1, \dots, w_d$ are generated according to a $\N(0,1/4)$ distribution, while $\sigma = 0.1$ and $\upsilon = 4$.
We consider two simulation setups: in the first one (setting A), we fix the number of parameters to $d = 10$ and we generate five datasets with sample size $n = 250$, $500$, $1000$, $2500$, $5000$; in the second one (setting B), we fix the sample size to $n = 500$ and we generate five datasets with number of parameters $d = 5, 10, 25, 50, 100$.
For each simulation setting, sample size and parameter dimension, we generate 100 datasets using the sampling mechanism described above.

For the estimation, we specify the linear predictor $\eta_{ij} = \bx_{ij}^\top \bbeta + \bz_{ij}^\top \bsu$, where $\bx_{ij} = (1, x_{ij})^\top$ is a covariate vector and $\bz_{ij}$ is a $d \times 1$ selection vector associated to the $j$th group, whose $j$th entry is equal to $1$ and all the others are $0$.
The prior distributions of $\bbeta, \bsu, \sigma_u^2$ are specified as in \eqref{eq:model}, where $\sigma_\beta^2 = 10^4$, $A_u = 2.0001$ and $B_u = 1.0001$.
These correspond to Inverse-Gamma distributions having $\E(\sigma_u^2) = 1$ and $\Var(\sigma_u^2) = 10^3$.

In the heteroscedastic data setting, we estimate a quantile model with $\tau = 0.9$; in the $t$-distributed data setting, we estimate a support vector regression model with $\eps = 0.05$; while in the Bernoulli data setting, we estimate support vector classification.

\begin{figure}[t]
	\centering
	\includegraphics[width = \textwidth]{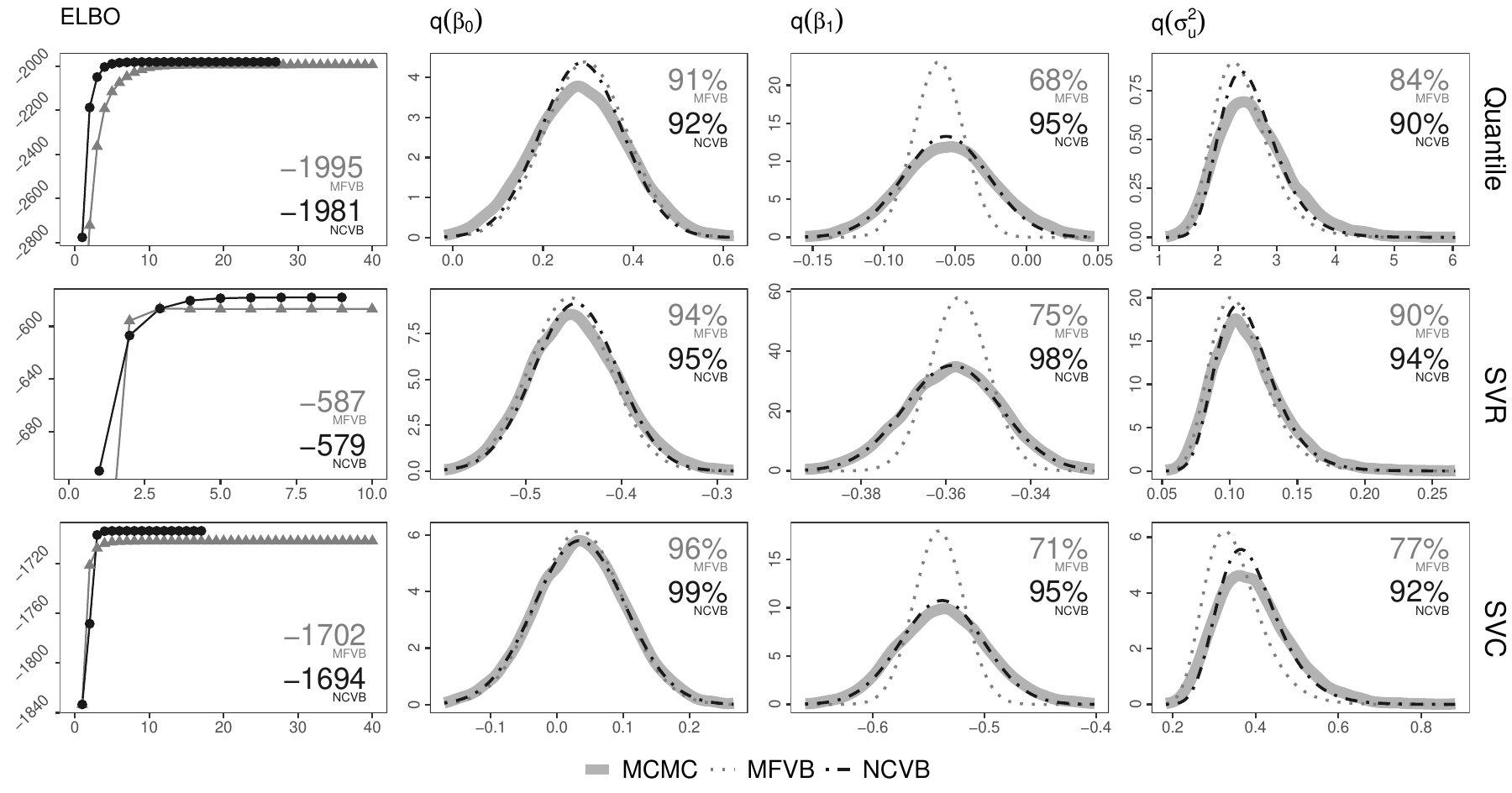}
	\caption{%
		\label{fig:simulation_pdf}
		Comparison between MCMC, MFVB and NCVB in terms of evidence lower bound and marginal posterior approximation.
		Each row corresponds to a different model.
		The first column shows the evidence lower bound evolution over the iterations.
		The other columns show the estimated marginal posterior densities for $\beta_0$, $\beta_1$ and $\sigma_u^2$ along with the associated accuracy scores for MFVB and NCVB.
	}
\end{figure}

We assess the quality of final variational approximations obtained via MFVB and NCVB by calculating the average accuracy score.  
For a generic parameter $\theta$, the accuracy score is defined as $\text{Accuracy}\{ q^*(\theta) \} = 100 \big( 1 - \half \int | \,q^*(\theta) - \pi(\theta | \by) \,| \,\d\theta \big) \%$.
The \emph{true} posterior density is here replaced by a kernel density estimate based on the MCMC samples from $\pi(\theta | \by)$ and the integral is calculated via adaptive Gauss-Kronrod quadrature. 

\begin{figure}[t]
	\centering
	\includegraphics[width = \textwidth]{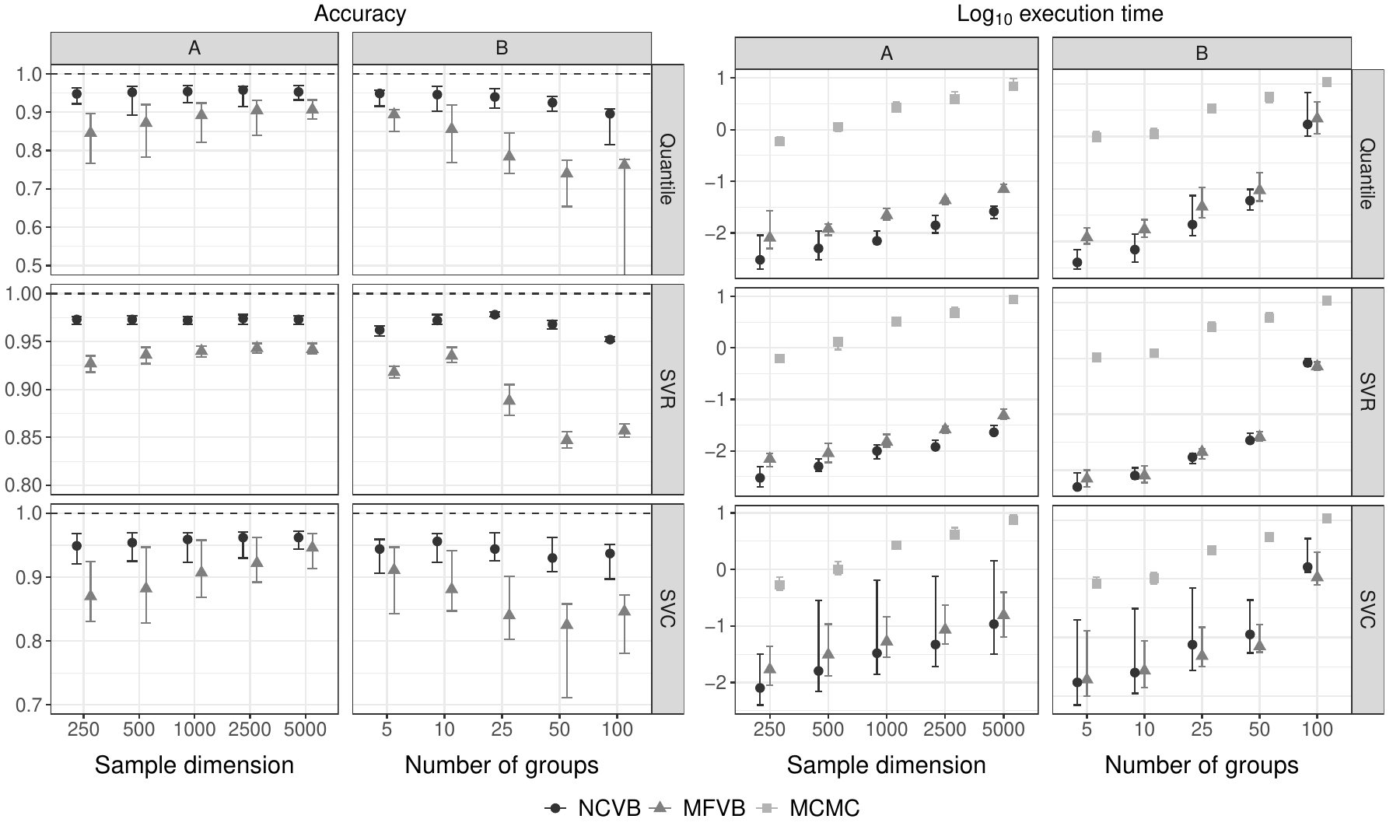}
	\caption{%
		\label{fig:simulation_summary}
		Median and $95\%$ confidence interval of the average marginal accuracy score (left) and of the $\log_{10}$-transformed elapsed execution times for the simulation setup described in the text.
		Each row corresponds to a model, each column corresponds to a simulation setting.
	}
\end{figure} 

Figure \ref{fig:simulation_pdf} shows a comparison between the marginal posterior approximations of $\beta_0$, $\beta_1$ and $\sigma_u^2$ for MCMC, MFVB and NCVB.
Additionally, it shows the evidence lower bound for each iteration of MFVB and NCVB.
Such approximations refer to the results obtained from one synthetic dataset in simulation setting B, where $n = 500$ and $d = 50$.
In this scenario, NCVB always reaches a higher evidence lower bound than MFVB, as predicted by Proposition \ref{prop:kullback_leibler_inequalities}.
Moreover, NCVB provides a better approximation of the marginal posteriors, in terms of accuracy score, while MFVB always underestimates the posterior variance of $\beta_1$. 

The left panel of Figure \ref{fig:simulation_summary} displays the distributions of the average accuracy scores obtained by averaging the marginal accuracy scores over all the parameters in the model.
At the increase of the sample size (setting A), the accuracy scores improve for all the considered approximations.
The opposite happens when the sample size is kept fixed and the number of parameters grows (setting B).
In both cases, we observe a uniform dominance of NCVB over MFVB in all the simulation setups and for all the models considered in this study.

The right panel of Figure \ref{fig:simulation_summary} displays the $\log_{10}$-transformed computation times measured in seconds.
The execution time increases with the dimension of the problem, either in terms of sample size or number of random effect parameters.
As we might expect, in all the considered scenarios, the deterministic approximation methods yield a significant speed gain over MCMC in terms of execution time.
In simulation setting A, NCVB provides a substantial speed gain over MFVB for both quantile and support vector regression.
For support vector classification, NCVB and MFVB show similar average performances, although NCVB exhibits slightly more variability while consistently achieving lower median execution times. 
In simulation setting B, the results are more balanced, with all the compared variational methods having similar execution times. However, NCVB still demonstrates some advantages in the quantile regression examples.

Overall, we conclude that MFVB and NCVB have comparable computational complexity per iteration and similar execution times. 
However, they differ in terms of posterior approximation accuracy, with NCVB consistently outperforming the competitors in terms of the evidence lower bound and marginal approximation accuracy.

\section{Real data application}%
\label{sec:real_data_application}

In this section, we present a real data problem concerning the forecast of the global electricity load consumption. In such a context, the data are usually highly dominated by non-stationary trends, multiple seasonal cycles of different lengths, like daily, weekly and monthly patterns, heteroscedasticity and by the presence of extreme values. Therefore, for the management of the power supply, it is of critical importance to understand and predict the behaviour of the distribution of the load consumption, especially during exceptional events. To this end, a nonparametric density forecasting approach can be taken by pooling the information of several quantile estimates. Each conditional quantile can then be express as a non-linear function of the available meteorological, economic and social factors using an additive model specification.

Here, we consider the data presented during the load forecasting track of the Global Energy Competition 2014 \citep{Hong2016}. 
The dataset collects half-hourly load consumption and temperatures over the period going from January 2005 to December 2011, resulting in a sample with $60600$ observations. We then estimate $5$ conditional quantiles, corresponding to $\tau = 0.05, 0.25, 0.5, 0.75, 0.95$, thus characterizing the conditional distribution of the load consumption without imposing any additional distributional assumption.

\begin{figure}
	\includegraphics[width = \textwidth]{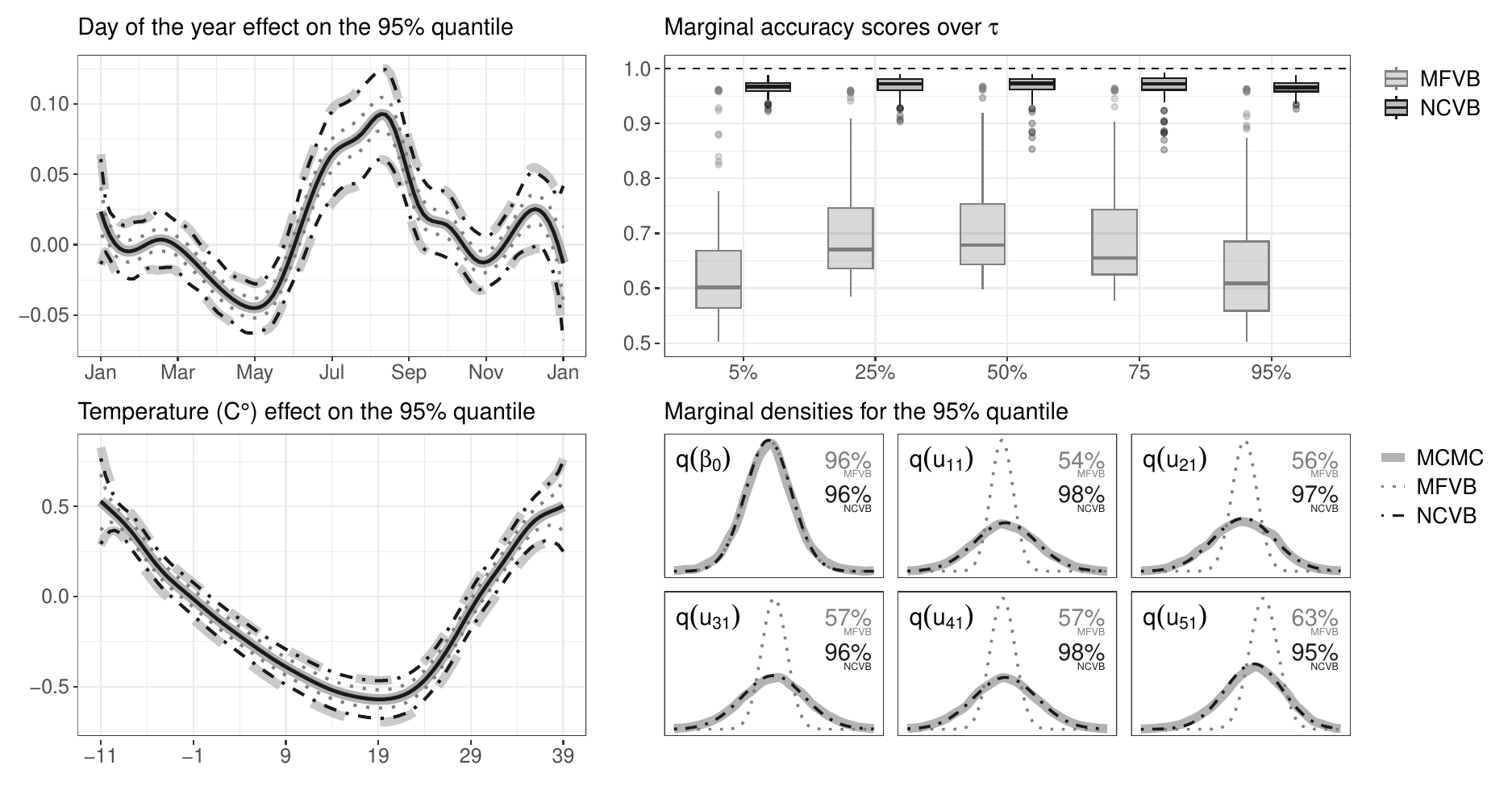}
	\caption{%
		\label{fig:gefcom_summary}
		Left: estimated posterior non-linear effects and $95\%$ credibility bands of the year cycle and temperature on the $95\%$ quantile of the power load consumption.
		Top right: boxplots of the marginal accuracy scores computed on the fixed and random effect parameters over the $5$ considered quantile levels.
		Bottom right: Marginal posterior distributions for some fixed and random effect parameters along with the associated accuracy scores.
	}
\end{figure}

As in \cite{Gaillard2016} and \cite{Fasiolo2021}, we model the $\tau$th quantile of load consumption as an additive function involving a long-term trend, daily, weekly, monthly, and yearly cycles, actual and smoothed temperature, as well as the lagged load consumption.
We employ non-linear basis expansions for all covariates, using cubic B-splines.
As a result, we obtain an additive quantile specification with $8$ smooth terms and $110$ basis coefficients, in total.
The coefficient vector $\bsu_h$ for each non-linear effect is treated as a random effect vector with its own variance parameter $\sigma_h^2$.
To incorporate regularization, we utilize classical O'Sullivan second-order differential penalization and construct the corresponding regularization matrix $\bR_h$ for B-spline basis coefficient vectors $\bsu_h$ \citep{Wand2008}.
We set diffuse prior distributions for all the parameters in the model, that is $\sigma_\beta^2 = 10^6$, $A_h=2.0001$, $B_h = 1.0001$. 
Under this model specification, we estimate the unknown parameters of the model using MCMC, MFVB, NCVB.

The results of our analysis demonstrate the excellent performance of the proposed non-conjugate variational approach in approximating the true posterior, as illustrated in Figure \ref{fig:gefcom_summary}. 
The accuracy scores across all considered quantile levels are consistently close to $1$, almost always exceeding 0.95 for non-conjugate variational Bayes. In contrast, mean field variational Bayes shows a median accuracy around $0.65$, with many values falling below $0.6$, particularly for the most extreme quantiles. 
Thus, even in real data applications, non-conjugate variational Bayes significantly outperforms the conjugate mean field approximation.

Regarding point estimation, both non-conjugate and mean-field approximations yield similar results. 
However, notable differences arise in quantifying posterior variance. 
Specifically, while mean-field approximations tend to systematically underestimate the posterior variance of the random effects, the proposed non-conjugate approximation captures posterior variability accurately, even for extreme quantiles. 
This effect is also evident in the posterior credibility bands of the non-linear effects, as shown in Figure~\ref{fig:gefcom_summary}.
Such empirical findings are further supported by additional experiments on real datasets, as described in the Appendix.

\section{Discussion and future extensions}%
\label{sec:discussion}

We have derived an efficient and easy-to-code variational method to estimate Bayesian pseudo-likelihood mixed models.
Our approach is highly versatile and can be applied to a wide range of models, extending far beyond the limited examples presented here. 
Our method effectively addresses challenges associated with non-regular loss functions and non-conjugate priors, ensuring high efficiency and outperforming the accuracy of conjugate mean field variational Bayes in numerous instances, both in simulations and real-data examples.

The proposed approach allows for several extensions and adaptations. 
For instance, specialized algorithms for models with multiple and nested random effects, dynamic linear models, latent spatial fields, and inducing shrinkage priors can be developed.
Frequentist mixed regression models are effectively addressed using a careful combination of variational expectation-maximization algorithms under Gaussian approximation, as demonstrated in studies like \cite{Ormerod2012} and \cite{Hui2019}, among others.
This approach provides a valuable alternative to the cumbersome Monte Carlo integration \citep{Geraci2007}, multivariate quadrature methods \citep{Geraci2014a}  and nested optimization  \citep{Geraci2019} techniques that are often required for models that lack the necessary regularity conditions for implementing Laplace-type approximations.

Another key contribution of our work is the introduction of variational loss smoothing to address non-regular minimization problems.
As Proposition ~\ref{prop:psi_function_properties} highlights, variational loss averaging is a method for constructing smooth majorizing objective functions from non-differentiable loss functions. Examples of loss smoothing can be found both in the quantile \citep{Yue2011, Fasiolo2021} and the support vector machine \citep{Lee2001, Staines2012} literature.
However, unlike other smoothing methods based on geometric considerations, our strategy is grounded in statistical reasoning, with a clear probabilistic interpretation akin to the expectation-maximization algorithm. 
Moreover, unlike arbitrary smoothing techniques, our approach incorporates a principled calibration rule, where the degree of smoothing is adaptively determined by the posterior variance of each linear predictor. 
This ensures that the level of approximation is tailored to each observation, balancing computational efficiency and approximation accuracy. 
Overall, our work advances the practical implementation of variational methods, offering a scalable and robust framework for complex Bayesian and frequentist mixed models.
\section*{Supplementary material}%
\label{sec:supplementary_material}

\begin{description}
	\item[Published article:] A peer-reviewed version of this article is published in the \emph{Journal of Computational and Graphical Statistics}, available at \href{https://doi.org/10.1080/10618600.2025.2527925}{$\mathtt{https://doi.org/10.1080/10618600.2025.2527925}$}.
	\item[Code and data:] The code and data for reproducing the results in Sections \ref{sec:simulation_studies} and \ref{sec:real_data_application} can be found as Supplementary Material in the published version of this article.
\end{description}

\clearpage

\begin{appendix}

\section{Evidence lower bound and optimal distributions}%
\label{app:evidence_lower_bound_and_optimal_distributions}

\subsection{Proof of Proposition~\ref{prop:evidence_lower_bound}: evidence lower bound}%
\label{proof:evidence_lower_bound}

First, we consider the definition of the evidence lower bound
and we notice that it can be written in terms of a sum of expected values calculated with respect to the $q$-density as:
\begin{equation*}
	\elbo (\by; q) 
	= \int_{\Theta} q(\btheta) \log \bigg\{ \frac{\pi(\by, \btheta)}{q(\btheta)} \bigg\} \,\d\btheta
	= \E_q\{\log \pi(\by, \btheta)\} - \E_q\{\log q(\btheta)\},
\end{equation*}
where $\log \pi(\by, \btheta) = \log \pi(\by | \btheta) + \log \pi(\btheta)$. From the model specification~\eqref{eq:model}, we have
\begin{align*}
	\log \pi(\by, \btheta) =
	\log \pi(\by | \btheta)
	+ \log \pi(\bbeta, \bsu | \sigma_1^2, \dots, \sigma_\H^2) 
	+ \sum_{h = 1}^\H \log \pi(\sigma_h^2).
\end{align*}
Similarly, thanks to the variational restriction~\eqref{eq:variational_restrictions}, for the variational density we have:
\begin{equation*}
	\log q(\btheta) = \log{q}(\bbeta, \bsu) + \sum_{h = 1}^{\H} \log{q}(\sigma_h^2) + \log{q}(\sigma_\veps^2).
\end{equation*}
Therefore, the lower bound can be decomposed as a sum of terms associated to different parameter blocks:
\begin{equation}
	\label{eq:lower_bound_additive_decomposition}
	\elbo(\by; q) =\;
	\underbrace{\phantom{\frac{|}{|}} \E_q \big[ \log \pi(\by | \btheta) \big]}_{\tsty T_1} 
	+ \underbrace{\E_q \bigg[ \log \frac{\pi(\bbeta, \bsu | \bsigma_u^2)}{q(\bbeta, \bsu)} \bigg]}_{\tsty T_2} 
	+ \sum_{h = 1}^{\H} \underbrace{\E_q \bigg[ \log \frac{\pi(\sigma_h^2)}{q(\sigma_h^2)} \bigg]}_{\tsty T_3}.
\end{equation}
We can thus evaluate each term separately and sum up the individual contributions $T_1$, $T_2$ and $T_3$.
The first term in~\eqref{eq:lower_bound_additive_decomposition} is the variational expectation of the log-pseudo-likelihood function:
\begin{align*}
	T_1
	= \E_q \big[ \log \pi(\by | \btheta) \big]
	= - \sum_{i=1}^{n} \E_q\{ \psi(y_i, \eta_i) \}.
\end{align*}
The second term in~\eqref{eq:lower_bound_additive_decomposition} is the expected contribution of $\bvtheta = (\bbeta, \bsu)$ to the lower bound and corresponds to the negative Kullback-Leibler divergence between the multivariate Gaussian laws $\N_d(\bmu, \bSigma)$ and $\N_d(\b0, \bS)$, where $\bS = \blockdiag \big[ \sigma_\beta^2 \bR_\beta^{-1}, \sigma_1^2 \bR_1^{-1}, \dots, \sigma_\H^2 \bR_\H^{-1} \big]$:
\begin{align*}
	T_2
	&= \E_q \Big[ - \half \log|\bS|_+ 
	- \half \bvtheta^\top \bS^{-1} \bvtheta 
	+ \half \log|\bSigma|
	+ \half (\bvtheta - \bmu)^\top \bSigma^{-1} (\bvtheta - \bmu) \Big] \\
	&= - \half \E_q \big( \log|\bS| \big)
	+ \half \log|\bSigma|
	- \half \E_q(\bvtheta^\top \bS^{-1} \bvtheta)
	+ \half \E_q \big[ (\bvtheta - \bmu)^\top \bSigma^{-1} (\bvtheta - \bmu) \big] \\
	&= - \half \Big[ d_\beta \log \sigma_\beta^2 - \log|\bR_\beta| \Big] 
	- \half \sum_{h = 1}^\H \Big[ d_h \E_q(\log \sigma_h^2) - \log|\bR_h| \Big] 
	+ \half \log|\bSigma|
	+ \xhalf{p+d} \\
	&\quad - \half (1/\sigma_\beta^2) \Big[ \bmu_\beta^\top \bR_\beta \bmu_\beta 
	+ \tr\big( \bR_\beta \bSigma_{\beta\beta} \big) \Big]
	+ \half \sum_{h = 1}^\H \E_q(1/\sigma_h^2) \Big[ \bmu_h^\top \bR_h \bmu_h + \tr \big( \bR_h \bSigma_{hh} \big) \Big].
\end{align*}
where we used factorization~\eqref{eq:variational_restrictions} and the identity $\E_q \big[ {(\bvtheta - \bmu)}^\top \bSigma^{-1} (\bvtheta - \bmu) \big] = p + d$.
The third term in~\eqref{eq:lower_bound_additive_decomposition} is the expected contribution of $\sigma_h^2$ to the lower bound, which corresponds to the negative Kullback-Leibler divergence between the Inverse-Gamma distributions $\IG(\alpha_h, \beta_h)$ and $\IG(A_h, B_h)$:
\begin{align*}
	T_3 
	&= \E_q \bigg[ \log \frac{B_h^{A_h}}{\Gamma(A_h)}
	- (A_h + 1) \log \sigma_h^2 - \frac{B_h}{\sigma_h^2}
	- \log \frac{\beta_h^{\alpha_h}}{\Gamma(\alpha_h)} 
	+ (\alpha_h + 1) \log \sigma_h^2 
	+ \frac{\beta_h}{\sigma_h^2} \,\bigg] \\
	&=	\log \frac{B_h^{A_h}}{\beta_h^{\alpha_h}} 
	- \log \frac{\Gamma(A_h)}{\Gamma(\alpha_h)} 
	- \frac{d_h}{2} \E_q (\log \sigma_h^2)
	- (B_h - \beta_h) \,\E_q (1 / \sigma_h^2),
\end{align*}
where we used $\alpha_h = A_h + d_h / 2$.

Finally, summing up the individual contributions $T_1$, $T_2$, $T_3$ and simplifying the redundant terms, we obtain the result:
\begin{align*}
	\elbo (\by; q) =
	& - \sum_{i=1}^{n} \E_q\{ \psi_0(y_i, \eta_i) \}
	- \frac{1}{2} \Big[ d_\beta \log \sigma_\beta^2 - \log|\bR_\beta| \Big] 
	- \frac{1}{2} \sum_{h = 1}^\H \Big[ d_h \E_q(\log \sigma_h^2) - \log|\bR_h| \Big] 
	+ \frac{1}{2} \log|\bSigma| \\
	& - \frac{1}{2 \sigma_\beta^2} \Big[ \bmu_\beta^\top \bR_\beta \bmu_\beta 
	+ \tr\big( \bR_\beta \bSigma_{\beta\beta} \big) \Big]
	+ \frac{1}{2} \sum_{h = 1}^\H \E_q(1/\sigma_h^2) \Big[ 
	\bmu_h^\top \bR_h \bmu_h + 
	\tr \big( \bR_h \bSigma_{hh} \big) \Big] \\
	& + \sum_{h = 1}^\H \bigg[ \log \frac{B_h^{A_h}}{\beta_h^{\alpha_h}} 
	- \log \frac{\Gamma(A_h)}{\Gamma(\alpha_h)}
	+ \frac{d_h}{2} \E_q(\log \sigma_h^2)
	- (B_h - \beta_h) \,\E_q (1 / \sigma_h^2) \bigg].
\end{align*}
The proof is concluded by noting that the second row in the formula above may be equivalently expressed as $\bmu^\top \bar\bR \bmu + \trace(\bar\bR \bSigma)$ by defining the block-diagonal matrix $\bar\bR = \blockdiag \big[ (1/\sigma_\beta^2) \bR_\beta, \E_q(1/\sigma_1^2) \bR_1, \dots, \E_q(1/\sigma_\H^2) \bR_\H \big]$.
\hfill $\square$

\subsection{Proof of Proposition~\ref{prop:optimal_natural_gradient_updates}: optimal distributions}%
\label{proof:optimal_natural_gradient_updates}

In this section, we prove Proposition~\ref{prop:optimal_natural_gradient_updates} and we derive the explicit solutions of the optimal variational distributions $q^{(t+1)}(\sigma_u^2)$ and $q^{(t+1)}(\bbeta, \bsu)$.

\subsubsection*{Optimal distribution of $\sigma_h^2$.}

Leveraging the Normal-Inverse-Gamma conjugacy, we here derive the optimal density of $\sigma_h^2$ using the mean field solution $q^{(t+1)}(\sigma_h^2) \propto \exp \big[ \E_{-\sigma_h^2} \{ \log \pi (\sigma_h^2 | \rest) \} \big]$, where $\pi(\sigma_h^2 | \rest)$ denotes the full-conditional distribution of $\sigma_h^2$ and $\E_{-\sigma_h^2}^{(t)}(\cdot)$ stands for the expected value calculated with respect to the density $q^{(t)}(\btheta) / q^{(t)}(\sigma_h^2)$.

Recalling that $\bsu_h | \sigma_h^2 \sim \N (\b0, \sigma_h^2 \bR_h^{-1})$ and $\sigma_h^2 \sim \IG(A_h, B_h)$, it is easy to show that the full-conditional log-density function of $\sigma_h^2$ is given by
\begin{equation*}
	\log \pi(\sigma_h^2 | \rest)
	= - (A_h + d_h / 2 + 1) \log \sigma_h^2 - \Big\{ B_h + (\bsu_h^\top \bR_h \,\bsu_h) / 2 \Big\} / \sigma_h^2 + \const,
\end{equation*}
which corresponds to the log-kernel of an Inverse-Gamma distribution with parameters $A_{h|\bullet} = A_h + d_h / 2$ and $B_{h|\bullet} = B_h + (\bsu_h^\top \bR_h \,\bsu_h) / 2$.
Thus, taking the variational expectation of $\log \pi(\sigma_h^2 | \rest)$ with respect to all the unknown parameters but $\sigma_h^2$, we get
\begin{equation*}
	\log q^{(t+1)}(\sigma_h^2) 
	= - (A_h + d_h / 2 + 1) \log \sigma_h^2 - \Big\{ B_h + \half \E_q^{(t)}(\bsu_h^\top \bR_h \,\bsu_h) / 2 \Big\} / \sigma_h^2 + \const.
\end{equation*}
The latter is the kernel of an Inverse-Gamma distribution with parameters $\alpha_h^{(t+1)} = A_h + d_h / 2$ and $\beta_h^{(t+1)} = B_h + \E_q^{(t)}(\bsu_h^\top \bR_h \,\bsu_h) / 2$.
This concludes the proof for $q^{(t+1)}(\sigma_h^2)$.

\subsubsection*{Optimal distribution of \texorpdfstring{$(\bbeta, \bsu)$}{(beta,u)}.}

According to Proposition~1, the evidence lower bound may be expressed as
\begin{align*}
	\elbo (\by; q, \bmu, \bSigma)
	&= - \E_q\{ \log \pi(\by | \btheta) \} - \KL\{ q(\bbeta, \bsu) \parallel \pi(\bbeta, \bsu | \bsigma_u^2) \} + \const \\
	&= - \sum_{i = 1}^{n} \E_q\{ \psi_0(y_i, \eta_i) \} + \frac{1}{2} \Big\{ \log|\bSigma| -  \bmu^\top \bar\bR \,\bmu - \tr(\bar\bR \bSigma) \Big\} + \const,
\end{align*}
where we discarded all the terms not depending on $\bmu$ and $\bSigma$.
Then, thanks to the Gaussian non-conjugate variational update proposed by \cite{Wand2014}, we can employ the natural gradient recursion~\eqref{eq:gaussian_canonical_natural_gradient} to optimize the evidence lower bound with respect to $\bmu$ and $\bSigma$.
The gradient and Hessian of $\elbo(\by; q, \bmu, \bSigma)$ are
\begin{equation*}
	\bg = - \sum_{i = 1}^{n} \frac{\partial}{\partial \bmu} \E_q\{ \psi_0(y_i, \eta_i) \} - \bar\bR \bmu,
	\quad \text{and} \quad
	\bH = - \sum_{i = 1}^{n} \frac{\partial^2}{\partial \bmu \,\partial \bmu^\top} \E_q\{ \psi_0(y_i, \eta_i) \} - \bar\bR.
\end{equation*}
We recall that the expected loss function may be written as
\begin{equation*}
	\Psi_0(y_i, \xi_i, \nu_i)
	= \E_q\{ \psi_0(y_i, \eta_i) \}
	= \int_{-\infty}^{+\infty} \psi_0(y_i, \eta) \,\phi(\eta; \xi_i, \nu_i^2) \,\d\eta,
\end{equation*}
which depends on $\bmu$ and $\bSigma$ only through the scalar variables $\xi_i = \bc_i^\top \bmu$ and $\nu_i^2 = \bc_i^\top \bSigma \,\bc_i$.
Therefore, thanks to the chain rule, we get
\begin{equation*}
	\frac{\partial \Psi_{0,i}}{\partial \bmu}
	= \frac{\partial \xi_i}{\partial \bmu} \frac{\partial \Psi_{0,i}}{\partial \xi_i}
	= \bc_i \Psi_{1,i},
	\quad \text{and} \quad
	\frac{\partial^2 \Psi_{0,i}}{\partial \bmu \,\partial \bmu^\top}
	= \frac{\partial \xi_i}{\partial \bmu} \frac{\partial \xi_i}{\partial \bmu^\top} \frac{\partial \Psi_{1,i}}{\partial \xi_i}
	= \bc_i \bc_i^\top \Psi_{2,i}.
\end{equation*}
In matrix form, we have $\bg = - \bC^\top \bPsi_1 - \bar\bR \bmu$ and $\bH = - \bC^\top \diag(\bPsi_2) \,\bC - \bar\bR$.
This concludes the proof for $q^{(t+1)}(\bbeta, \bsu)$ and also the proof of Proposition~\ref{prop:optimal_natural_gradient_updates}. 
\hfill $\square$

\subsection{Proof of Proposition~\ref{prop:psi_function_properties}: properties of the variational loss}%
\label{proof:psi_function_properties}

Let us assume that, for any $y \in \cY$, $\psi(y, \cdot) : \R \rarrow \R$ is a measurable function, having well-defined $r$th order weak derivative $\psi_r(y, \cdot)$, for $r = 1, \dots, R$.
Recall that the $r$th order weak derivative of $\psi(y,x)$ with respect to $x$, say $\psi_r(y,x)$, is defined as the integrable function satisfying the equation
\begin{equation}
	\label{eq:weak_derivative_definition}
	\int_a^b \frac{\d^r}{\d{x}^r} \vphi(x) \,\psi_0(y,x) \,\d{x}
	= (-1)^r \int_a^b \vphi(x) \,\psi_r(y,x) \,\d{x},
\end{equation}
for any infinitely differentiable test function $\vphi : [a,b] \rarrow \R$ such that $\vphi(a) = \vphi(b) = 0$.
We also recall that $\psi(y, x)$ is convex in $x$ if, and only if, $\psi_2(y,x) \geq 0$ almost everywhere.
Moreover, if $\psi(y, x)$ is $r$ times differentiable in $x$, then $\psi_r(y, x) = (\partial^r / \partial x^r) \psi(y, x)$.

\vspace{1em}\noindent\textbf{(a) Differentiability.}
The differentiability of $\Psi_0(y, \xi, \nu^2)$ with respect to $\xi$ and $\nu$ is guaranteed by the derivation under integral sign theorem and by the fact that $\phi(\,\cdot\,; \xi, \nu^2)$ is an analytic function having infinitely many continuous derivatives with respect to $\xi$ and $\nu$.
In fact, since $\Psi_0(y, \xi, \nu^2)$ depends on $\xi$ and $\nu$ only through $\phi(\,\cdot\,; \xi, \nu^2)$, we have
\begin{equation}
	\label{eq:derivation_under_integral_sign}
	\frac{\partial^{\,r}}{\partial \xi^r} \frac{\partial^{\,s}}{\partial \nu^s} \int_{-\infty}^{+\infty} \psi_0(y, x) \,\phi(x; \xi, \nu^2) \,\d{x} 
	= \int_{-\infty}^{+\infty} \psi_0(y, x) \frac{\partial^{\,r}}{\partial \xi^r} \frac{\partial^{\,s}}{\partial \nu^s} \,\phi(x; \xi, \nu^2) \,\d{x},
\end{equation}
for any non-negative integer $r$ and $s$.
Moreover, we observe that the right-hand side derivative may be written as $\nabla_\xi^r \nabla_\nu^s \,\phi(x; \xi, \nu^2) = p(x; \xi, \nu^2) \,\phi(x; \xi, \nu^2)$, where $p(x; \xi, \nu^2)$ is a linear combination of Hermite polynomials of finite order parametrized by $\xi$ and $\nu$.
Therefore, thanks to the Holder inequality, the right integral in~\eqref{eq:derivation_under_integral_sign} is bounded by
\begin{align*}
	\bigg| \int_{-\infty}^{+\infty} \psi_0(y, x) \,p(x; \xi, \nu) \,\phi(x; \xi, \nu^2) \,\d{x} \bigg|
	\leq \int_{-\infty}^{+\infty} |\psi_0(y, x) \,p(x; \xi, \nu)| \,\phi(x; \xi, \nu^2) \,\d{x} & \\
	\leq \bigg[ \int_{-\infty}^{+\infty} |\psi_0(y, x)| \,\phi(x; \xi, \nu^2) \,\d{x} \bigg] \bigg[ \int_{-\infty}^{+\infty} |p(x; \xi, \nu)| \,\phi(x; \xi, \nu^2) \,\d{x} \bigg] < \infty, & 
\end{align*}
which is finite, since $\psi(y,x)$ is integrable and the moments of a Gaussian distribution are all well-defined for any $\xi \in \R$ and $\nu > 0$.

\vspace{1em}\noindent\textbf{(b) Convergence.}
Let $x_t \sim \N(\xi, \nu_t^2)$ be a sequence of random variables such that $\nu_t^2 \rarrow 0$ as $t \rarrow \infty$.
Then, thanks to the closure of the convergence in probability with respect to continuous transformations, if $\psi(y, x)$ is continuous in $x$, we get $\dsty \psi (y, x_t) \parrow \psi (y, \xi)$ since $\dsty x_t \parrow \xi$.
Hence, $\Psi_0 (y, \xi, \nu^2) = \E \{ \psi(y, x_t) \} \rarrow \psi(y, \xi)$ pointwise as $t \rarrow \infty$.
Moreover, by the continuity of $\psi$ and $\Psi$, we have $\sup_{\xi \in \R} \big| \Psi_0(y,\xi,\nu^2) - \psi(y,\xi) \big| \rarrow 0$ uniformly for any $y \in \cY$.

\vspace{1em}\noindent\textbf{(c) Joint convexity.}
We recall that convex functions are closed with respect to convolution with positive measures. More formally, if $f(u, v)$ is a convex function in $u$ for any $v$ and if $w(v)$ is a non-negative weighting function of $v$, the integral transformation $g(u) = \int_{-\infty}^{+\infty} f(u, v) \,w(v) \,\d{v}$ is convex in $u$ (provided the integral exists); see, e.g., \cite{Lange2013}, Chapter 6.
Then, writing $\Psi_0(y, \xi, \nu^2) = \int_{-\infty}^{+\infty} \psi(y, \xi + \nu z) \,\phi(z) \,\d{z}$, the convexity of $\Psi_0(y, \xi, \nu^2)$ immediately follows from the linearity of $\xi + \nu z$ and the convexity of $\psi(y, \eta)$ with respect to $\eta$.

\vspace{1em}\noindent\textbf{(d) Majorization.}
The lower bound $\psi(y, \xi) \leq \Psi_0(y, \xi, \nu^2)$ immediately follows from the convexity of $\psi$ and Jensen's inequality: $\psi \{ y, \E(x) \} \leq \E \{ \psi(y, x) \} = \Psi_0 \{ y, \E(x), \Var(x) \}$.

\vspace{1em}\noindent\textbf{(e) Differentiation rule.}
Let consider the Gaussian random variable $x \sim \N(\xi, \nu^2)$, such that $\Psi_0(y, \xi, \nu^2) = \E\{ \psi(y, x) \}$.
To prove the identity $(\partial^r / \partial \xi^r) \,\E\{ \psi_0(y,x) \} = \E\{ \psi_r (y, x) \}$, for any $r = 1, \dots, R$, we use an induction argument.
Let us start from the initial step deriving under integral sign with respect to $\xi$:
\begin{equation*}
	\frac{\partial \Psi_0}{\partial \xi}
	= \int_{-\infty}^{+\infty} \frac{\psi_0(y,x)}{\nu} \frac{\partial}{\partial \xi} \phi \bigg(\frac{x - \xi}{\nu} \bigg) \,\d{x} \\
	= - \frac{1}{\nu} \int_{-\infty}^{+\infty} \,\frac{\psi_0(y,x)}{\nu} \,\dt\phi \bigg(\frac{x - \xi}{\nu} \bigg) \,\d{x}.
\end{equation*}
To lighten the notation, here and elsewhere, we adopt the time-derivative notation $\dt\phi(z) = \d\phi / \d{z}$.
Because of the location-scale representation of the Gaussian distribution, we write $\dsty x = \xi + \nu \, z$, where $z \sim \N(0,1)$ and $\dsty \d{x} = \nu \,\d{z}$. In this way, we have
\begin{equation*}
	\frac{\partial \Psi_0}{\partial \xi} 
	= - \frac{1}{\nu} \int_{-\infty}^{+\infty} \,\psi_0( y, \xi + \nu z) \,\dt\phi(z) \,\d{z}
	= \frac{1}{\nu} \int_{-\infty}^{+\infty} z\,\psi_0( y, \xi + \nu z) \,\phi(z) \,\d{z}.
\end{equation*}
Observing that $\dt\phi(z) = -z \,\phi(z)$ vanishes in the limit as $|z| \rarrow \infty$ for any $r$, we are allowed to integrate by parts with respect to $z$, to apply the definition of weak derivative of $\psi_0$, and finally to reparametrize back to $x$, obtaining
\begin{equation*}
	\frac{\partial \Psi_0}{\partial \xi}
	= \int_{-\infty}^{+\infty} \,\psi_1 (y, \xi + \nu z) \,\phi (z) \,\d{z}
	= \int_{-\infty}^{+\infty} \,\psi_1 (y, x) \,\phi (x; \xi, \nu) \,\d{x}
	= \E \{ \psi_1 (y, x) \},
\end{equation*}
where, thanks to the chain rule, $\d \psi_0 (y, \xi + \nu z) = \psi_1 (y, \xi + \nu z) \,\nu\,\d{z}$.
Such a result satisfy the identity in Proposition~3 for $r = 1$, hence the initial step is concluded.

Let us move to the induction step.
We consider the $r$th order derivative of $\Psi_0$, namely $\Psi_r$, and we differentiate again under integral sign with respect to $\xi$:
\begin{equation*}
	\frac{\partial}{\partial \xi} \frac{\partial^r \Psi_0}{\partial \xi^r}
	= \frac{\partial \Psi_r}{\partial \xi}
	= \int_{-\infty}^{+\infty} \psi_r (y, x) \frac{\partial}{\partial \xi} \phi(x; \xi, \nu^2) \,\d{x}.
\end{equation*}
Following the same arguments used before, sequentially applying derivation, location-scale transformation, integration by parts and back-transformation, we get
\begin{align*}
	\frac{\partial \Psi_r}{\partial \xi}
	&= \int_{-\infty}^{+\infty} \frac{\psi_r (y, x)}{\nu} \frac{\partial}{\partial \xi} \phi \bigg( \frac{x - \xi}{\nu} \bigg) \,\d{x}
	= - \frac{1}{\nu} \int_{-\infty}^{+\infty} \frac{\psi_r (y, x)}{\nu} \dt\phi \bigg( \frac{x - \xi}{\nu} \bigg) \,\d{x} \\
	&= - \frac{1}{\nu} \int_{-\infty}^{+\infty} \psi_r (y, \xi + \nu \,z) \,\dt\phi (z) \,\d{z}
	= \int_{-\infty}^{+\infty} \psi_{r+1} (y, \xi + \nu \,z ) \,\phi(z) \,\d{z} \\
	&= \int_{-\infty}^{+\infty} \psi_{r+1}(y,x) \,\phi(x; \xi, \nu^2)
	= \E\{ \psi_{r+1} (y,x) \}.
\end{align*}
This concludes the induction step and the prof for $k = 0$.

Now, we consider the following derivative $\Psi_r = (\partial^r / \partial \xi^r) \Psi_0$ for $r = u + v$ with $u = r-k$ and $v = k$:
\begin{align*}
	\Psi_r 
	= \frac{\partial^u}{\partial \xi^u} \frac{\partial^v \Psi_0}{\partial \xi^v} 
	= \frac{\partial^u \Psi_v}{\partial \xi^u}
	& = \int_{-\infty}^{+\infty} \frac{\psi_v(y,x)}{\nu} \frac{\partial^u}{\partial \xi^u} \phi \bigg( \frac{x - \xi}{\nu} \bigg) \,\d{x} \\
	& = \frac{1}{\nu^u} \int_{-\infty}^{+\infty} \frac{\psi_v(y,x)}{\nu} He_u \bigg( \frac{x - \xi}{\nu} \bigg) \,\phi \bigg( \frac{x - \xi}{\nu} \bigg) \,\d{x} \\
	& = \frac{1}{\nu^u} \int_{-\infty}^{+\infty} \psi_v(y, \xi + \nu z) \,He_u(z) \,\phi(z) \,\d{z} \\
	& = \nu^{-u} \,\E_q \{ \psi_v(y, \xi + \nu z) \,He_u(z) \}.
\end{align*}
Then, setting $u = r-k$ and $v = k$ we obtain the final result.
This concludes the proof.
\hfill $\square$

\subsection{Proof of Proposition~\ref{prop:kullback_leibler_inequalities}: Kullback-Leibler inequality}
\label{proof:kullback_leibler_inequality}

Let $q(\bomega, \btheta) = q(\bomega | \btheta) q(\btheta)$ be a density function over the product space $\Omega \times \Theta$, then we may write the Kullback-Leibler divergence between $q(\bomega, \btheta)$ and $\pi(\bomega, \btheta | \by)$ as follows
\begin{align*}
	& \KL\{ q(\bomega, \btheta) \parallel \pi(\bomega, \btheta | \by) \}
	= \KL\{ q(\bomega | \btheta) \,q(\btheta) \parallel \pi(\bomega | \btheta, \by) \,\pi(\btheta | \by) \} \\
	&\qquad = - \iint_{\Theta \times \Omega} q(\bomega | \btheta) \,q(\btheta) \log \bigg\{ \frac{\pi(\bomega | \btheta, \by) \,\pi(\btheta | \by)}{q(\bomega | \btheta) \,q(\btheta)} \bigg\} \,\d\btheta \,\d\bomega \\
	&\qquad = - \int_\Theta q(\btheta) \bigg[ \int_\Omega q(\bomega | \btheta) \log \bigg\{ \frac{\pi(\bomega | \btheta, \by) }{q(\bomega | \btheta)} \bigg\} \,\d\bomega \bigg] \,\d\btheta
	- \int_\Theta q(\btheta) \log \bigg\{ \frac{\pi(\btheta | \by) }{q(\btheta)} \bigg\} \,\d\btheta \\
	&\qquad = \int_\Theta q(\btheta) \,\KL\{ q(\bomega | \btheta) \parallel \pi(\bomega | \btheta, \by) \} \,\d\btheta + \KL\{ q(\btheta) \parallel \pi(\btheta | \by) \} \\
	&\qquad = \underbrace{\E_q \big[ \KL\{ q(\bomega | \btheta) \parallel \pi(\bomega | \btheta, \by) \} \big]}_{\geq 0} \;+\; \underbrace{\KL\{ q(\btheta) \parallel \pi(\btheta | \by) \}}_{\geq 0}.
\end{align*}
Notice that the first term nullifies if and only if $\KL\{ q(\bomega | \btheta) \parallel \pi(\bomega | \btheta, \by) \} = 0$ for almost every $\btheta \in \Theta$, while the second term  corresponds to the Kullback-Leibler divergence calculated over the marginal posterior distribution of $\btheta$, which does not depend on $q(\bomega | \btheta)$.
The optimal posterior approximation of $\pi(\bomega, \btheta | \by)$ is the solution of the following variational problem:
\begin{align*}
	& \min_{q(\btheta) \in \cQ_\M, q(\bomega | \btheta) \in \cQ_\C} \E_q \big[ \KL\{ q(\bomega | \btheta) \parallel \pi(\bomega | \btheta, \by) \} \big] + \KL\{ q(\btheta) \parallel \pi(\btheta | \by) \} \\
	& \quad =  \min_{q(\btheta) \in \cQ_\M} \bigg[ \min_{q(\bomega | \btheta) \in \cQ_\C} \E_q \big[ \KL\{ q(\bomega | \btheta) \parallel \pi(\bomega | \btheta, \by) \} \big] + \KL\{ q(\btheta) \parallel \pi(\btheta | \by) \}  \bigg],
\end{align*}
which involve a first profiling with respect to $q(\bomega | \btheta) \in \cQ_\C$ and an external optimization with respect to $q(\btheta) \in \cQ_\M$.
Therefore, denoting by $q_\A^*(\bomega | \btheta) \in \cQ_\C$ the optimal approximation of $\pi(\bomega | \btheta, \by)$ for any $\btheta \in \Theta$, which is the minimizer of $\E_q \big[ \KL\{ q(\bomega | \btheta) \parallel \pi(\bomega | \btheta, \by) \} \big]$, we obtain
\begin{equation}
	\label{eq:marginal_augmented_optimal_density}
	q_\A^*(\btheta) = \argmin_{q(\btheta) \in \cQ_\M} \;\KL\{ q(\btheta) \parallel \pi(\btheta | \by) \} + \E_q \big[ \KL\{ q_\A^*(\bomega | \btheta) \parallel \pi(\bomega | \btheta, \by) \} \big].
\end{equation}
We now consider two scenarios, respectively, $\pi(\bomega | \btheta, \by) \in \cQ_\C$ and $\pi(\bomega | \btheta, \by) \notin \cQ_\C$.
In the first scenario, thanks to the fundamental properties of the Kullback-Leibler divergence \citep{Kullback1951}, the unique solution $q_\A^*(\bomega | \btheta)$ is given by 
\begin{align*}
	q_\A^*(\bomega | \btheta)
	& = \argmin_{q(\bomega | \btheta) \in \cQ_\C} \E_q \big[ \KL\{ q(\bomega | \btheta) \parallel \pi(\bomega | \btheta, \by) \} \big] \\
	& = \argmin_{q(\bomega | \btheta) \in \cQ_\C} \KL\{ q(\bomega | \btheta) \parallel \pi(\bomega | \btheta, \by) \} = \pi(\bomega | \btheta, \by),
\end{align*}
since the Kullback-Leibler divergence is strictly convex with respect to $q(\bomega | \btheta)$, and $q_\A^*(\bomega | \btheta) = \pi(\bomega | \btheta, \by)$ is the only density function such that $\KL\{ q_\A^*(\bomega | \btheta) \parallel \pi(\bomega | \btheta, \by) \} = 0$, for any $\btheta \in \Theta$. 
As a consequence, the optimal variational approximation of $\pi(\btheta | \by)$ is the solution of
\begin{align*}
	q_\A^*(\btheta) 
	& = \argmin_{q(\btheta) \in \cQ_\M} \KL\{ q(\btheta) \parallel \pi(\btheta | \by) \} + \E_q \big[ \KL\{ q_\A^*(\bomega | \btheta) \parallel \pi(\bomega | \btheta, \by) \} \big] \\
	& = \argmin_{q(\btheta) \in \cQ_\M} \KL\{ q(\btheta) \parallel \pi(\btheta | \by) \} = q_\M^*(\btheta),
\end{align*}
which, by definition, leads to $q_\A^*(\btheta) = q_\M^*(\btheta)$ almost everywhere.
The equality of the Kullback-Leibler divergences directly follows from the correspondence between $q_\A^*(\btheta)$ and $q_\M^*(\btheta)$ and by the global optimality of $q_\A^*(\bomega | \btheta) = \pi(\bomega | \btheta, \by)$.
Then, the resulting approximation for $\pi(\bomega, \btheta | \by)$ corresponds to $q_\A^*(\bomega, \btheta) = \pi(\bomega | \btheta, \by) \,q_\A^*(\btheta)$.
This concludes the first part of the proof.

The second scenario we consider is $\pi(\bomega | \btheta, \by) \neq \cQ_\C$.
In this case, for any $q(\bomega | \btheta) \in \cQ_\C$ and $\btheta \in \Theta$ we have $\KL\{ q(\bomega | \btheta) \parallel \pi(\bomega | \btheta, \by) \} > 0$ , in fact $q_\A^*(\bomega | \btheta) \neq \pi(\bomega | \btheta, \by)$ almost everywhere.
As a consequence, the optimal distribution $q_\A^*(\btheta)$ is the solution of the regularized variational problem in \eqref{eq:marginal_augmented_optimal_density}.
By strict convexity and by definition of global minimum, $\KL\{ q_\M^*(\btheta) \parallel \pi(\btheta | \by) \} \leq \KL\{ \bar{q}(\btheta) \parallel \pi(\btheta | \by) \}$ for any $\bar{q}(\btheta) \in \cQ_\M$.
In particular, the inequality holds for $\bar{q}(\btheta) = q_\A^*(\btheta)$, and
\begin{align*}
	\KL\{ q_\M^*(\btheta) \parallel \pi(\btheta | \by) \} 
	& \leq \KL\{ q_\A^*(\btheta) \parallel \pi(\btheta | \by) \} \\
	& < \KL\{ q_\A^*(\btheta) \parallel \pi(\btheta | \by) \} + \E_{q_\A^*} \big[ \KL\{ q_\A^*(\bomega | \btheta) \parallel \pi(\bomega | \btheta, \by) \} \big].
\end{align*}
This concludes the proof of Proposition~\ref{prop:kullback_leibler_inequalities}.
\hfill $\square$

\begin{remark}
	An interesting consequence of the above proof is that, in the marginal approximation setting, the implied joint approximation for the augmented posterior $\pi(\bomega, \btheta | \by)$ is given by $q_\M^*(\bomega, \btheta) = \pi(\bomega | \btheta, \by) \,q_\M^*(\btheta)$.
	This leads to the inequality $\KL\{ q_\M^*(\bomega, \btheta) \parallel \pi(\bomega, \btheta | \by) \} \leq \KL\{ q_\A^*(\bomega, \btheta) \parallel \pi(\bomega, \btheta | \by) \}$.
	Also, we have $\KL\{ q_\M^*(\btheta) \parallel \pi(\btheta | \by) \} = \KL\{ q_\M^*(\bomega, \btheta) \parallel \pi(\bomega, \btheta | \by) \}$.
	Then, under suitable compatibility conditions, the marginal approximation $q_\M^*$ dominates the augmented approximation $q_\A^*$ both in the marginal and augmented Kullback-Leibler metrics.
\end{remark}

\begin{remark}
	\label{rmrk:optimality}
	The key assumption in Proposition~\ref{prop:kullback_leibler_inequalities} is that $q_\M^*(\btheta)$ and $q_\A^*(\btheta)$ belong in the same functional space $\cQ_\M$.
	Whenever such a compatibility assumption is not satisfied there are no guaranties that Proposition~\ref{prop:kullback_leibler_inequalities} holds true.
	In the following, we show this fact by using a counterexample. 
\end{remark}

\begin{example}
	Let us consider a generic augmented space $\cQ_\A$ such that $\pi(\bomega, \btheta | \by) \in \cQ_\A$. On the opposite, we consider $\cQ_\M$ such that $\pi(\btheta | \by) \notin \cQ_\M$. This way, $\cQ_\A$ and $\cQ_\M$ are not compatible by construction, since there exists at least one element of $\cQ_\A$ whose marginal density does not belong to $\cQ_\M$, namely $\pi(\btheta | \by) = \int_\Omega \pi(\bomega, \btheta | \by) \,\d\bomega$.
	Recall that the Kullback-Leibler divergence is always non-negative, i.e. $\KL(q \parallel \pi) \geq 0$, and reaches 0 if and only if $q = \pi$ almost everywhere, namely $\KL(\pi \parallel \pi) = 0$. Then, there exists at least one element of $\cQ_\A$ such that $\KL \{ q_\A (\bomega, \btheta) \parallel \pi(\bomega, \btheta | \by) \} = 0$, which corresponds to $q_\A (\bomega, \btheta) = \pi(\bomega, \btheta | \by)$, whereas $\KL \{ q_\M (\btheta) \parallel \pi(\btheta | \by) \} > 0$ for any $q_\M (\btheta) \in \cQ_\M$, since $\pi(\btheta | \by) \notin \cQ_\M$. This means that
	\begin{align*}
		\underbrace{ 
			\KL \{ q_\M(\btheta) \parallel \pi(\btheta | \by) \}
		}_{\textstyle > 0} 
		\;\nleq\; 
		\underbrace{
			\KL \{ \pi(\bomega, \btheta | \by) \parallel \pi(\bomega, \btheta | \by) \}
		}_{\textstyle = 0}, \qquad \forall\, q_\M(\btheta) \in \cQ_\M,
	\end{align*}
	which contradicts the inequality in Proposition~\ref{prop:kullback_leibler_inequalities}.
\end{example}

\section{Variational loss function derivation}%
\label{app:psi_function_derivation}

We here derive the explicit closed-form expressions of the variational loss functions $\Psi_r(y, \xi, \nu^2)$ under several pseudo-likelihood specifications.
To do so, we first define $\delta_0$ as the Dirac delta function at $0$, we denote by $x \sim \N(\xi, \nu^2)$ the generic Gaussian random variables with mean $\xi \in \R$ and variance $\nu^2 > 0$ and we set the non-random constants $a, b \in \R$ and $c > 0$.
Then, we recall the following identities:
\begin{align}
	\label{eq:expected_delta_function}
	\E \{ \delta_0(x) \} &= \phi(0; \xi, \nu^2), \\
	\label{eq:expected_sign_function}
	\E \{ \sign(x) \} &= 1 - 2 \Phi(0; \xi, \nu^2), \\
	\label{eq:expected_absolute_value}
	\E \{ |x| \} &= \xi - 2 \xi \Phi(0; \xi, \nu^2) + 2 \nu^2 \phi(0; \xi, \nu^2), \\
	\label{eq:expected_truncated_first_moment}
	\E\{ x \,\bbI_{[a,b]}(x) \} &= \xi \big[ \Phi(b; \xi, \nu^2) - \Phi(a; \xi, \nu^2) \big] - \nu^2 \big[ \phi(b; \xi, \nu^2) - \phi(a; \xi, \nu^2) \big], \\
	\label{eq:expected_truncated_second_moment}
	\E\{ x^2 \,\bbI_{[a,b]}(x) \} &= (\xi^2 + \nu^2) \big[ \Phi(b; \xi, \nu^2) - \Phi(a; \xi, \nu^2) \big] - 2 \nu^2 \xi \big[ \phi(b; \xi, \nu^2) - \phi(a; \xi, \nu^2) \big] \\
	\nonumber & \qquad - \nu^2 \big[ (b - \xi) \phi(b; \xi, \nu^2) - (a - \xi) \phi(a; \xi, \nu^2) \big].
\end{align}
Moreover, in the following derivations, we will make use of a simplified a notation for the identity functions of left and right unbounded intervals, that is: $\bbI_{<b}(x) = \bbI_{(-\infty,b]}(x)$ and $\bbI_{>a}(x) = \bbI_{[a,+\infty)}(x)$.

\subsection{Support vector classification}

By the definition of Hinge loss function $\psi^\SVC (y,\eta) = \psi^\SVC (1 - y \eta) = 2 \max(0, 1 - y \eta)$, we may write
\begin{align*}
	\psi_0^\SVC (x) &= 2 \max(0,x) = |x| + (x), \\
	\psi_1^\SVC (x) &= - y \,\sign(x) - y = - 2 y \big[ 1 - \bbI_{<0}(x) \big], \\
	\psi_2^\SVC (x) &= 2 \,\delta_0(x),
\end{align*}
where $x = 1 - y \eta \sim \N(1 - y \xi, \nu^2)$.
Then, defining $z = (1 - y \xi) / \nu$, taking the expectation and using~\eqref{eq:expected_sign_function} and~\eqref{eq:expected_absolute_value}, we get
\begin{align*}
	\Psi_0^\SVC (y, \xi, \nu^2) = \E \{ \psi_0^\SVC (x) \}
	&= \E \,|x| + \E(x)
	= 2 \nu z \Phi(z) + 2 \nu \phi(z), \\
	\Psi_1^\SVC (y, \xi, \nu^2) = \E \{ \psi_1^\SVC (x) \}
	&= - 2 y \big[ 1 - \,\E\{\bbI_{< 0}(x)\} \big]
	= - 2 y \Phi(z), \\
	\Psi_2^\SVC (y, \xi, \nu^2) = \E \{ \psi_2^\SVC (x) \}
	&= 2 \,\E \{\delta_0(x)\} 
	= 2 \nu^{-1} \phi(z).
\end{align*}
This concludes the derivation.

\subsection{Support vector regression}

By the definition of $\eps$-insensitive loss function $\psi^\SVR (y,\eta) = \psi ^\SVR(y - \eta) = \max(0, |y - \eta| - \eps)$, we may write
\begin{align*}
	\psi_0^\SVR (x) &= \half |x - \eps| + \half (x - \eps) + \half |x + \eps| - \half (x + \eps), \\
	\psi_1^\SVR (x) &= - \half \sign(x - \eps) - \half \sign(x + \eps), \\
	\psi_2^\SVR (x) &= 2 \delta_0 (x - \eps) + 2 \delta_0 (x + \eps),
\end{align*}
where $x = y - \eta \sim \N(y - \xi, \nu^2)$.
Then, defining $z_\eps^- = (y - \xi - \eps) / \nu$ and $z_\eps^+ = (y - \xi + \eps) / \nu$, and using identity~\eqref{eq:expected_absolute_value}, we have
\begin{align*}
	\half \E \,|x - \eps| + \half \E(x - \eps) &= \nu z_\eps^- \Phi(z_\eps^-) + \nu \phi(z_\eps^-), \\
	\half \E \,|x + \eps| - \half \E(x + \eps) &= \nu z_\eps^+ (\Phi(z_\eps^+) - 1) + \nu \phi(z_\eps^+),
\end{align*}
and
\begin{align*}
	\Psi_0^\SVR (y, \xi, \nu^2) = \E\{ \psi_0^\SVR (x) \} 
	& = \half \big[ \E \,|x - \eps| + \E(x - \eps) + \E \,|x + \eps| - \E(x + \eps) \big] \\
	& = \nu \big[ z_\eps^- \Phi(z_\eps^-) + \phi(z_\eps^-) + z_\eps^+ (\Phi(z_\eps^+) - 1) + \phi(z_\eps^+) \big].
\end{align*}
Similarly, using~\eqref{eq:expected_sign_function}, we have
\begin{align*}
	\Psi_1^\SVR(y, \xi, \nu^2) = \E \{ \psi_1^\SVR (x) \}
	& = - \half \E\{ \sign(x - \eps) \} - \half \E\{ \sign(x + \eps) \} = 1 - \Phi(z_\eps^+) - \Phi(z_\eps^-), \\
	\Psi_2^\SVR(y, \xi, \nu^2) = \E \{ \psi_2^\SVR (x) \}
	& = \E\{ \delta_{0}(x - \eps) \} + 2 \, \E\{ \delta_{0}(x + \eps) \} = \nu^{-1} \phi(z_\eps^+) + \nu^{-1} \phi(z_\eps^-).
\end{align*}
This concludes the derivation.

\subsection{Quantile regression}

By the definition of quantile check loss $\psi^\QR(y,\eta) = \psi(y - \eta) = (y - \eta) [ \tau - \bbI_{\leq 0}(y - \eta)]$, we may write
\begin{align*}
	\psi_0^\QR (x) &= \half |x| + (\tau - \half) (x), \\
	\psi_1^\QR (x) &= - \half \sign(x) - (\tau - \half), \\
	\psi_2^\QR (x) &= \delta_0(x),
\end{align*}
where $x = y - \eta \sim \N(y - \xi, \nu^2)$.
Then, defining $z = (y - \xi) / \nu$, taking the expectation and using~\eqref{eq:expected_sign_function} and~\eqref{eq:expected_absolute_value}, we get
\begin{align*}
	\Psi_0^\QR (y, \xi, \nu^2) = \E \big\{ \psi_0^\QR (x) \big\}
	&= \half \E(|x|) + (\tau - \half) \E(x)
	= \nu z (\Phi(z) - 1 + \tau) + \nu \,\phi(z), \\
	\Psi_1^\QR (y, \xi, \nu^2) = \E \big\{ \psi_1^\QR (x) \big\}
	&= - \half \E\{\sign(x)\} - (\tau - \half)
	= 1 - \tau - \Phi(z) , \\
	\Psi_2^\QR (y, \xi, \nu^2) = \E \big\{ \psi_2^\QR (x) \big\}
	&= \E \{ \delta_0(x) \}
	= \nu^{-1} \phi(z).
\end{align*}
This concludes the derivation.

\subsection{Expectile regression}

By the definition of expectile loss $\psi^\ER (y,\eta) = \psi^\ER (y - \eta) = \half (y - \eta)^2 |\tau - \bbI_{\leq 0}(y - \eta)|$, we may write
\begin{align*}
	\psi_0^\ER (x) &= \half x^2 \big[ \tau + (1-2 \tau) \bbI_{< 0}(x) \big], \\
	\psi_1^\ER (x) &= - x \big[ \tau + (1-2 \tau) \bbI_{< 0}(x) \big], \\
	\psi_2^\ER (x) &= \tau + (1-2 \tau) \,\bbI_{< 0}(x),
\end{align*}
where $x = y - \eta \sim \N(y - \xi, \nu^2)$.
Then, defining $z = (y - \xi) / \nu$, taking the expectation and using~\eqref{eq:expected_truncated_second_moment}, we have
\begin{align*}
	\Psi_0^\ER (y, \xi, \nu^2) = \E \{ \psi_0^\ER (x) \}
	&= \half \tau \E (x^2) + \half (1-2 \tau) \,\E \{ x^2 \,\bbI_{<0}(x) \} \\
	&= \half \tau \nu^2 (z^2 + 1) + \half (1-2 \tau) \big[ \nu^2 (z^2 + 1) (1 - \Phi(z)) - \nu^2 z \phi(z) \big] \\
	&= \half \nu^2 (z^2 + 1) \big[ \tau + (1-2 \tau) (1 - \Phi(z)) \big] - \half (1-2 \tau) \nu^2 z \phi(z).
\end{align*}
In the same way, using~\eqref{eq:expected_truncated_first_moment}, we get
\begin{align*}
	\Psi_1^\ER (y, \xi, \nu^2) = \E \{ \psi_1^\ER (x) \}
	&= - \tau \E(x) - (1-2 \tau) \,\E \{ x \,\bbI_{<0}(x) \} \\
	&= - \nu z \big[ \tau + (1-2 \tau) (1 - \Phi(z)) \big] + (1-2 \tau) \nu \phi(z),
\end{align*}
and 
\begin{equation*}
	\Psi_2^\ER (y, \xi, \nu^2) = \E \{ \psi_2^\ER (x) \}
	= \tau + (1-2 \tau) \,\E \{ \bbI_{< 0}(x) \}
	= \tau + (1-2 \tau) (1 - \Phi(z)).
\end{equation*}
This concludes the derivation.

\subsection{Huber regression}

By the definition of Huber regression loss, we may write $\psi^\HR (y, \eta) = \psi^\HR (y - \eta)$ so that
\begin{align*}
	\psi_0^\HR (x) &= \xfrac{1}{2\eps} x^2 \bbI_{\leq\eps}(|x|) - x \bbI_{< -\eps}(x) + x \bbI_{> \eps}(x) - \xfrac{\eps}{2} [ \bbI_{< -\eps}(x) + \bbI_{> \eps}(x) ], \\
	\psi_1^\HR (x) &= - \xfrac{1}{\eps} x \,\bbI_{\leq\eps}(|x|) - \bbI_{< -\eps}(x) + \bbI_{> \eps}(x), \\
	\psi_2^\HR (x) &= \xfrac{1}{\eps} \bbI_{\leq\eps}(|x|),
\end{align*}
where $x = y - \eta \sim \N(y - \xi, \nu^2)$.
Then, defining defining $z = (y - \xi) / \nu$, $z_\eps^- = (y - \xi - \eps) / \nu$ and $z_\eps^+ = (y - \xi + \eps) / \nu$, and using identities~\eqref{eq:expected_truncated_first_moment} and~\eqref{eq:expected_truncated_second_moment}, we get
\begin{equation*}
	\begin{aligned}
		\Psi_0^\HR (y, \xi, \nu^2) = \; & \E \{ \psi_0^\HR (x) \} = \\
		= \; & \half \nu^2 (z^2 + 1) \big[ \Phi(z_\eps^+) - \Phi(z_\eps^-) \big] + \nu^2 z \big[ \phi(z_\eps^+) - \phi(z_\eps^-) \big]
		- \half \nu^2 \big[ z_\eps^+ \phi(z_\eps^+) - z_\eps^- \phi(z_\eps^-) \big] \\
		& - \eps \nu \big[ z (1 - \Phi(z_\eps^+)) - \phi(z_\eps^+) \big]
		+ \eps \nu \big[ z \Phi(z_\eps^-) + \phi(z_\eps^-) \big]
		- \half \eps^2 \big[ 1 - \Phi(z_\eps^+) + \Phi(z_\eps^-) \big].
	\end{aligned}
\end{equation*}
In the same way, we use~\eqref{eq:expected_truncated_first_moment} to find the expectations of $\psi_1^\HR (x)$ and $\psi_2^\HR (x)$
\begin{align*}
	\Psi_1^\HR (y, \xi, \nu^2) = \E\{ \psi_1^\HR (x) \}
	&= - \nu z \big[ \Phi(z_\eps^+) - \Phi(z_\eps^-) \big] - \nu \big[ \phi(z_\eps^+) - \phi(z_\eps^-) \big] + \eps \big[ 1 - \Phi(z_\eps^+) - \Phi(z_\eps^-) \big], \\
	\Psi_2^\HR (y, \xi, \nu^2) = \E\{ \psi_2^\HR (x) \} &= \xfrac{1}{\eps} \big[ \Phi(z_\eps^+) - \Phi(z_\eps^-) \big].
\end{align*}
This concludes the derivation.

\subsection{Huber classification}

By the definition of Huber classification loss, we may write $\psi^\HC(y,\eta) = \psi^\HC(1 - y \eta)$ so that
\begin{align*}
	\psi_0^\HC (x) &= \xfrac{1}{4\eps}  (\eps + x)^2 \,\bbI_{\leq\eps}(|x|) + (x) \bbI_{>\eps}(1 - y\eta), \\
	\psi_1^\HC (x) &= - \xfrac{1}{2\eps} y (\eps + x) \,\bbI_{\leq\eps}(|x|) - y \,\bbI_{>\eps}(x), \\
	\psi_2^\HC (x) &= \xfrac{1}{2\eps} \bbI_{\leq\eps}(|x|),
\end{align*}
where $x = 1 - y \eta \sim \N(1 - y \xi, \nu^2)$.
Then, defining $z = (1 - y \xi) / \nu$, $z_\eps^+ = (1 - y \xi + \eps) / \nu$ and $z_\eps^+ = (1 - y \xi - \eps) / \nu$, and using identities~\eqref{eq:expected_truncated_first_moment} and~\eqref{eq:expected_truncated_second_moment}, we get
\begin{equation*}
	\begin{aligned}
		\Psi_0^\HC (y, \xi, \nu^2) = \; & \E \{ \psi_0^\HC (x) \} =
		\xfrac{1}{4\eps} \E \{ x^2 \bbI_{\leq\eps}(|x|) \}
		+ \xfrac{1}{2} \E \{ x \bbI_{\leq\eps}(|x|) \}
		+ \xfrac{\eps}{4} \E \{ \bbI_{\leq\eps}(|x|) \}
		+ \E \{ x \bbI_{>\eps}(x) \} \\
		=\;
		& \xfrac{1}{4\eps} \big[ \nu^2 (z^2+1) (\Phi(z_\eps^+) - \Phi(z_\eps^-)) + y \,\nu^2 z (\phi(z_\eps^+) - \phi(z_\eps^-)) - \eps \nu (\phi(z_\eps^+) + \phi(z_\eps^-))\big] \\
		& - \half \big[ y \nu z (\Phi(z_\eps^+) - \Phi(z_\eps^-)) + \nu (\phi(z_\eps^+) - \phi(z_\eps^-)) \big]
		+ \xfrac{\eps}{4} \big[ \Phi(z_\eps^+) - \Phi(z_\eps^-) \big].
	\end{aligned}
\end{equation*} 
In the same way, we use~\eqref{eq:expected_truncated_first_moment} to find the expectations of $\psi_1^\HC (y, \eta)$ and $\psi_2^\HC (y, \eta)$
\begin{align*}
	\Psi_1^\HC (y, \xi, \nu^2) = \E \{ \psi_1^\HC (x) \} &= - \xfrac{y}{2 \eps} \big[ \mu (\Phi_\eps^+ - \Phi_\eps^-) - \nu^2 (\phi_\eps^+ - \phi_\eps^-) \big] - \xfrac{y}{2} (2 - \Phi_\eps^+ - \Phi_\eps^-), \\
	& = \xfrac{y}{2 \eps} \big[ \nu z (\Phi(z_\eps^+) - \Phi(z_\eps^-)) + \nu (\phi(z_\eps^+) - \phi(z_\eps^-)) \big] - \xfrac{y}{2} \big[ \Phi(z_\eps^+) + \Phi(z_\eps^-) \big] \\
	\Psi_2^\HC (y, \xi, \nu^2) = \E \{ \psi_2^\HC (x) \} &= \xfrac{1}{2 \eps} \big[ \Phi(z_\eps^+) - \Phi(z_\eps^-) \big].
\end{align*}
This concludes the derivation.

\section{Non-conjugate variational algorithms}%
\label{app:non_conjugate_variational_algorithms}

\subsection{Batch optimization algorithm}

The recursive refinement of the variational parameters in Proposition~\ref{prop:optimal_natural_gradient_updates} of the main paper gives rise to the non-conjugate variational Bayes optimization routine summarized in Algorithm \ref{alg:ncvb_uf_alg}.
We assess the algorithm convergence by monitoring the relative change of the lower bound, which is defined as $\big| \,\elbo \{ \by; q^{(t+1)}(\btheta) \} / \elbo \{ \by; q^{(t)}(\btheta) \} - 1 \big|$.

\begin{algorithm}[h!]
	
	\While{convergence is not reached}{
		
		$\bxi^{(t)} \gets \bC \bmu^{(t)}$; \quad
		$\bnu^{(t)} \gets {\dsty \stack_{i = 1, \dots, n}} \big[ \bc_i^\top \bSigma^{(t)} \bc_i \big]^{1/2}$; \hfill $\cO(n K_1 + nK_1^2)$ \\
		
		Evaluate $\bPsi_r^{(t)} = \Psi_r(\by, \bxi^{(t)}, \bnu^{(t)})$, $r = 0, 1, 2$; 
		\hfill $\cO(n)$ \\
		
		$\bW^{(t)} \gets \diag \big[ \bPsi_2^{(t)} \big]$; \quad
		$\bp^{(t)} \gets - \bPsi_1^{(t)} / \bPsi_2^{(t)} + \bxi^{(t)}$; \hfill
		$\cO(n)$ \\
		
		\For{$h$ from $1$ to $H$}{
			$\beta_h^{(t+1)} \gets  B_h + \frac{1}{2} \big\{ \bmu_h^{(t)\top} \bR_h \,\bmu_h^{(t)} + \tr(\bR_h \bSigma_{hh}^{(t)}) \big\}$; \hfill
			$\cO(d_h^2)$ \\
			
			$\gamma_h^{(t+1)} \gets (A_h + d_h / 2) / \beta_h^{(t+1)}$; \hfill
			$\cO(1)$
		}
		
		$\bar\bR^{(t)} \gets \blockdiag \big[ \sigma_\beta^{-2} \bR_\beta, \gamma_1^{(t+1)} \bR_1, \dots, \gamma_\H^{(t+1)} \bR_\H \big]$; \hfill
		$\cO(K_2)$ \\
		$\bSigma^{(t+1)} \gets \big[ \bar\bR^{(t)} - \bC^\top \bW^{(t)} \bC \big]^{-1}$; \hfill $\cO(n K_1^2 + K_1^3)$ \\
		$\bmu^{(t+1)} \gets \bSigma^{(t+1)} \bC^\top \bW^{(t)} \bp^{(t)}$; \hfill $\cO(n K_1^2)$ \\
	} 
	
	\caption{%
		\label{alg:ncvb_uf_alg}
		NCVB-UF -- Pseudo-code description of the proposed \emph{non-conjugate variational Bayes} algorithm for \emph{unfactorized} approximate Bayesian inference in pGLMMs. On the right, we report the computational cost in number of floating point operations for each step.
	}
\end{algorithm}

As discussed in the main paper, Section \ref{subsec:psi_function_computation}, the variational loss derivatives, say $\Psi_r(y_i, \xi_i, \nu_i)$, can be evaluated using analytic expressions or numerical quadrature.
In both cases, such a computation can be performed in parallel for each observational unit, where a single evaluation require $\cO(1)$ in the analytic case and $\cO(N)$ for the numerical quarature, where $N$ is the number of the considered integration knots.
Then, defining $K_r = p^r + \sum_{h = 1}^{\H} d_h^r$, for $r \in \bbN$, one iteration of Algorithm~\ref{alg:ncvb_uf_alg} requires $\cO(K_2 + nK_1^2 + K_1^3)$ flops and $\cO(nK_1 + K_1^2)$ memory allocations per iteration.
This is equivalent to standard implementations of expectation-maximization, Gibbs sampling and mean field variational Bayes for the estimation of (conjugate) mixed regression models.
Under the same variational restriction, cheaper iterations may be obtained by exploiting problem-specific sparsity structures of the design and prior matrices; see, e.g., \cite{Tan2018}, \cite{Nolan2020a}, \cite{Nolan2020b} and \cite{Menictas2023}.
Alternatively, additional factorization restrictions can be considered at the cost of losing precision in the posterior approximation.

\subsection{Restricted non-conjugate algorithms}

As discussed in Section~\ref{subsec:alternative_variational_factorizations} of the main text, restricted approximation can be obtained by projecting the unfactorized Gaussian approximation obtained at each iteration of the natural gradient update using a Kullback-Leibler minimization within the restricted space $\cQ_\RF$.
Here, we discuss two alternative restricted variational approximations corresponding to the mean field assumptions in \eqref{eq:mean_field_assumption}. Throughout, we suppress the iteration counter to lighten the notation.
Before deriving the fully and partially factorized approximations and the corresponding algorithm, we recall the following results.

\begin{result}
	\label{res:optimal_density_ff}
	Let $q_\UF(\btheta)$ be the density of $\N(\bmu^\UF, \bSigma^\UF)$ and denote with $\bOmega^\UF = (\bSigma^\UF)^{-1}$ the corresponding precision matrix.
	Let $q_\FF(\btheta) = \prod_{h = 1}^{\H} q_\FF(\btheta_h)$ be the \emph{fully factorized} approximation of $q_\UF(\btheta)$.
	Then, the coordinate-wise minimizer
	\begin{equation*}
		q_\FF^*(\btheta_h) 
		= \argmin_{q(\btheta_h)} \;\KL \bigg\{ \prod_{h = 1}^{\H} q_\FF(\btheta_h) \,\bigg\Vert\, q_\UF(\btheta) \bigg\}
		\propto \exp\big[ \E_{-\btheta_h} \{ \log q_\UF(\btheta_h | \rest) \}\big]
	\end{equation*}
	is the Gaussian density of $\N(\bmu_h^\FF, \bSigma_h^\FF)$ with mean and variance
	\begin{equation*}
		\bmu_h^\FF = \bSigma_{hh}^\FF \bigg( \sum_{\ell = 0}^{\H} \bOmega_{h\ell}^\UF \bmu_\ell^\UF - \sum_{\ell \neq h} \bOmega_{h\ell}^\UF \bmu_\ell^\FF \bigg), \qquad
		\bSigma_{hh}^\FF = \big( \bOmega_{hh}^\UF \big)^{-1}.
	\end{equation*}
\end{result}

\begin{result}
	\label{res:optimal_density_pf}
	Let $q_\UF(\btheta)$ be the density of $\N(\bmu^\UF, \bSigma^\UF)$ and denote with $\bOmega^\UF = (\bSigma^\UF)^{-1}$ the corresponding precision matrix.
	Let $q_\PF(\btheta) = q_\PF(\btheta_{\cC} | \btheta_{\cH}) \prod_{h \in \cH}^{} q_\PF(\btheta_h)$ be the \emph{partially factorized} approximation of $q_\UF(\btheta)$, where $(\cC, \cH)$ is an arbitrary partition of the parameter indices.
	Then, the coordinate-wise minimizers 
	\begin{align*}
		q_\PF^*(\btheta_{\cC} | \btheta_{\cH}) &
		= \argmin_{q(\btheta_{\cC} | \btheta_{\cH})} \;\KL \bigg\{ q_\PF^{}(\btheta_{\cC} | \btheta_{\cH}) \prod_{h \in \cH}^{} q_\PF(\btheta_h) \,\bigg\Vert\, q_\UF^{}(\btheta) \bigg\}
		= q_\UF(\btheta_{\cC} | \btheta_{\cH}),
		\quad\text{and}\quad \\
		q_\PF^*(\btheta_h) &
		= \argmin_{q(\btheta_h)} \;\KL \bigg\{ q_\PF(\btheta_{\cC} | \btheta_{\cH}) \prod_{h \in \cH}^{} q_\PF(\btheta_h) \,\bigg\Vert\, q_\UF^{}(\btheta) \bigg\}
		\propto \exp \bigg[ \E_{-\btheta_h} \bigg\{ \log \int q_\UF^{}(\btheta_{\cC}, \btheta_{\cH}) \,\d\btheta_{\cC} \bigg\} \bigg],
	\end{align*}
	are, respectively, the densities of $\N(\bmu_{\cC|\cH}^\PF, \bSigma_{\cC|\cH}^\PF)$ and $\N(\bmu_h^\PF, \bSigma_h^\PF)$, for $h \in \cH$, with mean and variance
	\begin{eqnarray*}
		\bmu_{\cC|\cH}^\PF = \bSigma_{\cC|\cH}^\PF \bigg( \sum_{h \in \cH}^{} \bOmega_{\cC h}^\UF \bmu_h^\UF - \sum_{h \in \cH} \bOmega_{\cC h}^\UF \btheta_h \bigg), & & 
		\bSigma_{\cC|\cH}^\PF = \big( \bOmega_{\cC\cC}^{\UF} \big)^{-1}, \\
		\bmu_h^\PF = \bSigma_{hh}^\PF \Big( \bdelta_h - \bOmega_{h\cC}^\UF \big( \bOmega_{\cC\cC}^\UF \big)^{-1} \bdelta_{\cC} \Big), & & \bSigma_{hh}^\PF = \Big( \bOmega_{hh}^\UF - \bOmega_{h\cC}^\UF \big(\bOmega_{\cC\cC}^\UF\big)^{-1} \bOmega_{\cC h}^\UF \Big)^{-1},
	\end{eqnarray*}
	where $\bdelta_h = \sum_{\ell \in \cH}^{} \bOmega_{h\ell}^\UF \bmu_\ell^\UF - \sum_{\ell \neq h}^{} \bOmega_{h\ell}^\UF \bmu_\ell^\PF$ and $\bdelta_{\cC} = \sum_{\ell = 1}^{\H} \bOmega_{\cC \ell}^\UF \bmu_\ell^\UF - \sum_{\ell \neq h}^{} \bOmega_{\cC \ell}^\UF \bmu_\ell^\PF$.
	Moreover, the marginal variational distribution of $\btheta_{\cC}$ is $\N(\bmu_{\cC}^\PF, \bSigma_{\cC}^\PF)$ with mean and variance
	\begin{equation*}
		\bmu_{\cC}^\PF = \bSigma_{\cC|\cH}^\PF \bigg( \sum_{h \in \cH}^{} \bOmega_{\cC h}^\UF \bmu_h^\UF - \sum_{h \in \cH} \bOmega_{\cC h}^\UF \bmu_h^\PF \bigg), \qquad
		\bSigma_{\cC\cC}^\PF = \bSigma_{\cC|\cH}^\PF + \bSigma_{\cC|\cH}^\PF \bigg( \sum_{h \in \cH}^{} \bOmega_{\cC h}^\UF \bSigma_{hh}^\PF \bOmega_{h \cC}^\UF \bigg) \bSigma_{\cC|\cH}^\PF.
	\end{equation*}
\end{result}

\noindent Result~\ref{res:optimal_density_ff} is standard in the literature, while Result~\ref{res:optimal_density_pf} follows by Propositions 1 and 4 of \cite{Goplerud2024}.

\subsubsection*{Fully factorized approximation}

We start by deriving the projected non-conjugate update for the fully factorized case, say $q(\btheta) \in \cQ_\FF$.
Under this setting, the optimal fully factorized approximation of $q_\UF^*(\btheta) \in \cQ_\UF$, say $q_\FF^*(\btheta) \in \cQ_\FF$, is the minimum of $\KL\{ q_\FF(\btheta) \parallel q_\UF^*(\btheta) \} = \KL \big\{ q_\FF(\bbeta) {\tsty \prod_{h = 1}^{\H}} q_\FF(\bsu_h) \,q_\FF(\sigma_h^2) \parallel q_\UF^*(\btheta) \big\}$.
Such a minimizer can be obtained by iterating over the coordinate variational solutions
\begin{equation}
	\label{eq:optimal_density_ff}
	q_\FF^*(\bbeta) \propto \exp\big[ \E_{-\bbeta} \{ \log q_\UF^*(\bbeta | \rest) \}\big], \quad
	q_\FF^*(\bsu_h) \propto \exp\big[ \E_{-\bsu_h} \{ \log q_\UF^*(\bbeta | \rest) \}\big], \quad
	q_\FF^*(\sigma_h^2) = q_\UF^*(\sigma_h^2).
\end{equation}
Thanks to Result~\ref{res:optimal_density_ff}, the variational parameter updates for $q_\FF^*(\bbeta)$ and $q_\FF^*(\bsu_h)$ are
\begin{eqnarray*}
	\bmu_\beta^\FF = \bSigma_{\beta\beta}^\FF \bX^\top \bW \bigg(\bp - \sum_{h = 1}^{\H} \bZ_h \bmu_h^\FF \bigg), & & 
	\bSigma_{\beta\beta}^\FF = \big( \bX^\top \bW \bX + \sigma_\beta^{-2} \bR_\beta \big)^{-1}, \\
	\bmu_h^\FF = \bSigma_{hh}^\FF \bZ_h^\top \bW \bigg( \bp - \bX \bmu_\beta - \sum_{\ell \neq h}^{} \bZ_\ell \bmu_\ell^\FF \bigg), & & 
	\bSigma_{hh}^\FF = \bigg( \bZ_h^\top \bW \bZ_h + \frac{\alpha_h}{\beta_h} \bR_h \bigg)^{-1}.
\end{eqnarray*}
The recursive refinement of the variational parameters using such updates give rise to Algorithm~\ref{alg:ncvb_ff_alg}.
Using only block diagonal information, Algorithm~\ref{alg:ncvb_ff_alg} requires $\cO(n K_2 + K_3)$ flops and $\cO(n K_1 + K_2)$ memory allocations per iteration, thus yielding a significant advantage in terms of computational resources over Algorithm~\ref{alg:ncvb_uf_alg}.

\begin{algorithm}[h!]
	
	\While{convergence is not reached}{
		
		$\bxi^{(t)} \gets \bX \bmu_\beta^{(t)} + \sum_{h =1}^{\H} \bZ_h \bmu_h^{(t)}$; \hfill $\cO(nK_1)$ \\
		$\bnu^{(t)} \gets {\dsty \stack_{i = 1, \dots, n}} \big[ \bx_i^\top \bSigma_{\beta\beta}^{(t)} \bx_i + \sum_{h = 1}^{\H} \bz_{ih}^\top \bSigma_{hh}^{(t)} \bz_{ih} \big]^{1/2}$; \hfill $\cO(nK_2)$ \\
		
		Evaluate $\bPsi_r^{(t)} = \Psi_r(\by, \bxi^{(t)}, \bnu^{(t)})$, $r = 0, 1, 2$; 
		\hfill $\cO(n)$ \\
		
		$\bW^{(t)} \gets \diag \big[ \bPsi_2^{(t)} \big]$; \quad
		$\bp^{(t)} \gets - \bPsi_1^{(t)} / \bPsi_2^{(t)} + \bxi^{(t)}$; \hfill
		$\cO(n)$ \\
		
		\For{$h$ from $1$ to $H$}{
			$\beta_h^{(t+1)} \gets  B_h + \frac{1}{2} \big\{ \bmu_h^{(t)\top} \bR_h \,\bmu_h^{(t)} + \tr(\bR_h \bSigma_{hh}^{(t)}) \big\}$; \hfill
			$\cO(d_h^2)$ \\
			
			$\gamma_h^{(t+1)} \gets (A_h + d_h / 2) / \beta_h^{(t+1)}$; \hfill
			$\cO(1)$
		}
		
		$\bp_\beta^{(t)} \gets \bp^{(t)} - \sum_{h=1}^{\H} \bZ_h \bmu_h^{(t)}$; \hfill $O(n)$ \\
		$\bSigma_{\beta\beta}^{(t+1)} \gets \big[ \sigma_\beta^{-2} \bR_\beta - \bX^\top \bW^{(t)} \bX \big]^{-1}$; \hfill $\cO(np^2 + p^3)$ \\
		$\bmu_\beta^{(t+1)} \gets \bSigma_{\beta\beta}^{(t+1)} \bX^\top \bW^{(t)} \bp_\beta^{(t)}$; \hfill $\cO(np + p^2)$ \\
		
		\For{$h$ from $1$ to $H$}{
			$\bp_{h}^{(t)} \gets \bp^{(t)} - \bX \bmu_\beta^{(t+1)} - \sum_{\ell = 1}^{h-1} \bZ_\ell \bmu_\ell^{(t+1)} + \sum_{\ell = h+1}^{\H} \bZ_\ell \bmu_\ell^{(t)}$; \hfill $\cO(nK_1)$ \\
			$\bSigma_{hh}^{(t+1)} \gets \big[ \gamma_h^{(t+1)} \bR_h - \bZ_h^\top \bW^{(t)} \bZ_h \big]^{-1}$; \hfill $\cO(nd_h^2 + d_h^3)$ \\
			$\bmu_h^{(t+1)} \gets \bSigma_{hh}^{(t+1)} \bZ_h^\top \bW^{(t)} \bp^{(t)}$; \hfill $\cO(nd_h + d_h^2)$ \\
		}
	}
	
	\caption{%
		\label{alg:ncvb_ff_alg}
		NCVB-FF -- Pseudo-code description of the proposed \emph{non-conjugate variational Bayes} algorithm for \emph{fully factorized} approximate Bayesian inference in pGLMMs. On the right, we report the computational cost in number of floating point operations for each step.
	}
\end{algorithm}

\subsubsection*{Partially factorized approximation}

Similarly to the fully factorized setting, to obtain the partially factorized update for $q_\PF(\btheta) \in \cQ_\PF$, we minimize $\KL\{ q_\PF(\btheta) \parallel q_\UF^*(\btheta) \} = \KL \big\{ q_\PF(\bbeta | \bsu) {\tsty \prod_{h = 1}^{\H}} q_\PF(\bsu_h) \,q_\PF(\sigma_h^2) \parallel q_\UF^*(\btheta) \big\}$.
Thanks to Result~\ref{res:optimal_density_pf}, we can iterate over the closed-form updates
\begin{equation}
	\label{eq:optimal_density_pf}
	q_\PF^*(\bbeta | \bsu) = q_\UF^*(\bbeta | \bsu), \quad
	q_\PF^*(\bsu_h) \propto \exp \bigg[ \E_{-\bsu_h} \bigg\{ \log \int q_\UF^*(\bbeta, \bsu) \,\d\bbeta \bigg\} \bigg], \quad
	q_\PF^*(\sigma_h^2) = q_\UF^*(\sigma_h^2).
\end{equation}
Then, defining $\bM = \bW - \bW \bX (\bX^\top \bW \bX + \sigma_\beta^{-2} \bR_\beta)^{-1} \bX^\top \bW$, using Result~\ref{res:optimal_density_pf} and applying straightforward simplifications, we obtain the analytic updates
\begin{eqnarray*}
	\bmu_{\beta | u}^\PF = \bSigma_{\beta | u}^\PF \bX^\top \bW \bigg( \bp - \sum_{h = 1}^{\H} \bZ_h \bsu_h \bigg), & &
	\bSigma_{\beta | u}^\PF = \big( \bX^\top \bW \bX + \sigma_\beta^{-2} \bR_\beta \big)^{-1}, \\
	\bmu_h^\PF = \bSigma_{hh}^\PF \bZ_h^\top \bM \bigg( \bp - \sum_{\ell \neq h}^{} \bZ_\ell \bmu_\ell^\PF \bigg), & & 
	\bSigma_{hh}^\PF = \bigg( \bZ_h^\top \bM \bZ_h + \frac{\alpha_h}{\beta_h} \bR_h \bigg)^{-1}.
\end{eqnarray*}
Then, after the block-wise updates of $\bsu_h$, the resulting variational marginal mean and variance of $\bbeta$ can be obtained as
\begin{equation*}
	\bmu_\beta^\PF = \bSigma_{\beta | u}^\PF \bigg( \bX^\top \bW \bp - \sum_{h = 1}^{\H} \bZ_h \bmu_h^\PF \bigg), \qquad
	\bSigma_{\beta\beta}^\PF = \bSigma_{\beta | u}^\PF + \bSigma_{\beta | u}^\PF \bigg( \sum_{h = 1}^{\H} \bX^\top \bW \,\bZ_h \bSigma_{hh}^\PF \bZ_h^\top \bW \,\bX \bigg) \bSigma_{\beta | u}^\PF.
\end{equation*}
The iterative application of such variational updates give rise to Algorithm~\ref{alg:ncvb_pf_alg}.
Using only low dimensional block information, Algorithm~\ref{alg:ncvb_pf_alg} needs $\cO(n K_2 + K_3)$ flops and $\cO(n K_1 + K_2)$ memory allocations per iteration, which is much cheaper than Algorithm~\ref{alg:ncvb_uf_alg} and comparable with Algorithm~\ref{alg:ncvb_ff_alg}.

\begin{algorithm}[h!]
	
	\While{convergence is not reached}{
		
		$\bxi^{(t)} \gets \bX \bmu_\beta^{(t)} + \sum_{h =1}^{\H} \bZ_h \bmu_h^{(t)}$; \hfill $\cO(nK_1)$ \\
		$\bnu^{(t)} \gets {\dsty \stack_{i = 1, \dots, n}} \big[ \bx_i^\top \bSigma_{\beta\beta}^{(t)} \bx_i + \sum_{h = 1}^{\H} \big( \bz_{ih}^\top \bSigma_{hh}^{(t)} \bz_{ih} - 2 \,\bx_i^\top \bSigma_{\beta h}^{(t)} \bz_{ih} \big) \big]^{1/2}$; \hfill $\cO(nK_2)$ \\
		
		Evaluate $\bPsi_r^{(t)} = \Psi_r(\by, \bxi^{(t)}, \bnu^{(t)})$, $r = 0, 1, 2$; 
		\hfill $\cO(n)$ \\
		
		$\bW^{(t)} \gets \diag \big[ \bPsi_2^{(t)} \big]$; \quad
		$\bp^{(t)} \gets - \bPsi_1^{(t)} / \bPsi_2^{(t)} + \bxi^{(t)}$; \hfill
		$\cO(n)$ \\
		
		\For{$h$ from $1$ to $H$}{
			$\beta_h^{(t+1)} \gets  B_h + \frac{1}{2} \big\{ \bmu_h^{(t)\top} \bR_h \,\bmu_h^{(t)} + \tr(\bR_h \bSigma_{hh}^{(t)}) \big\}$; \hfill
			$\cO(d_h^2)$ \\
			
			$\gamma_h^{(t+1)} \gets (A_h + d_h / 2) / \beta_h^{(t+1)}$; \hfill
			$\cO(1)$
		}
		
		$\bOmega_{\beta\beta}^{(t+1)} \gets \bX^\top \bW^{(t)} \bX + \sigma_\beta^{-2} \bR_\beta$; \quad
		$\bomega_{\beta\beta}^{(t+1)} \gets \bX^\top \bW^{(t)} \bp^{(t)}$; \hfill $\cO(np^2 + p^3)$ \\
		
		$\bSigma_{\beta|u}^{(t+1)} \gets \big[ \bOmega_{\beta\beta}^{(t+1)} \big]^{-1}$; \quad
		$\bar\bOmega_{\beta\beta}^{(t+1)} \gets \bOmega_{\beta\beta}^{(t+1)}$; \quad
		$\bar\bomega_\beta^{(t+1)} \gets \bomega_{\beta\beta}^{(t+1)}$; \hfill $\cO(p + p^2 + p^3)$ \\
		
		\For{$h$ from $1$ to $H$}{
			$\bp_h^{(t)} \gets \bp^{(t)} - \sum_{\ell = 1}^{h-1} \bZ_\ell \bmu_\ell^{(t+1)} - \sum_{\ell = h+1}^{\H} \bZ_\ell \bmu_\ell^{(t)}$; \hfill $\cO(n)$ \\
			
			$\bOmega_{\beta h}^{(t+1)} \gets \bX^\top \bW^{(t)} \bZ_h$; \quad
			$\bOmega_{hh}^{(t+1)} \gets \bZ_h^\top \bW^{(t)} \bZ_h + \gamma_h^{(t+1)} \bR_h$; \hfill $\cO(npd_h + n d_h^2 + d_h^2)$ \\
			
			$\bomega_{\beta h}^{(t+1)} \gets \bX \bW^{(t)} \bp_h^{(t)}$; \quad
			$\bomega_{hh}^{(t+1)} \gets \bZ_h \bW^{(t)} \bp_h^{(t)}$; \hfill $\cO(np + nd_h)$ \\
			
			$\bSigma_{hh}^{(t+1)} \gets \big[ \bOmega_{hh}^{(t+1)} - \bOmega_{h \beta}^{(t+1)} \bSigma_{\beta|u}^{(t+1)} \bOmega_{\beta h}^{(t+1)} \big]^{-1}$; \hfill $\cO(p^2 d_h + p d_h^2 + d_h^2 + d_h^3)$ \\
			
			$\bmu_h^{(t+1)} \gets \bSigma_{hh}^{(t+1)} \big[ \bomega_{hh}^{(t+1)} - \bOmega_{h\beta}^{(t+1)} \bSigma_{\beta|u}^{(t+1)} \bomega_{\beta h}^{(t+1)} \big]$; \hfill $\cO(p^2 + p d_h + d_h + d_h^2)$ \\
			
			$\bSigma_{\beta h}^{(t+1)} \gets \bSigma_{\beta|u}^{(t+1)} \bOmega_{\beta h}^{(t+1)} \bSigma_{hh}^{(t+1)}$; \hfill $\cO(d_h p^2 + d_h^2 p)$ \\
			
			$\bar\bOmega_{\beta\beta}^{(t+1)} \gets \bar\bOmega_{\beta\beta} + \bOmega_{\beta h}^{(t+1)} \bSigma_{hh}^{(t+1)} \bOmega_{h \beta}^{(t+1)}$; \hfill $\cO(d_h p^2 + d_h^2 p + p^2 + p^3)$ \\
			
			$\bar\bomega_\beta^{(t+1)} \gets \bar\bomega_\beta + \bOmega_{\beta h}^{(t+1)} \bmu_h^{(t+1)}$; \hfill $\cO(p d_h + p + p^2)$ \\
		}
		
		$\bSigma_{\beta\beta}^{(t+1)} \gets \bSigma_{\beta|u}^{(t+1)} \bar\bOmega_{\beta\beta}^{(t+1)} \bSigma_{\beta|u}^{(t+1)}$; \hfill $\cO(p^3)$ \\
		
		$\bmu_\beta^{(t+1)} \gets \bSigma_{\beta|u}^{(t+1)} \bar\bomega_\beta^{(t+1)}$; \hfill $\cO(p^2)$ \\
	}
	
	\caption{%
		\label{alg:ncvb_pf_alg}
		NCVB-PF -- Pseudo-code description of the proposed \emph{non-conjugate variational Bayes} algorithm for \emph{partially factorized} approximate Bayesian inference in pGLMMs. On the right, we report the computational cost in number of floating point operations for each step.
	}
\end{algorithm}

\subsection{Stochastic optimization algorithm}

\subsubsection*{Unfactorized approximation}

The iterative refinement of the variational parameters outlined in Section~\ref{subsec:stochastic_and_online_variational_approximation} of the main paper gives rise to Algorithm~\ref{alg:sncvb_uf_alg}. 
The overall computational complexity of one recursion of such a stochastic update requires $\cO(n_\scb K_1 + n_\scb K_1^2 + K_1^3)$ flops and $\cO(n_\scb K_1 + K_1^2)$ memory allocations, thus making it independent on the sample size $n$ and improving its scalability in massive data problems.

\begin{algorithm}[h!]
	
	\While{convergence is not reached}{
		
		\vspace{0.5em}
		Sample a minibatch $B_t \subset \{ 1, \dots, n \}$ of dimension $n_\scb$;
		
		$\bxi_\scb^{(t)} \gets \bC_\scb \bmu^{(t)}$; \quad
		$\bnu_\scb^{(t)} \gets {\dsty \stack_{i \in B_t}} \big[ \bc_i^\top \bSigma^{(t)} \bc_i \big]^{1/2}$; \hfill
		$\cO(n_\scb K_1^2)$\\
		
		Evaluate $\bPsi_{r,\scb}^{(t)} = \Psi_r(\by_\scb, \bxi_\scb^{(t)}, \bnu_\scb^{(t)})$, for $r = 0, 1, 2$, on $B_t$; 
		\hfill $\cO(n_\scb)$ \\
		
		$\bW_\scb^{(t)} \gets \diag\big[ \bPsi_{2,\scb}^{(t)} \big]$; \quad
		$\bp_\scb^{(t)} \gets - \bPsi_{1,\scb}^{(t)} / \bPsi_{2,\scb}^{(t)} + \bxi_\scb^{(t)}$; \hfill $\cO(n_\scb)$ \\
		
		\For{$h$ from $1$ to $H$}{
			$\beta_h^{(t+1)} \gets (1 - \rho_t) \beta_\veps^{(t)} + \rho_t \big[ B_h + \frac{1}{2} \big\{ \bmu_h^{(t)\top} \bR_h \,\bmu_h^{(t)} + \tr(\bR_h \bSigma_{hh}^{(t)}) \big\} \big]$; \hfill
			$\cO(d_h^2)$ \\
			$\gamma_h^{(t+1)} \gets (A_h + d_h / 2) / \beta_\veps^{(t+1)}$; \hfill
			$\cO(1)$ \\
		}
		
		$\bar\bR^{(t)} \gets \blockdiag \big[ \sigma_\beta^{-2} \bR_\beta, \gamma_1^{(t+1)} \bR_1, \dots, \gamma_\H^{(t+1)} \bR_\H \big]$; \hfill
		$\cO(K_2)$ \\
		
		$\bomega^{(t+1)} \gets (1 - \rho_t) \bomega^{(t)} + \rho_t \big[ \frac{n}{n_\scb} \bC_\scb^\top \bW_\scb^{(t)} \bp_\scb^{(t)} \big]$; \hfill
		$\cO(n_\scb K_1)$ \\
		$\bOmega^{(t+1)} \gets (1 - \rho_t) \bOmega^{(t)} + \rho_t \big[ \frac{n}{n_\scb} \,\bC_\scb^\top \bW_\scb^{(t)} \bC_\scb + \bar\bR^{(t)} \big]$; \hfill
		$\cO(n_\scb K_1^2)$ \\
		
		$\bSigma^{(t+1)} \gets \big[ \bOmega^{(t+1)} \big]^{-1}$; \quad
		$\bmu^{(t+1)} \gets \bSigma^{(t+1)} \bomega^{(t+1)}$; \hfill
		$\cO(K_1^3)$ \\
	}
	
	\caption{%
		\label{alg:sncvb_uf_alg}
		sNCVB-UF -- Pseudo-code description of the proposed \emph{stochastic non-conjugate variational Bayes} algorithm for \emph{unfactorized} approximate Bayesian inference in pGLMMs. On the right, we report the computational cost in number of floating point operations for each step.
	}
\end{algorithm}

\noindent In its vein, Algorithm \ref{alg:sncvb_uf_alg} generalizes the stochastic non-conjugate variational message passing of \cite{Tan2014} and \cite{Tan2017} to more general mixed effect models, allowing for non-exponential family models and non-smooth pseudo-likelihoods, say the pGLMM family.

\subsubsection*{Fully factorized approximation}

In the \emph{fully factorized} setting, we can obtain as well a stochastic update of the variational parameters.
Thanks to Result~\ref{res:optimal_density_ff}, and exploiting the block structure of $\bOmega^\UF$ and $\bmu^\UF$ under the stochastic natural update described in Section 3.2 of the main paper, we obtain 
\begin{eqnarray*}
	\bOmega_{hh}^{(t+1)} \gets (1 - \rho_t) \bOmega_{hh}^{(t)} + \rho_t \bigg[ \frac{n}{n_\scb} \,\bZ_{\scb,h}^\top \bW_\scb^{(t)} \bZ_{\scb,h} + \frac{\alpha_h^{(t)}}{\beta_h^{(t)}} \bR_h \bigg], & &
	\bSigma_{hh}^{(t+1)} \gets \big[ \bOmega_{hh}^{(t+1)} \big]^{-1}, \\
	\bomega_{h}^{(t+1)} \gets (1 - \rho_t) \bomega_h^{(t)} + \rho_t \bigg[ \frac{n}{n_\scb} \bZ_{\scb,h}^\top \bW_\scb^{(t)} \bigg( \bp_\scb^{(t)} - \sum_{\ell \neq h}^{} \bZ_\ell \bmu_\ell^{(t)} \bigg) \bigg], & &
	\bmu_h^{(t+1)} \gets \bSigma_{hh}^{(t+1)} \bomega_{h}^{(t+1)}.
\end{eqnarray*}
Iterating such a stochastic update until convergence, we obtain Algorithm~\ref{alg:sncvb_ff_alg}.
In terms of computation complexity, Algorithm~\ref{alg:sncvb_ff_alg} requires $\cO(n_\scb K_1 + n_\scb K_2 + K_3)$ flops and $\cO(n_\scb K_1 + K_2)$ memory allocations per iteration, which is much cheaper than Algorithm~\ref{alg:sncvb_uf_alg}.

\begin{algorithm}[h!]
	
	\While{convergence is not reached}{
		
		\vspace{0.5em}
		Sample a minibatch $B_t \subset \{ 1, \dots, n \}$ of dimension $n_\scb$;
		
		$\bxi_\scb^{(t)} \gets \bX_\scb \bmu_\beta^{(t)} + \sum_{h =1}^{\H} \bZ_{\scb,h} \bmu_h^{(t)}$; \hfill $\cO(n_\scb K_1)$ \\
		
		$\bnu_\scb^{(t)} \gets {\dsty \stack_{i \in B_t}} \big[ \bx_i^\top \bSigma_{\beta\beta}^{(t)} \bx_i + \sum_{h = 1}^{\H} \bz_{ih}^\top \bSigma_{hh}^{(t)} \bz_{ih} \big]^{1/2}$ \hfill
		$\cO(n_\scb K_2)$ \\
		
		Evaluate $\bPsi_{r,\scb}^{(t)} = \Psi_r(\by_\scb, \bxi_\scb^{(t)}, \bnu_\scb^{(t)})$, for $r = 0, 1, 2$, on $\cB_t$; 
		\hfill $\cO(n_\scb)$ \\
		
		$\bW_\scb^{(t)} \gets \diag\big[ \bPsi_{2,\scb}^{(t)} \big]$; \quad
		$\bp_\scb^{(t)} \gets - \bPsi_{1,\scb}^{(t)} / \bPsi_{2,\scb}^{(t)} + \bxi_\scb^{(t)}$; \hfill $\cO(n_\scb)$ \\
		
		\For{$h$ from $1$ to $H$}{
			$\beta_h^{(t+1)} \gets (1 - \rho_t) \beta_\veps^{(t)} + \rho_t \big[ B_h + \frac{1}{2} \big\{ \bmu_h^{(t)\top} \bR_h \,\bmu_h^{(t)} + \tr(\bR_h \bSigma_{hh}^{(t)}) \big\} \big]$; \hfill
			$\cO(d_h^2)$ \\
			$\gamma_h^{(t+1)} \gets (A_h + d_h / 2) / \beta_\veps^{(t+1)}$; \hfill
			$\cO(1)$ \\
		}
		
		$\bp_{\scb,\beta}^{(t)} \gets \bp_\scb^{(t)} - \sum_{h = 1}^{\H} \bZ_{\scb,h} \bmu_h^{(t)}$; \hfill $O(n_\scb)$ \\
		$\bomega_{\beta}^{(t+1)} \gets (1 - \rho_t) \bomega_{\beta}^{(t)} + \rho_t \big[ \frac{n}{n_\scb} \bX_\scb^\top \bW_\scb^{(t)} \bp_{\scb,\beta}^{(t)} \big]$; \hfill
		$\cO(n_\scb p + n_\scb p^2)$ \\
		
		$\bOmega_{\beta\beta}^{(t+1)} \gets (1 - \rho_t) \bOmega_{\beta\beta}^{(t)} + \rho_t \big[ \frac{n}{n_\scb} \,\bX_\scb^\top \bW_\scb^{(t)} \bX_\scb + \sigma_\beta^{-2} \bR_\beta \big]$; \hfill
		$\cO(n_\scb  p^2)$ \\
		
		$\bSigma_{\beta\beta}^{(t+1)} \gets \big[ \bOmega_{\beta\beta}^{(t+1)} \big]^{-1}$; \quad
		$\bmu_\beta^{(t+1)} \gets \bSigma_{\beta\beta}^{(t+1)} \bomega_{\beta}^{(t+1)}$; \hfill
		$\cO(p^3)$ \\
		
		\For{$h$ from $1$ to $H$}{
			$\bp_{\scb,h}^{(t)} \gets \bp_\scb^{(t)} - \bX \bmu_\beta^{(t+1)} - \sum_{\ell = 1}^{h-1} \bZ_{\scb,\ell} \bmu_\ell^{(t+1)} - \sum_{\ell = h+1}^{\H} \bZ_{\scb,\ell} \bmu_\ell^{(t)}$; \hfill $(n_\scb)$ \\
			$\bomega_h^{(t+1)} \gets (1 - \rho_t) \bomega_h^{(t)} + \rho_t \big[ \frac{n}{n_\scb} \bZ_{\scb,h}^\top \bW_\scb^{(t)} \bp_{\scb,h}^{(t)} \big]$; \hfill
			$\cO(n_\scb d_h + n_\scb d_h^2)$ \\
			$\bOmega_{hh}^{(t+1)} \gets (1 - \rho_t) \bOmega_{hh}^{(t)} + \rho_t \big[ \frac{n}{n_\scb} \,\bZ_{\scb,h}^\top \bW_\scb^{(t)} \bZ_{\scb,h} + \gamma_h^{(t+1)} \bR_h \big]$; \hfill
			$\cO(n_\scb d_h^2)$ \\
			
			$\bSigma_{hh}^{(t+1)} \gets \big[ \bOmega_{hh}^{(t+1)} \big]^{-1}$; \quad
			$\bmu_h^{(t+1)} \gets \bSigma_{hh}^{(t+1)} \bomega_h^{(t+1)}$; \hfill
			$\cO(d_h^3)$ \\
		}
	}
	
	\caption{%
		\label{alg:sncvb_ff_alg}
		sNCVB-FF -- Pseudo-code description of the proposed \emph{stochastic non-conjugate variational Bayes} algorithm for \emph{fully factorized} approximate Bayesian inference in pGLMMs. On the right, we report the computational cost in number of floating point operations for each step.
	}
\end{algorithm}

\subsubsection*{Partially factorized approximation}

Similarly to the \emph{fully factorized} case, in the \emph{partially factorized} setting, we use Result~\ref{res:optimal_density_pf} together with the block structure of $\bOmega^\UF$ and $\bmu^\UF$ under the stochastic natural parameter recursion to obtain the following stochastic updates
\begin{eqnarray*}
	\bOmega_{\beta\beta}^{(t+1)} &\gets& (1 - \rho_t) \bOmega_{\beta\beta}^{(t)} + \rho_t \bigg[ \dsty\frac{n}{n_\scb} \,\bX_\scb^\top \bW_\scb^{(t)} \bX_\scb + \sigma_\beta^{-2} \bR_\beta \bigg], \\ \rule{0pt}{18pt} 
	\bOmega_{hh}^{(t+1)} &\gets& (1 - \rho_t) \bOmega_{hh}^{(t)} + \rho_t \bigg[ \dsty\frac{n}{n_\scb} \,\bZ_{\scb,h}^\top \bW_\scb^{(t)} \bZ_{\scb,h} + \frac{\alpha_h^{(t)}}{\beta_h^{(t)}} \bR_h \bigg], \\ \rule{0pt}{18pt} 
	\bOmega_{\beta h}^{(t+1)} &\gets& (1 - \rho_t) \bOmega_{\beta h}^{(t)} + \rho_t \bigg[ \dsty\frac{n}{n_\scb} \,\bX_\scb^\top \bW_\scb^{(t)} \bZ_{\scb,h} \bigg], \\ \rule{0pt}{18pt} 
	\bomega_{\beta\beta}^{(t+1)} &\gets& (1 - \rho_t) \bomega_{\beta\beta}^{(t)} + \rho_t \bigg[ \dsty\frac{n}{n_\scb} \bX_\scb^\top \bW_\scb^{(t)} \bp_\scb^{(t)} \bigg], \\ \rule{0pt}{18pt} 
	\bomega_{hh}^{(t+1)} &\gets& (1 - \rho_t) \bomega_{hh}^{(t)} + \rho_t \bigg[ \dsty\frac{n}{n_\scb} \bZ_{\scb h}^\top \bW_\scb^{(t)} \bigg( \bp_\scb^{(t)} - \sum_{\ell \neq h} \bZ_{\scb \ell} \bmu_\ell^{(t)} \bigg) \bigg], \\ \rule{0pt}{18pt} 
	\bomega_{\beta h}^{(t+1)} &\gets& (1 - \rho_t) \bomega_{\beta h}^{(t)} + \rho_t \bigg[ \dsty\frac{n}{n_\scb} \bX_\scb^\top \bW_\scb^{(t)} \bigg( \bp_\scb^{(t)} + \sum_{\ell \neq h} \bZ_{\scb \ell} \bmu_\ell^{(t)} \bigg) \bigg],
\end{eqnarray*}
and
\begin{eqnarray*}
	\bSigma_{hh}^{(t+1)} &\gets& \Big[ \bOmega_{hh}^{(t+1)} - \bOmega_{h\beta}^{(t+1)} (\bOmega_{\beta\beta}^{(t+1)})^{-1} \bOmega_{\beta h}^{(t+1)} \Big]^{-1}, \\ 
	\bmu_h^{(t+1)} &\gets& \bSigma_{hh}^{(t+1)} \Big[ \bomega_{hh}^{(t+1)} - \bOmega_{h\beta}^{(t+1)} (\bOmega_{\beta\beta}^{(t+1)})^{-1} \bOmega_{\beta h}^{(t+1)} \Big], \\
	\bSigma_{\beta\beta}^{(t+1)} &\gets& (\bOmega_{\beta\beta}^{(t+1)})^{-1} \bigg[ \bOmega_{\beta\beta}^{(t+1)} + \sum_{h = 1}^{\H} \bOmega_{\beta h}^{(t+1)} \bSigma_{hh}^{(t+1)} \bOmega_{h \beta}^{(t+1)} \bigg] (\bOmega_{\beta\beta}^{(t+1)})^{-1}, \\ 
	\bmu_\beta^{(t+1)} &\gets& (\bOmega_{\beta\beta}^{(t+1)})^{-1} \bigg[ \bomega_{\beta\beta}^{(t+1)} + \sum_{h = 1}^{\H} \bOmega_{\beta h}^{(t+1)} \bmu_h^{(t+1)} \bigg].
\end{eqnarray*}
Iterating such a stochastic updates until convergence, we obtain Algorithm~\ref{alg:sncvb_pf_alg}.
In terms of computation complexity,  Algorithm~\ref{alg:sncvb_pf_alg} requires $\cO(n_\scb K_1 + n_\scb K_2 + K_3)$ flops and $\cO(n_\scb K_1 + K_2)$ memory allocations per iteration, which is much cheaper than Algorithm~\ref{alg:sncvb_uf_alg} and comparable with Algorithm~\ref{alg:sncvb_ff_alg}.

\begin{algorithm}[h!]
	
	\While{convergence is not reached}{
		
		\vspace{0.5em}
		Sample a minibatch $B_t \subset \{ 1, \dots, n \}$ of dimension $n_\scb$;
		
		$\bxi_\scb^{(t)} \gets \bX_\scb \bmu_\beta^{(t)} + \sum_{h =1}^{\H} \bZ_{\scb,h} \bmu_h^{(t)}$; \hfill $\cO(n_\scb K_1)$ \\
		
		$\bnu_\scb^{(t)} \gets {\dsty \stack_{i \in B_t}} \big[ \bx_i^\top \bSigma_{\beta\beta}^{(t)} \bx_i + \sum_{h = 1}^{\H} \big( \bz_{ih}^\top \bSigma_{hh}^{(t)} \bz_{ih} - 2 \,\bx_i^\top \bSigma_{\beta h}^{(t)} \bz_{ih} \big) \big]^{1/2}$; \hfill
		$\cO(n_\scb K_2)$ \\
		
		Evaluate $\bPsi_{r,\scb}^{(t)} = \Psi_r(\by_\scb, \bxi_\scb^{(t)}, \bnu_\scb^{(t)})$, for $r = 0, 1, 2$, on $B_t$; 
		\hfill $\cO(n_\scb)$ \\
		
		$\bW_\scb^{(t)} \gets \diag\big[ \bPsi_{2,\scb}^{(t)} \big]$; \quad
		$\bp_\scb^{(t)} \gets - \bPsi_{1,\scb}^{(t)} / \bPsi_{2,\scb}^{(t)} + \bxi_{\scb}^{(t)}$; \hfill $\cO(n_\scb)$ \\
		
		\For{$h$ from $1$ to $H$}{
			$\beta_h^{(t+1)} \gets (1 - \rho_t) \beta_\veps^{(t)} + \rho_t \big[ B_h + \frac{1}{2} \big\{ \bmu_h^{(t)\top} \bR_h \,\bmu_h^{(t)} + \tr(\bR_h \bSigma_{hh}^{(t)}) \big\} \big]$; \hfill
			$\cO(d_h^2)$ \\
			$\gamma_h^{(t+1)} \gets (A_h + d_h / 2) / \beta_\veps^{(t+1)}$; \hfill
			$\cO(1)$ \\
		}
		
		$\bomega_{\beta\beta}^{(t+1)} \gets (1 - \rho_t) \bomega_{\beta\beta}^{(t)} + \rho_t \big[ \frac{n}{n_\scb} \bX_\scb^\top \bW_\scb^{(t)} \bp_\scb^{(t)} \big]$; \hfill $\cO(p + n_\scb p)$ \\
		
		$\bOmega_{\beta\beta}^{(t+1)} \gets (1 - \rho_t) \bOmega_{\beta\beta}^{(t)} + \rho_t \big[ \frac{n}{n_\scb} \,\bX_\scb^\top \bW_\scb^{(t)} \bX_\scb + \sigma_\beta^{-2} \bR_\beta \big]$; \hfill
		$\cO(p^2 + n_\scb p^2)$ \\
		
		$\bSigma_{\beta|u}^{(t+1)} \gets \big[ \bOmega_{\beta\beta}^{(t+1)} \big]^{-1}$; \quad
		$\bar\bOmega_{\beta\beta}^{(t+1)} \gets \bOmega_{\beta\beta}^{(t+1)}$; \quad
		$\bar\bomega_\beta^{(t+1)} \gets \bomega_{\beta\beta}^{(t+1)}$; \hfill $\cO(p + p^2 + p^3)$ \\
		
		\For{$h$ from $1$ to $H$}{
			$\bp_{\scb,h}^{(t)} \gets \bp_\scb^{(t)} - \sum_{\ell = 1}^{h-1} \bZ_{\scb \ell} \bmu_\ell^{(t+1)} - \sum_{\ell = h+1}^{\H} \bZ_{\scb \ell} \bmu_\ell^{(t)}$; \hfill 
			$\cO(n_\scb)$ \\
			
			$\bomega_{hh}^{(t+1)} \gets (1 - \rho_t) \bomega_{hh}^{(t)} + \rho_t \big[ \frac{n}{n_\scb} \bZ_{\scb h}^\top \bW_\scb^{(t)} \bp_{\scb,h}^{(t)} \big]$; \hfill $\cO(d_h + n_\scb d_h)$ \\
			
			$\bomega_{\beta h}^{(t+1)} \gets (1 - \rho_t) \bomega_{\beta h}^{(t)} + \rho_t \big[ \frac{n}{n_\scb} \bX_\scb^\top \bW_\scb^{(t)} \bp_{\scb,h}^{(t)} \big]$; \hfill $\cO(p^2 + n_\scb p^2)$ \\
			
			$\bOmega_{hh}^{(t+1)} \gets (1 - \rho_t) \bOmega_{hh}^{(t)} + \rho_t \big[ \frac{n}{n_\scb} \,\bZ_{\scb,h}^\top \bW_\scb^{(t)} \bZ_{\scb,h} + \gamma_h^{(t+1)} \bR_h \big] $; \hfill $\cO(d_h^2 + n_\scb d_h^2)$ \\
			
			$\bOmega_{\beta h}^{(t+1)} \gets (1 - \rho_t) \bOmega_{\beta h}^{(t)} + \rho_t \big[ \frac{n}{n_\scb} \,\bX_\scb^\top \bW_\scb^{(t)} \bZ_{\scb,h} \big] $; \hfill $\cO(d_h p + n_\scb d_h p)$ \\
			
			$\bSigma_{hh}^{(t+1)} \gets \big[ \bOmega_{hh}^{(t+1)} - \bOmega_{h \beta}^{(t+1)} \bSigma_{\beta|u}^{(t+1)} \bOmega_{\beta h}^{(t+1)} \big]^{-1}$; \hfill $\cO(d_h p^2 + d_h^2 + d_h^3)$ \\
			
			$\bSigma_{\beta h}^{(t+1)} \gets \bSigma_{\beta|u}^{(t+1)} \bOmega_{\beta h}^{(t+1)} \bSigma_{hh}^{(t+1)}$; \hfill $\cO(d_h p^2 + d_h^2 p)$ \\
			
			$\bmu_h^{(t+1)} \gets \bSigma_{hh}^{(t+1)} \big[ \bomega_{hh}^{(t+1)} - \bOmega_{h \beta}^{(t+1)} \bSigma_{\beta|u}^{(t+1)} \bomega_{\beta h}^{(t+1)} \big]$; \hfill $\cO(d_h + d_h p + p^2 + d_h^2)$ \\
			
			$\bar\bOmega_{\beta\beta}^{(t+1)} \gets \bar\bOmega_{\beta\beta} + \bOmega_{\beta h}^{(t+1)} \bSigma_{hh}^{(t+1)} \bOmega_{h \beta}^{(t+1)}$; \hfill $\cO(d_h p^2 + d_h^2 p + p^2 + p^3)$ \\
			
			$\bar\bomega_\beta^{(t+1)} \gets \bar\bomega_\beta + \bOmega_{\beta h}^{(t+1)} \bmu_h^{(t+1)}$; \hfill $\cO(d_h p + p + p^2)$ \\
		}
		
		$\bSigma_{\beta\beta}^{(t+1)} \gets \bSigma_{\beta|u}^{(t+1)} \bar\bOmega_{\beta\beta}^{(t+1)} \bSigma_{\beta|u}^{(t+1)}$; \hfill $\cO(p^3)$ \\
		
		$\bmu_\beta^{(t+1)} \gets \bSigma_{\beta|u}^{(t+1)} \bar\bomega_\beta^{(t+1)}$; \hfill $\cO(p^2)$ \\
	}
	
	\caption{%
		\label{alg:sncvb_pf_alg}
		sNCVB-PF -- Pseudo-code description of the proposed \emph{stochastic non-conjugate variational Bayes} algorithm for \emph{partially factorized} approximate Bayesian inference in pGLMMs. On the right, we report the number of floating point operations corresponding to each step.
	}
\end{algorithm}

\section{Additional numerical results}%
\label{app:additional_numerical_results}

Here, we present additional results and application on real data problems relative to distributional regression (power load consumption), robust regression (beijing air quality) and unbalanced classification (bank marketing and forest cover type) tasks.

Likewise the simulations studies in Section~\ref{sec:simulation_studies}, we estimate the \emph{true} posterior via MCMC, and we approximate it via data augmented MFVB and NCVB.
In addition, we also consider the stochastic non-conjugate method outlined in Algorithm~\ref{alg:sncvb_uf_alg}, named sNCVB.
Doing so, we consider the learning rate schedule $\rho_t = \rho_0 / (1 + \rho_0 \,t)^{3/4}$, which satisfy the Robbins-Monro conditions \citep{Robbins1951}, where $\rho_0 > 0$ controls both the initial learning rate value and the decay speed to 0 of $\rho_t$ as $t$ diverges.
In particular, in our experiments we set $\rho_0 = 0.05$ and we use minibatch samples with $100$ observations at each iteration of the algorithm.
We run MCMC and sNCVB for 10000 iterations, while MFVB and NCVB are stopped when the relative change of the evidence lower bound falls below $10^{-6}$, with a maximum number of iterations set to 500.
We set diffuse priors for all the parameters in the model, that is: $\sigma_\beta^2 = 10^6$, $A_h = 2.0001$, $B_h = 1.0001$.

To evaluate the computational efficiency and the posterior approximation quality of the considered methods, we report for each algorithm the number of iterations to convergence, the memory usage in GB, the execution time in seconds, the speed gain over MCMC, the average accuracy score and the evidence lower bound obtained at the end of the optimization.
The speed gain over MCMC is calculated as the MCMC execution time divided by the execution time of the considered variational approximations, namely MFVB, NCVB and sNCVB.

Before analyzing the numerical results, we introduce all the datasets and models that we considered in this study.

\begin{description}
	\item[Power load consumption.] The first dataset we examine has been utilized in the \emph{load forecasting track of the Global Energy Competition 2014} and is readily accessible in the supplementary material of \cite{Hong2016}.
	Further details on the dataset are provided in Section~\ref{sec:real_data_application} of the main paper.
	We model the $\tau$th quantile of the load consumption as an additive function of the available covariates obtaining the linear predictor
	\begin{align*}
		\eta_i = \beta_0 +
		&	f_1^{\scb,15}(\mathtt{day\_hour}_i) +
		f_2^{\scd,7 }(\mathtt{week\_day}_i) + 
		f_3^{\scb,10}(\mathtt{month\_day}_i) + 
		f_4^{\scb,15}(\mathtt{year\_day}_i) \,+ \\ 
		&	f_5^{\scb,5 }(\mathtt{trend}_i) +
		f_6^{\scb,15}(\mathtt{temperature}_i) +
		f_7^{\scb,10}(\mathtt{smooth\_temp}_i) +
		f_8^{\scb,10}(\mathtt{lagged\_load}_i),
	\end{align*}
	where $f_h^{\scb,r}(\cdot) = \bsu_h^\top \bz_h^{\scb}(\cdot)$ is a cubic B-spline basis expansion with $r$ equally spaced knots, i.e. $d_h = 3 + r$ basis functions, and and $f_h^{\scd,r} = \bsu_h^\top \bz_h^{\scd}(\cdot)$ is a dummy encoding of a discrete variable with $r$ categories.
	The considered covariates are: the hour of the day ($\mathtt{day\_hour}$), the day of the week ($\mathtt{week\_day}$), the day of the month $(\mathtt{month\_day}$), the day of the year ($\mathtt{year\_day}$), the time index ($\mathtt{trend}$), the temperature in Celsius scale ($\mathtt{temperature}$), the smoothed temperature in Celsius scale ($\mathtt{smooth\_temp}$) and the 1-day lagged power load consumption ($\mathtt{lagged\_load}$).
	In total, we obtain a completed covariate vector, $\bc_i^\top = (1, \bx_i^\top, \bz_1^\top, \dots, \bz_8^\top)$, of dimension 110.
	
	\item[Beijing air quality.] In the second application, we explore the Beijing PM2.5 dataset, which is freely available in the UCI machine learning repository \citep{Chen2017}.
	This dataset comprises 43824 hourly measurements of PM2.5 concentration, collected by the US Embassy in Beijing from January 1st, 2010 to December 31st, 2014.
	It also includes 7 meteorological variables recorded at the Beijing Capital International Airport, namely dew point ($\mathtt{dew\_point}$), temperature ($\mathtt{temperature}$), pressure ($\mathtt{pressure}$), combined wind direction ($\mathtt{direction}$), cumulated wind speed ($\mathtt{speed}$), cumulated hours of snow ($\mathtt{snow}$), and cumulated hours of rain ($\mathtt{rain}$).
	Additionally, the dataset provides information about the hour ($\mathtt{hour}$), day ($\mathtt{day}$), month ($\mathtt{month}$), and year ($\mathtt{year}$) of each measurement.
	The objective of our analysis is to estimate the behavior of PM2.5 concentration in Beijing based on the available covariates.
	To achieve this, we employ a robust support vector regression model \citep{Vapnik1998} with an insensitivity parameter of $\eps = 0.01$.
	We adopt an additive specification for the linear predictor, accounting for potential non-linear effects of the covariates using appropriate regularized Fourier and B-spline basis expansions:
	\begin{align*}
		\eta_i = \; & \beta_0 +
		f_{1}^{\scd,5}(\mathtt{year}_i) +
		f_{2}^{\scf,6}(\mathtt{month}_i) + 
		f_{3}^{\scf,10}(\mathtt{day}_i) + 
		f_{4}^{\scf,10}(\mathtt{hour}_i) + \\ &
		f_{5}^{\scb,10}(\mathtt{dew\_point}_i) +
		f_{6}^{\scb,10}(\mathtt{temperature}_i) +
		f_{7}^{\scb,10}(\mathtt{pressure}_i) + \\ &
		f_{8}^{\scd,4}(\mathtt{direction}_i) +
		f_{9}^{\scb,10}(\mathtt{speed}_i) +
		f_{10}^{\scb,10}(\mathtt{snow}_i) +
		f_{11}^{\scb,10}(\mathtt{rain}_i),
	\end{align*}
	where $f_h^{\scb,r}(\cdot) = \bsu_h^\top \bz_h^{\scb}(\cdot)$ is a cubic B-spline basis expansion with $r$ equally spaced knots, $f_h^{\scf,r} = \bsu_h^\top \bz_h^{\scf}(\cdot)$ is a Fourier basis expansion with $r$ frequencies and $2r$ basis functions, and $f_h^{\scd,r} = \bsu_h^\top \bz_h^{\scd}(\cdot)$ is a dummy encoding of a discrete variable with $r$ categories.
	This yields a 140-dimensional completed vector $\bc_i^\top$ for each observation $y_i$.
	We model the $h$th basis coefficient vector $\bsu_h$ as a random effect vector, allowing it to have its own variance parameter $\sigma_h^2$, with $h$ ranging from $1$ to $11$.
	
	\item[Bank marketing.] In the third application, we examine a bank marketing dataset obtained from the UCI machine learning repository \citep{Moro2012}.
	This dataset comprises 45211 instances involving various marketing campaigns conducted by a Portuguese banking institution. 
	The objective of our analysis is to predict the likelihood of subscribing a financial product based on 16 covariates, including campaign descriptions, last contact reports, and client information such as age, job position, education, marital status, and financial position.
	Following necessary data preprocessing steps, such as dichotomization of categorical covariates and transformation of numeric variables, we obtain 48 predictors for our analysis.
	We adopt a support vector classification model \citep{Vapnik1998} to predict the subscriptions, incorporating an adaptive ridge penalty on the regression parameters to mitigate overfitting.
	In this approach, we estimate the posterior distribution of the ridge regularization parameter from the available data, along with the other unknown coefficients.
	
	\item[Forest cover types.] For the fourth application, we consider the cover-type dataset provided by the UCI machine learning repository \citep{Blackard1998}.
	It collects 581012 observations on forest cover-type and cartographic attributes measured on a $30 \times 30$ meters grid from four wildness areas located in the Roosevelt National Forest of northern Colorado.
	The goal of the analysis is to predict the forest cover-type encoded in 7 categories given the elevation, aspect, slope, hillshade, soil-type and other topological features of the observed regions.
	In this application, we predict the \emph{lodgepole pine} category against the others using a support vector machine classifier \citep{Vapnik1998} with a grouped ridge penalty, such that each group of covariates has its own shrinkage parameter. 
	Doing so, we end up having a 52-dimensional vector of fixed and random effect covariates, which is divided in 11 subgroups reflecting the structure of the grouped ridge penalty.
\end{description}

\begin{table}
	\centering
	\small
	\caption{%
		\label{tab:application_results}
		Summary results for the four applications described in the text.
		From left to right column: model, approximation method, number of iterations, execution time in seconds (s), speed gain, average marginal accuracy (standard deviation), evidence lower bound.
	}
	\begin{tabular}{lrrrrrlr}
		\toprule
		Model & Method & Iterations & Exe. time & Speed gain & \multicolumn{2}{c}{Accuracy} & ELBO \\
		\midrule
		\rowcolor{lightgray!75} \multicolumn{8}{l}{Power load consumption ($n = 60600$, $p + d = 94$)} \\
		Qauntile      & MCMC  & 10000 & 323.29 &        &      &        &           \\ 
		$\tau = 0.05$ & MFVB  &   173 &  14.05 &  23.01 & 0.64 & (0.13) & 218645.08 \\ 
		& NCVB  &    19 &   2.88 & 112.25 & 0.97 & (0.01) & 218694.54 \\ 
		& sNCVB & 10000 &   7.89 &  40.97 & 0.85 & (0.11) & 218681.28 \\ 
		\midrule
		Qauntile      & MCMC  & 10000 & 330.11 &        &      &        &           \\ 
		$\tau = 0.25$ & MFVB  &    75 &   6.68 &  49.42 & 0.71 & (0.10) & 145486.29 \\ 
		& NCVB  &    18 &   3.01 & 109.67 & 0.97 & (0.02) & 145522.77 \\ 
		& sNCVB & 10000 &   8.64 &  38.21 & 0.88 & (0.09) & 145515.79 \\ 
		\midrule
		Qauntile      & MCMC  & 10000 & 358.50 &        &      &        &           \\ 
		$\tau = 0.50$ & MFVB  &    59 &   5.71 &  62.78 & 0.72 & (0.10) & 129560.45 \\ 
		& NCVB  &    19 &   3.32 & 107.98 & 0.97 & (0.02) & 129593.65 \\ 
		& sNCVB & 10000 &   8.90 &  40.28 & 0.89 & (0.09) & 129587.54 \\ 
		\midrule
		Qauntile      & MCMC  & 10000 & 388.67 &        &      &        &           \\ 
		$\tau = 0.75$ & MFVB  &    83 &   7.88 &  49.32 & 0.70 & (0.11) & 142784.65 \\ 
		& NCVB  &    20 &   3.44 & 112.99 & 0.96 & (0.03) & 142820.49 \\ 
		& sNCVB & 10000 &   8.42 &  46.16 & 0.88 & (0.09) & 142814.12 \\ 
		\midrule
		Qauntile      & MCMC  & 10000 & 374.69 &        &      &        &           \\ 
		$\tau = 0.95$ & MFVB  &   145 &  13.58 &  27.59 & 0.65 & (0.13) & 213858.06 \\ 
		& NCVB  &    14 &   2.66 & 140.86 & 0.96 & (0.01) & 213906.94 \\ 
		& sNCVB & 10000 &   9.12 &  41.08 & 0.84 & (0.11) & 213888.24 \\
		\rowcolor{lightgray!75} \multicolumn{8}{l}{Beijing air quality ($n = 43824$, $p + d = 140$)} \\
		SVM Reg.      & MCMC  & 10000 & 559.36 &        &       &         &          \\
		$\eps = 0.05$ & MFVB  &    43 &   4.69 & 119.34 & 76.66 & (12.74) & -6357.40 \\
		& NCVB  &    22 &   2.36 & 236.12 & 96.88 &  (1.62) & -6274.64 \\
		& sNCVB & 10000 &  10.22 &  54.75 & 89.39 &  (8.48) & -6281.89 \\
		\rowcolor{lightgray!75} \multicolumn{8}{l}{Bank marketing ($n = 45211$, $p + d = 48$)} \\
		SVM Class.    & MCMC  & 10000 & 121.69 &        &      &         &         \\
		& MFVB  &   722 &  18.75 &  6.49 & 89.46 & (14.66) & -504.91 \\
		& NCVB  &    68 &   2.24 & 54.33 & 95.50 &  (3.59) & -493.45 \\
		& sNCVB & 10000 &   5.43 & 22.39 & 93.50 &  (7.12) & -495.27 \\
		\rowcolor{lightgray!75} \multicolumn{8}{l}{Forest cover-type ($n = 581012$, $p + d = 52$)} \\
		SVM Class.    & MCMC  & 10000 & 2469.34 &        &       &         &        \\
		& MFVB  &   128 &   39.81 &  62.02 & 55.96 & (13.08) & -430.34 \\
		& NCVB  &    99 &   28.42 &  86.88 & 90.31 & (16.61) & -430.31 \\
		& sNCVB & 10000 &    9.56 & 258.30 & 75.05 & (20.92) & -430.32 \\
		\bottomrule
	\end{tabular}
\end{table}

\noindent Table~\ref{tab:application_results} presents the computational efficiency and posterior approximation accuracy results obtained in the four applications described so far.
Among the considered approaches, the proposed batch NCVB algorithm demonstrates excellent performance, exhibiting both a good efficiency and a superior accuracy.
It converges rapidly and achieves a substantial speed gain over MCMC, conjugate MFVB and sNCVB in the first three applications, while in the fourth one, it is the second fastest method after sNCVB outperforming MCMC sampling and conjugate MFVB.
In comparison to conjugate mean field variational Bayes and sNCVB, batch NCVB showcases the highest average accuracy and evidence lower bound across all scenarios.
Notably, it consistently surpasses $95\%$ accuracy in all settings, with minor deviations from the central value.

Conversely, sNCVB requires more iterations to converge, but, thanks to minibatch subsampling, each iteration is computationally cheaper than both MFVB and NCVB.
Consequently, sNCVB exhibits favorable scalability, making it well-suited for handling massive data problems, where batch optimization would necessitate an impractical amount of memory to store the dataset and approximate the posterior distribution.
Actually, in the forest cover-type application, which has the highest number of observations among the considered examples, sNCVB yields a substantial speed gain over MCMC, MFVB and batch NCVB.

Although the proposed stochastic method does not attain the same precision as batch NCVB in terms of accuracy, it consistently achieves superior average accuracy and evidence lower bound compared to conjugate MFVB.
This behavior is not surprising, since the approximation obtained via stochastic optimization is, by definition, a noisy version of the deterministic approximation yielding from batch NCVB. 
The precision of the proposed stochastic optimization technique thus depends on the amount of randomness injected in the algorithm via minibatch subsampling and by the number of iterations set before to stop the execution.

Overall, we can conclude that the proposed non-conjugate variational Bayes methods are competitive with alternative conjugate mean field variational Bayes both in terms of computational speed and posterior approximation accuracy, systematically outperforming data-augmentation based approximations in all the simulation studies and real-data examples discussed in Sections~\ref{sec:simulation_studies}~and~\ref{sec:real_data_application} of the main paper and, here, in Appendix \ref{app:additional_numerical_results}.
\end{appendix}

\clearpage

\bibliographystyle{apalike}
\bibliography{./bib/biblio}

\begin{thebibliography}{}

\bibitem[Armagan and Zaretzki, 2011]{Armagan2011}
Armagan, A. and Zaretzki, R.~L. (2011).
\newblock A note on mean-field variational approximations in {B}ayesian probit
  models.
\newblock {\em Computational Statistics \& Data Analysis}, 55(1):641--643.

\bibitem[Bissiri et~al., 2016]{Bissiri2016}
Bissiri, P.~G., Holmes, C.~C., and Walker, S.~G. (2016).
\newblock A general framework for updating belief distributions.
\newblock {\em Journal of the Royal Statistical Society. Series B. Statistical
  Methodology}, 78(5):1103--1130.

\bibitem[Blackard, 1998]{Blackard1998}
Blackard, J. (1998).
\newblock {Covertype}.
\newblock UCI Machine Learning Repository.
\newblock {DOI}: https://doi.org/10.24432/C50K5N.

\bibitem[Blei et~al., 2017]{Blei2017}
Blei, D.~M., Kucukelbir, A., and McAuliffe, J.~D. (2017).
\newblock Variational inference: A review for statisticians.
\newblock {\em Journal of the American Statistical Association},
  112(518):859--877.

\bibitem[Chen, 2017]{Chen2017}
Chen, S. (2017).
\newblock {Beijing PM2.5 Data}.
\newblock UCI Machine Learning Repository.
\newblock {DOI}: https://doi.org/10.24432/C5JS49.

\bibitem[Duan et~al., 2018]{Duan2018}
Duan, L.~L., Johndrow, J.~E., and Dunson, D.~B. (2018).
\newblock Scaling up data augmentation {MCMC} via calibration.
\newblock {\em Journal of Machine Learning Research}, 19:1--34.

\bibitem[Duchi et~al., 2011]{Duchi2011}
Duchi, J., Hazan, E., and Singer, Y. (2011).
\newblock Adaptive subgradient methods for online learning and stochastic
  optimization.
\newblock {\em Journal of Machine Learning Research}, 12:2121--2159.

\bibitem[Durante and Rigon, 2019]{Durante2019}
Durante, D. and Rigon, T. (2019).
\newblock Conditionally conjugate mean-field variational {B}ayes for logistic
  models.
\newblock {\em Statistical Science. A Review Journal of the Institute of
  Mathematical Statistics}, 34(3):472--485.

\bibitem[Fasano et~al., 2022]{Fasano2022}
Fasano, A., Durante, D., and Zanella, G. (2022).
\newblock Scalable and accurate variational {B}ayes for high-dimensional binary
  regression models.
\newblock {\em Biometrika}, 109(4):901--919.

\bibitem[Fasiolo et~al., 2021]{Fasiolo2021}
Fasiolo, M., Wood, S.~N., Zaffran, M., Nedellec, R., and Goude, Y. (2021).
\newblock Fast calibrated additive quantile regression.
\newblock {\em Journal of the American Statistical Association},
  116(535):1402--1412.

\bibitem[Gaillard et~al., 2016]{Gaillard2016}
Gaillard, P., Goude, Y., and Nedellec, R. (2016).
\newblock Additive models and robust aggregation for gefcom2014 probabilistic
  electric load and electricity price forecasting.
\newblock {\em International Journal of forecasting}, 32(3):1038--1050.

\bibitem[Gelman et~al., 2013]{Gelman2013}
Gelman, A., Carlin, J.~B., Stern, H.~S., Dunson, D.~B., Vehtari, A., and Rubin,
  D.~B. (2013).
\newblock {\em Bayesian data analysis, third edition}.
\newblock CRC press.

\bibitem[Geraci, 2019]{Geraci2019}
Geraci, M. (2019).
\newblock Modelling and estimation of nonlinear quantile regression with
  clustered data.
\newblock {\em Computational Statistics \& Data Analysis}, 136:30--46.

\bibitem[Geraci and Bottai, 2007]{Geraci2007}
Geraci, M. and Bottai, M. (2007).
\newblock Quantile regression for longitudinal data using the asymmetric
  {L}aplace distribution.
\newblock {\em Biostatistics}, 8(1):140--154.

\bibitem[Geraci and Bottai, 2014]{Geraci2014a}
Geraci, M. and Bottai, M. (2014).
\newblock Linear quantile mixed models.
\newblock {\em Statistics and Computing}, 24(3):461--479.

\bibitem[Goplerud, 2022]{Goplerud2022}
Goplerud, M. (2022).
\newblock Fast and accurate estimation of non-nested binomial hierarchical
  models using variational inference.
\newblock {\em Bayesian Analysis}, 17(2):623--650.

\bibitem[Goplerud et~al., 2025]{Goplerud2024}
Goplerud, M., Papaspiliopoulos, O., and Zanella, G. (2025).
\newblock Partially factorized variational inference for high-dimensional mixed
  models.
\newblock {\em Biometrika}, 112(2):asae067.

\bibitem[Hoffman and Blei, 2015]{Hoffman2015}
Hoffman, M. and Blei, D. (2015).
\newblock Stochastic structured variational inference.
\newblock In {\em Artificial Intelligence and Statistics}, pages 361--369.
  PMLR.

\bibitem[Hoffman et~al., 2013]{Hoffman2013}
Hoffman, M.~D., Blei, D.~M., Wang, C., and Paisley, J. (2013).
\newblock Stochastic variational inference.
\newblock {\em Journal of Machine Learning Research}, 14:1303--1347.

\bibitem[Hong et~al., 2016]{Hong2016}
Hong, T., Pinson, P., Fan, S., Zareipour, H., Troccoli, A., and Hyndman, R.~J.
  (2016).
\newblock Probabilistic energy forecasting: Global energy forecasting
  competition 2014 and beyond.
\newblock {\em International Journal of Forecasting}, 32(3):896--913.

\bibitem[Hui et~al., 2019]{Hui2019}
Hui, F. K.~C., You, C., Shang, H.~L., and M\"{u}ller, S. (2019).
\newblock Semiparametric regression using variational approximations.
\newblock {\em Journal of the American Statistical Association},
  114(528):1765--1777.

\bibitem[Johndrow et~al., 2019]{Johndrow2019}
Johndrow, J.~E., Smith, A., Pillai, N., and Dunson, D.~B. (2019).
\newblock {MCMC} for imbalanced categorical data.
\newblock {\em Journal of the American Statistical Association},
  114(527):1394--1403.

\bibitem[Khan and Lin, 2017]{Khan2017a}
Khan, M.~E. and Lin, W. (2017).
\newblock Conjugate-computation variational inference: Converting variational
  inference in non-conjugate models to inferences in conjugate models.
\newblock In {\em Artificial Intelligence and Statistics}, pages 878--887.
  PMLR.

\bibitem[Khan and Nielsen, 2018]{Khan2018b}
Khan, M.~E. and Nielsen, D. (2018).
\newblock Fast yet simple natural-gradient descent for variational inference in
  complex models.
\newblock In {\em 2018 International Symposium on Information Theory and Its
  Applications (ISITA)}, pages 31--35. IEEE.

\bibitem[Knowles and Minka, 2011]{Knowles2011}
Knowles, D. and Minka, T. (2011).
\newblock Non-conjugate variational message passing for multinomial and binary
  regression.
\newblock {\em Advances in Neural Information Processing Systems},
  24:1701--1709.

\bibitem[Koenker, 2005]{Koenker2005}
Koenker, R. (2005).
\newblock {\em Quantile regression}, volume~38 of {\em Econometric Society
  Monographs}.
\newblock Cambridge University Press, Cambridge.

\bibitem[Koenker and Bassett, 1978]{Koenker1978}
Koenker, R. and Bassett, Jr., G. (1978).
\newblock Regression quantiles.
\newblock {\em Econometrica}, 46(1):33--50.

\bibitem[Kozumi and Kobayashi, 2011]{Kozumi2011}
Kozumi, H. and Kobayashi, G. (2011).
\newblock Gibbs sampling methods for {B}ayesian quantile regression.
\newblock {\em Journal of Statistical Computation and Simulation},
  81(11):1565--1578.

\bibitem[Kucukelbir et~al., 2017]{Kucukelbir2017}
Kucukelbir, A., Tran, D., Ranganath, R., Gelman, A., and Blei, D.~M. (2017).
\newblock Automatic differentiation variational inference.
\newblock {\em J. Mach. Learn. Res.}, 18(14):1--45.

\bibitem[Kullback and Leibler, 1951]{Kullback1951}
Kullback, S. and Leibler, R.~A. (1951).
\newblock On information and sufficiency.
\newblock {\em Annals of Mathematical Statistics}, 22:79--86.

\bibitem[Lange, 2013]{Lange2013}
Lange, K. (2013).
\newblock {\em Optimization}.
\newblock Springer, New York, second edition.

\bibitem[Lee and Mangasarian, 2001]{Lee2001}
Lee, Y.-J. and Mangasarian, O.~L. (2001).
\newblock {SSVM}: a smooth support vector machine for classification.
\newblock {\em Computational Optimization and Applications. An International
  Journal}, 20(1):5--22.

\bibitem[Lewandowski et~al., 2010]{Lewandowski2010}
Lewandowski, A., Liu, C., and Wiel, S.~V. (2010).
\newblock Parameter expansion and efficient inference.
\newblock {\em Statistical Science}, 25(4):533 -- 544.

\bibitem[Luts and Ormerod, 2014]{Luts2014a}
Luts, J. and Ormerod, J.~T. (2014).
\newblock Mean field variational {B}ayesian inference for support vector
  machine classification.
\newblock {\em Computational Statistics \& Data Analysis}, 73:163--176.

\bibitem[Luts and Wand, 2015]{Luts2015}
Luts, J. and Wand, M.~P. (2015).
\newblock Variational inference for count response semiparametric regression.
\newblock {\em Bayesian Analysis}, 10(4):991--1023.

\bibitem[McCullagh and Nelder, 1989]{McCullagh1989}
McCullagh, P. and Nelder, J.~A. (1989).
\newblock {\em Generalized linear models. Second edition}.
\newblock Chapman \& Hall, London.

\bibitem[McCulloch et~al., 2008]{McCulloch2008}
McCulloch, C.~E., Searle, S.~R., and Neuhaus, J.~M. (2008).
\newblock {\em Generalized, linear, and mixed models. Second edition}.
\newblock Wiley Series in Probability and Statistics. John Wiley \& Sons, Inc.,
  Hoboken, NJ.

\bibitem[McLean and Wand, 2019]{McLean2019}
McLean, M.~W. and Wand, M.~P. (2019).
\newblock Variational message passing for elaborate response regression models.
\newblock {\em Bayesian Analysis}, 14(2):371--398.

\bibitem[Menictas et~al., 2023]{Menictas2023}
Menictas, M., Credico, G.~D., and Wand, M.~P. (2023).
\newblock Streamlined variational inference for linear mixed models with
  crossed random effects.
\newblock {\em Journal of Computational and Graphical Statistics},
  32(1):99--115.

\bibitem[Menictas and Wand, 2015]{Menictas2015}
Menictas, M. and Wand, M.~P. (2015).
\newblock Variational inference for heteroscedastic semiparametric regression.
\newblock {\em Australian \& New Zealand Journal of Statistics},
  57(1):119--138.

\bibitem[Minka, 2013]{Minka2013}
Minka, T.~P. (2013).
\newblock Expectation propagation for approximate {B}ayesian inference.
\newblock {\em arXiv preprint arXiv:1301.2294}.

\bibitem[Moro et~al., 2012]{Moro2012}
Moro, S., Rita, P., and Cortez, P. (2012).
\newblock Bank marketing.
\newblock UCI Machine Learning Repository.
\newblock {DOI}: https://doi.org/10.24432/C5K306.

\bibitem[Neville et~al., 2014]{Neville2014}
Neville, S.~E., Ormerod, J.~T., and Wand, M.~P. (2014).
\newblock Mean field variational {B}ayes for continuous sparse signal
  shrinkage: pitfalls and remedies.
\newblock {\em Electronic Journal of Statistics}, 8(1):1113--1151.

\bibitem[Nolan et~al., 2020]{Nolan2020b}
Nolan, T.~H., Menictas, M., and Wand, M.~P. (2020).
\newblock Streamlined computing for variational inference with higher level
  random effects.
\newblock {\em Journal of Machine Learning Research}, 21:Paper No. 157, 62.

\bibitem[Nolan and Wand, 2017]{Nolan2017}
Nolan, T.~H. and Wand, M.~P. (2017).
\newblock Accurate logistic variational message passing: algebraic and
  numerical details.
\newblock {\em Stat}, 6:102--112.

\bibitem[Nolan and Wand, 2020]{Nolan2020a}
Nolan, T.~H. and Wand, M.~P. (2020).
\newblock Streamlined solutions to multilevel sparse matrix problems.
\newblock {\em The ANZIAM Journal}, 62(1):18--41.

\bibitem[Ormerod and Wand, 2010]{Ormerod2010}
Ormerod, J.~T. and Wand, M.~P. (2010).
\newblock Explaining variational approximations.
\newblock {\em The American Statistician}, 64(2):140--153.

\bibitem[Ormerod and Wand, 2012]{Ormerod2012}
Ormerod, J.~T. and Wand, M.~P. (2012).
\newblock Gaussian variational approximate inference for generalized linear
  mixed models.
\newblock {\em Journal of Computational and Graphical Statistics}, 21(1):2--17.

\bibitem[Polson and Scott, 2011]{Polson2011}
Polson, N.~G. and Scott, S.~L. (2011).
\newblock Data augmentation for support vector machines.
\newblock {\em Bayesian Analysis}, 6(1):1--23.

\bibitem[Ranganath et~al., 2014]{Ranganath2014}
Ranganath, R., Gerrish, S., and Blei, D. (2014).
\newblock Black box variational inference.
\newblock In {\em Artificial intelligence and statistics}, pages 814--822.
  PMLR.

\bibitem[Richardson, 1997]{Richardson1997}
Richardson, A.~M. (1997).
\newblock Bounded influence estimation in the mixed linear model.
\newblock {\em Journal of the American Statistical Association},
  92(437):154--161.

\bibitem[Robbins and Monro, 1951]{Robbins1951}
Robbins, H. and Monro, S. (1951).
\newblock A stochastic approximation method.
\newblock {\em Annals of Mathematical Statistics}, 22:400--407.

\bibitem[Rohde and Wand, 2016]{Rohde2016}
Rohde, D. and Wand, M.~P. (2016).
\newblock Semiparametric mean field variational {B}ayes: general principles and
  numerical issues.
\newblock {\em Journal of Machine Learning Research}, (172):1--47.

\bibitem[Rue et~al., 2009]{Rue2009}
Rue, H., Martino, S., and Chopin, N. (2009).
\newblock Approximate {B}ayesian inference for latent {G}aussian models by
  using integrated nested {L}aplace approximations.
\newblock {\em Journal of the Royal Statistical Society. Series B. Statistical
  Methodology}, 71(2):319--392.

\bibitem[Staines and Barber, 2012]{Staines2012}
Staines, J. and Barber, D. (2012).
\newblock Variational optimization.
\newblock {\em arXiv preprint arXiv:1212.4507}.

\bibitem[Tan, 2017]{Tan2017}
Tan, L. S.~L. (2017).
\newblock Stochastic variational inference for large-scale discrete choice
  models using adaptive batch sizes.
\newblock {\em Statistics and Computing}, 27(1):237--257.

\bibitem[Tan, 2025]{Tan2025}
Tan, L. S.~L. (2025).
\newblock Analytic natural gradient updates for {C}holesky factor in {G}aussian
  variational approximation.
\newblock {\em Journal of the Royal Statistical Society. Series B. Statistical
  Methodology}.

\bibitem[Tan and Nott, 2013]{Tan2013}
Tan, L. S.~L. and Nott, D.~J. (2013).
\newblock Variational inference for generalized linear mixed models using
  partially noncentered parametrizations.
\newblock {\em Statistical Science. A Review Journal of the Institute of
  Mathematical Statistics}, 28(2):168--188.

\bibitem[Tan and Nott, 2014]{Tan2014}
Tan, L. S.~L. and Nott, D.~J. (2014).
\newblock A stochastic variational framework for fitting and diagnosing
  generalized linear mixed models.
\newblock {\em Bayesian Analysis}, 9(4):963--1004.

\bibitem[Tan and Nott, 2018]{Tan2018}
Tan, L. S.~L. and Nott, D.~J. (2018).
\newblock Gaussian variational approximation with sparse precision matrices.
\newblock {\em Statistics and Computing}, 28(2):259--275.

\bibitem[Tierney and Kadane, 1986]{Tierney1986}
Tierney, L. and Kadane, J.~B. (1986).
\newblock Accurate approximations for posterior moments and marginal densities.
\newblock {\em Journal of the American Statistical Association},
  81(393):82--86.

\bibitem[Vapnik, 1998]{Vapnik1998}
Vapnik, V.~N. (1998).
\newblock {\em Statistical learning theory}.
\newblock John Wiley \& Sons, Inc., New York.

\bibitem[Wand, 2014]{Wand2014}
Wand, M.~P. (2014).
\newblock Fully simplified multivariate normal updates in non-conjugate
  variational message passing.
\newblock {\em Journal of Machine Learning Research}, 15:1351--1369.

\bibitem[Wand, 2017]{Wand2017}
Wand, M.~P. (2017).
\newblock Fast approximate inference for arbitrarily large semiparametric
  regression models via message passing.
\newblock {\em Journal of the American Statistical Association},
  112(517):137--156.

\bibitem[Wand and Ormerod, 2008]{Wand2008}
Wand, M.~P. and Ormerod, J.~T. (2008).
\newblock On semiparametric regression with {O}'{S}ullivan penalized splines.
\newblock {\em Australian \& New Zealand Journal of Statistics},
  50(2):179--198.

\bibitem[Wand et~al., 2011]{Wand2011a}
Wand, M.~P., Ormerod, J.~T., Padoan, S.~A., and Fr{\"{u}}hrwirth, R. (2011).
\newblock Mean field variational {B}ayes for elaborate distributions.
\newblock {\em Bayesian Analysis}, 6(4):847--900.

\bibitem[Yue and Rue, 2011]{Yue2011}
Yue, Y.~R. and Rue, H. (2011).
\newblock Bayesian inference for additive mixed quantile regression models.
\newblock {\em Computational Statistics \& Data Analysis}, 55(1):84--96.

\bibitem[Zeiler, 2012]{Zeiler2012}
Zeiler, M.~D. (2012).
\newblock Adadelta: an adaptive learning rate method.
\newblock {\em arXiv preprint arXiv:1212.5701}.

\end{thebibliography}

\end{document}